\documentclass[12pt]{article}
\usepackage{bbm}
\usepackage{color}
\usepackage{authblk}
\usepackage{latexsym}
\usepackage{epsfig,amssymb,euscript, mathrsfs,cite}
\usepackage{amsmath}
\usepackage{slashed}
\usepackage{cancel}
\usepackage{verbatim}
\usepackage{mathrsfs} 
\usepackage[normalem]{ulem}
\usepackage{enumerate}
\definecolor{MyDarkBlue}{rgb}{0.15,0.15,0.45}
\usepackage[linktocpage=true]{hyperref}
\hypersetup{
colorlinks=true,
citecolor=MyDarkBlue,
linkcolor=MyDarkBlue,
urlcolor=MyDarkBlue,
pdfauthor={},
pdftitle={},
pdfsubject={hep-th}
}
\newsavebox{\ns}
\newsavebox{\dbrane}
\newsavebox{\dbshort}

\def\be{\begin{equation}}
\def\ee{\end{equation}}
\def\bea{\begin{eqnarray}}
\def\eea{\end{eqnarray}}

\newcommand{\nn}{\nonumber\\}

\newcommand{\hook}{\mathbin{\rule[.2ex]{.4em}{.03em}\rule[.2ex]{.03em}{.9ex}}}
\newcommand\cA{\mathcal{A}}
\newcommand\cB{\mathcal{B}}
\newcommand\cC{\mathcal{C}}
\newcommand\cK{\mathcal{K}}
\newcommand\cG{\mathcal{G}}
\newcommand\cF{\mathcal{F}}
\newcommand\tcF{\widetilde{\mathcal{F}}}
\newcommand\cL{\mathcal{L}}
\newcommand\cM{\mathcal{M}}
\newcommand\cN{\mathcal{N}}
\newcommand\cO{\mathcal{O}}
\newcommand\cR{\mathcal{R}}
\newcommand\cV{\mathcal{V}}
\newcommand\cW{\mathcal{W}}
\newcommand\cI{\mathcal{I}}
\newcommand{\e}{\mathrm{e}}
\newcommand{\ii}{\mathrm{i}}
\newcommand{\dd}{\mathrm{d}}
\newcommand{\zbar}{\bar{z}}
\newcommand{\ubar}{\overline{u}}
\newcommand{\tz}{\tilde{z}}
\newcommand{\tepsilon}{\tilde{\epsilon}}

\newcommand{\barX}{\overline{X}}
\newcommand{\tX}{\widetilde{X}}
\newcommand{\barL}{\overline{L}}
\newcommand{\tL}{\widetilde{L}}

\newcommand{\tW}{\widetilde{W}}
\newcommand{\tcalF}{\widetilde{\mathcal{F}}}
\newcommand{\tC}{\widetilde{C}}

\newcommand{\smallspace}{\mskip2mu}
\newcommand{\tinyspace}{\mskip1mu}
\newcommand{\xinew}{\zeta}
\newcommand{\Pnew}{\mathbb{P}}
\newcommand{\R}{\mathbb{R}}
\newcommand{\Z}{\mathbb{Z}}
\newcommand{\Fgrav}{F_{\rm grav}}

\newcommand\diff{\mathrm{d}}

\newcommand{\E}{\text{E}}

\newcommand{\ex}{\mathrm{e}}
\newcommand{\vol}{\mathrm{vol}}
\newcommand{\Vol}{\mathrm{Vol}}

\newlength{\sswidth}

\newcommand{\abs}[1]{\left| #1 \right|}

\numberwithin{equation}{section}       % equation numbers in each section

\newcommand{\cutoff}{\delta}
\newcommand{\cutoffspace}{M_\delta}
\newcommand{\cutoffbdry}{\partial M_\delta}
\newcommand{\cutoffdue}{\delta}

\newcommand{\newphi}{\omega}

\begin{document}

\begin{titlepage}

%\begin{flushright}
%Imperial/TP/2023/JG/**
%\end{flushright}

\vskip 1cm

\begin{center}

%\today

{\Large \bf Equivariant localization for $D=4$ \\ 
\vskip 0.2cm
 gauged supergravity}

\vskip 1cm
{Pietro Benetti Genolini$^{\mathrm{a}}$, Jerome P. Gauntlett$^{\mathrm{b}}$,
Yusheng Jiao$^{\mathrm{b}}$,\\
\vskip 0.1cm
 Alice L\"uscher$^{\mathrm{c}}$  and James Sparks$^{\mathrm{c}}$}

\vskip 1cm

${}^{\mathrm{a}}$\textit{D\'epartment de Physique Th\'eorique, Universit\'e de Gen\`eve,\\
24 quai Ernest-Ansermet, 1211 Gen\`eve, Suisse\\}

\vskip 0.2cm

${}^{\mathrm{b}}$\textit{Blackett Laboratory, Imperial College, \\
Prince Consort Rd., London, SW7 2AZ, U.K.\\}

\vskip 0.2cm

${}^{\mathrm{c}}$\textit{Mathematical Institute, University of Oxford,\\
Andrew Wiles Building, Radcliffe Observatory Quarter,\\
Woodstock Road, Oxford, OX2 6GG, U.K.\\}

\vskip 0.2 cm

\end{center}

\vskip 0.5 cm

\begin{abstract}
\noindent  
We consider supersymmetric solutions of $D=4$, $\mathcal{N}=2$ Euclidean gauged supergravity coupled to
an arbitrary number of vector multiplets. Such solutions admit an R-symmetry Killing vector, $\xi$, constructed as a bilinear in the 
Killing spinor. The Killing spinor bilinears can also be used to construct polyforms that are equivariantly closed
under the action of the equivariant exterior derivative $\dd_\xi=\dd-\xi\hook$\,. This allows one to compute various flux integrals and the on-shell action
using localization, without solving any supergravity equations, just assuming the supersymmetric solutions exist. 
The flux integrals allow one to obtain important UV-IR relations, relating fixed point data in the bulk to data on the
asymptotic AdS boundary, allowing one to express the gravitational free energy in terms of boundary SCFT data.
We illustrate the formalism with a number of examples, including classes of solutions which are unlikely to 
be ever constructed in closed form. 

\end{abstract}

\end{titlepage}

\pagestyle{plain}
\setcounter{page}{1}
\newcounter{bean}
\baselineskip18pt

\tableofcontents

\newpage

\section{Introduction}

Supersymmetric solutions of supergravity theories play a pivotal role in many aspects of string and M-theory. 
In a recent development 
it has been shown that supersymmetric solutions with an R-symmetry Killing vector have a set of canonically defined equivariantly closed forms \cite{BenettiGenolini:2023kxp}. Moreover, the integrals of these forms can be used to compute various physical observables without solving any partial differential equations, instead just  
inputting some topological information and assuming the solution exists.

By definition a supersymmetric solution
solves both the equations of motion and the Killing spinor equations. The R-symmetry Killing vector is constructed
as a Killing spinor bilinear, while the equivariantly closed forms are constructed using various Killing spinor bilinears
as well as the supergravity fields. Integrals of these equivariantly closed forms can then be evaluated using ``localization," 
i.e. using the Berline--Vergne--Atiyah--Bott (BVAB) fixed point formula \cite{BV:1982, Atiyah:1984px}, which expresses the integrals in terms of the fixed point set of the R-symmetry 
Killing vector. The precise details of these constructions, as well as the type of physical observables one can compute,
depend on which particular supergravity theory one is studying and which class of solutions one is considering. 
A number of different examples of have now been explored from this point of view \cite{BenettiGenolini:2023kxp,BenettiGenolini:2023yfe,BenettiGenolini:2023ndb,BenettiGenolini:2024kyy,Suh:2024asy,Couzens:2024vbn,Hristov:2024cgj,BenettiGenolini:2024xeo,Cassani:2024kjn,BenettiGenolini:2024hyd,Crisafio:2024fyc}
and in each case the formalism allows one to profitably analyze 
general classes of solutions, most of which are unlikely to be ever found in explicit form.
Equivariant localization has also been used to study supergravity solutions from other points of view in\cite{Hristov:2021qsw,Martelli:2023oqk,Colombo:2023fhu}. 

In this paper we expand upon and extend the results presented in \cite{BenettiGenolini:2024xeo}. The general setting
is Euclidean $D=4$, $\mathcal{N}=2$ gauged supergravity coupled to an arbitrary number of vector multiplets. This
includes the STU model, whose solutions uplift on a seven-sphere to give $D=11$ solutions holographically dual to ABJM theory.
By setting all the vector multiplets to zero, it also includes minimal $\mathcal{N}=2$ gauged supergravity,
whose solutions can be uplifted on a Sasaki--Einstein seven-manifold $SE_7$, as well as other spaces, which are dual to other $\mathcal{N}=2$
SCFTs in $d=3$. The equivariant formalism is also relevant in holography
even when the $D=4$ gauged supergravity does not arise from a consistent Kaluza--Klein truncation, as
we discuss in section \ref{secdisc}.

Associated with the most general supersymmetric solutions of $D=4$, $\mathcal{N}=2$  gauged supergravity, 
there are two types of equivariantly closed forms. The first and simplest is the equivariant completion of the two-form gauge field strengths,
with the zero-form component canonically expressed in terms of the scalar Killing spinor bilinears along with the
supergravity scalar fields. 
Using the BVAB theorem, these forms can be used to obtain expressions for the integrated fluxes in terms of the fixed point set, which 
is assumed to lie in the bulk of the solution (i.e. not on the boundary).
In particular this allows one to obtain important UV-IR relations which relate
IR fixed point data in the bulk of the solution to UV data associated with deformations of the boundary SCFT.
The second is the equivariant extension of a four-form whose integral
gives the bulk contribution to the Euclidean on-shell action of a given solution. Using the BVAB formula this gives a contribution
at the fixed point set plus a boundary contribution. The full on-shell action, which we refer to as the gravitational free energy, is supplemented by additional, standard boundary terms: the Gibbons--Hawking--York term, counterterms and, in order to ensure supersymmetry, a Legendre transform associated with implementing an alternative quantization for
half of the scalar fields.  A key result is that when evaluated on a supersymmetric solution, 
these boundary terms exactly cancel the boundary contribution coming from the bulk integral, leading 
to an expression for the gravitational free energy which is just expressed in terms of the fixed point set in the bulk.
From a holographic perspective it is more natural to express the gravitational free energy in terms of UV data, and
this can be achieved using the UV-IR relations. 

We illustrate the formalism by discussing various examples. These include cases
associated with known supergravity solutions as well as cases where the solutions
are unlikely to be found in explicit form. 
Amongst the results we shall obtain are proofs of some conjectures for the free energy of mass-deformed three-dimensional holographic $\cN=2$ SCFTs in the large $N$ limit as well as extensions thereof. Notably, we provide a general proof of a conjecture \cite{Zan:2021ftf} for the free energy on $S^3$ (see \eqref{eq:Rsymmetry_sigmaI_S3_Simpler2}, \eqref{FP2}), and extend a conjecture of \cite{Hosseini:2016tor} for the free energy on the direct product of a circle and a Riemann surface to a much broader family of Seifert manifolds (see \eqref{eq:Fgrav_TaubBolt_Saddles}). 
Another result is that we obtain the on-shell action for the supersymmetric monodromy defect solutions of ABJM theory
studied in \cite{Arav:2024wyg}, which were also used to compute supersymmetric R\'enyi entropies.
In addition, we discuss solutions with a ``cigar'' $\R^2$ factor that appear as supersymmetric non-extremal deformations of Wick-rotated Lorentzian black holes and in each case we show that the on-shell action is largely independent of the deformation parameters, as expected from their holographic interpretation as gravity duals to supersymmetric indices. 
In a companion paper \cite{BenettiGenolini:2024hyd} we further
analyze toric cases, which have two linearly independent Killing vectors, allowing us
to develop the formalism further and to consider additional classes of examples.

We will also make a connection between equivariantly closed forms
and an expression that arises in computing the on-shell action using the ``Wald formalism" for conserved quantities \cite{Wald:1993nt} in settings 
without supersymmetry.\footnote{Another approach for computing on-shell Euclidean actions, generalizing
\cite{Gibbons:1976ue}, using an approach based on Kaluza--Klein reduction appears in \cite{Taylor:1998fc}.} In particular, in a theory of gravity in dimension $k$, associated with a top-form $\Phi_k$ which integrates to give the on-shell action,
one can construct a $\Phi_{k-2}$ form satisfying $\xi\hook \Phi_k=\dd \Phi_{k-2}$, where $\xi$ is a Killing vector. In two-derivative theories of gravity we have 
$\Phi_{k-2}=-\frac{1}{2}*\dd\xi^\flat+\dots$, where $\xi^\flat$ is the one-form dual to $\xi$ and the dots are additional contributions that depend
on the matter content of the theory. If the theory includes gauge fields these additional contributions depend on a choice of gauge.
This structure allows one to compute the on-shell action for Euclidean solutions which have a single co-dimension two fixed point surface and the normal bundle is trivial, as for example in certain Wick-rotated black hole solutions. However, if the normal bundle is not trivial and/or the fixed point set includes components with higher co-dimension, this structure is not sufficient to evaluate the on-shell action. On the other hand, if one were able to construct a continuation
$\Phi=\Phi_k+\Phi_{k-2}+\dots +\Phi_{0/1}$ that is equivariantly closed, and with all forms globally defined and gauge invariant, then for general solutions one can deploy
the BVAB fixed point theorem to evaluate the on-shell action. 
In the $D=4$ supergravity context this is precisely what we construct for supersymmetric solutions, with all components of
the polyform $\Phi$ globally defined and canonically expressed in terms of Killing spinor bilinears and supergravity fields.

The plan of the rest of the paper is as follows.
In section \ref{sec2} we introduce the 
Euclidean $\mathcal{N}=2$, $D=4$ gauged supergravity theory that we study and explain how the equivariantly closed forms can be constructed using
the spinor bilinears. In section \ref{evalactsec} we explain how the on-shell action can be expressed in
terms of the fixed point data using the BVAB formula. In $D=4$ the fixed point set consists of isolated points (``nuts") or two-dimensional
submanifolds (``bolts") and including both cases leads to our final expression for the gravitational free energy given
in \eqref{Fgrav}.
In section \ref{sec:uvirrels} we use equivariant localization to compute various flux integrals. By considering
flux integrals associated with certain non-compact two-dimensional submanifolds which intersect the boundary,
we can  relate UV boundary data to IR fixed point data.
We consider various different examples to illustrate the formalism in section \ref{sec:examples} and \ref{sec:six}. Some of these were briefly considered
in \cite{BenettiGenolini:2024xeo}, and, as well as expanding 
on some of the details, we also treat 
new examples. 
Solutions with orbifold singularities, including 
spindle solutions and gravitational fillings of Seifert three-manifolds, 
are considered in 
section~\ref{sec:orbifoldexamples},
and we conclude with a discussion in section~\ref{secdisc}. Various material is collected in six appendices.
In appendix \ref{localnosusy} we make some general comments on equivariant forms as well as make a connection with the Wald formalism
for general theories of gravity
(not assuming supersymmetry).
In appendix \ref{app:Euclidean_KSE} we review some details of the derivation of the supersymmetry equations in Euclidean signature. In appendix \ref{app:bilinears} we have collected some technical details regarding the spinor
bilinears. A brief discussion regarding how one can obtain results for minimal gauged supergravity is presented in appendix
\ref{mingaugedsugraapp}, which allows one to connect with the results of \cite{BenettiGenolini:2019jdz}
which were obtained using different methods. Appendix \ref{app:holrenn} derives the important result regarding the cancellation of all boundary
terms in computing the on-shell action, which leads to our final result just expressed in terms of fixed point data.
Finally, in appendix \ref{subsubsec:FP} we discuss the explicit supergravity solutions of \cite{Freedman:2013oja}
that are dual to real mass deformations of ABJM theory on $S^3$ and analyze the Killing spinor bilinears.

\section{\texorpdfstring{$\mathcal{N}=2$}{N=2} gauged supergravity in \texorpdfstring{$D=4$}{D=4}}\label{sec2}

We will study a $D=4$ Euclidean theory that is obtained by a Wick rotation of Lorentzian $\mathcal{N}=2$ gauged supergravity. 
We show that supersymmetric solutions of this theory admit equivariantly closed forms constructed from spinor bilinears.
Later, in section \ref{evalactsec}, we will impose additional reality constraints on the Euclidean theory to further utilise the formalism. 

\subsection{Lorentzian theory}

In Lorentzian signature, the bosonic fields consist of the metric $g_{\mu\nu}$, Abelian gauge fields $A^I$ with $I=0,\dots, n$, and complex scalars $z^i$ with $i=1, \dots, n$. The scalars parametrize a special K\"ahler manifold with K\"ahler potential $\cK$ and Hermitian metric $\cG_{i\overline{j}}\equiv \partial_i\partial_{\bar j}\cK$, where we denote 
$\partial_i \equiv\partial_{z^i}, \partial_{\bar{j}} \equiv\partial_{\bar{z}^{\bar{j}}}$. The special K\"ahler manifold is also the base of a symplectic bundle with covariantly holomorphic sections
\begin{equation}
	\begin{pmatrix}
		L^I \\ M_I
	\end{pmatrix} \equiv \e^{\cK/2} \begin{pmatrix}
		X^I \\ \cF_I
	\end{pmatrix} \, ,
\end{equation}
where $X^I=X^I(z)$ are holomorphic. We also have
the symplectic constraint
\begin{equation}\label{symconlorentzian}
	\e^{\cK}( X^I \overline{\cF}_I - \cF_I \overline{X}^I ) = \ii \,.
\end{equation}
We will assume there exists a prepotential $\cF=\cF(X^I)$, which is a holomorphic function of $X^I$, homogeneous of degree 2, such that
\begin{equation}
	\cF_I = \frac{\partial \cF}{\partial X^I} \, .
\end{equation}
We then define the kinetic matrix
\begin{equation}\label{cNdef}
	\cN_{IJ} = \overline{\cF}_{IJ} + \ii \frac{N_{IK}X^K N_{JL}X^L}{N_{NM} X^N X^M} \equiv \cR_{IJ} + \ii \tinyspace \cI_{IJ} \, ,
\end{equation}
where $\cF_{IJ} = \partial_I \partial_J \cF$ and $N_{IJ} = 2\smallspace \text{Im} \smallspace \cF_{IJ}$, and we have defined $\partial_I\equiv \partial_{X^I}$, $\cR_{IJ} \equiv \text{Re} \smallspace \cN_{IJ}$ and $\cI_{IJ} \equiv \text{Im}\smallspace \cN_{IJ}$. We also observe that
\begin{align}\label{calNrel}
\cN_{IJ}X^J=\cF_I\,.
\end{align}

The action for the bosonic fields, in the conventions of \cite{BenettiGenolini:2024kyy}, is given by
\begin{align}
\label{eq:Lorentzian_I4}
S&= \frac{1}{16\pi G_4} \int \left[ R - 2 \cG_{i\overline{j}} \partial^\mu z^i \partial_\mu \zbar^{\overline{j}} -  \cV (z, \zbar)  \right.  \nn
& \qquad \qquad \qquad \qquad \left. + \frac{1}{4} \cI_{IJ} F^I_{\mu\nu} F^{J\mu\nu} - \frac{1}{8}  \cR_{IJ} \varepsilon^{\mu\nu\rho\sigma} F^I_{\mu\nu} F^J_{\rho\sigma} \right] \vol_4 \, .
\end{align}
Here $G_4$ is the Newton constant, and the scalar potential is
\begin{equation}
\label{eq:Lorentzian_ScalarPotential}
	\cV \equiv \xinew_I \xinew_J \e^\cK \left( \cG^{i\overline{j}} \nabla_i X^I \nabla_{\overline{j}} \overline{X}^J - 3 X^I \overline{X}^J \right) \, , 
\end{equation}
where $\xinew_I\in\mathbb{R}$ are the Fayet--Iliopoulos gauging parameters and we have defined
	\begin{equation}\label{nablaXdef}
	\nabla_i X^I \equiv ( \partial_i + \partial_i \cK) X^I \, .
\end{equation}
Note that
\begin{align}\label{nablaLdef}
\nabla_i L^I &\equiv \left( \partial_i + \frac{1}{2}\partial_i \cK \right) L^I=\e^{\cK/2}\nabla_i X^I \,, \nn
\nabla_{\bar i} {L}^I &\equiv \left( \partial_{\bar i} - \frac{1}{2}\partial_{
\bar i} \cK \right) {L}^I=0 \, .
\end{align}
Under a K\"ahler transformation $\cK\to \cK+h(z)+\bar h(\bar z)$ we have $X^I\to \ex^{-h}X^I$ and $L^I\to \ex^{-\frac{h}{2}+\frac{\bar h}{2}}L^I$.
We can also introduce the holomorphic and real superpotentials 
\begin{equation}\label{wsuperpots}
	W \equiv \xinew_I X^I \, , \qquad \cW \equiv - \sqrt{2}\smallspace \e^{\cK/2} |{W}| \, ,
\end{equation}
in terms of which the scalar potential \eqref{eq:Lorentzian_ScalarPotential} takes the form
\begin{equation}\label{curlyvpot}
\begin{split}
	\cV &= \e^{\cK} \left( \cG^{i\overline{j}} \nabla_i W \nabla_{\overline{j}} \overline{W} - 3 W \overline{W} \right) 
	= 2  \cG^{i\overline{j}} \partial_i \cW \partial_{\overline{j}} \cW - \frac{3}{2} \cW^2 \, .
\end{split}
\end{equation}

A solution is supersymmetric if there are solutions to the Killing spinor equations 
\begin{align}
\label{eq:Lorentzian_4d_GravitinoVariation}
	0 &= \nabla_\mu \epsilon +  \frac{\ii}{2} \cA_\mu  \Gamma_5 \epsilon - \frac{\ii}{2} A^R \epsilon + \frac{1}{2\sqrt{2}} \Gamma_\mu \e^{\cK/2} \left( W \tinyspace\Pnew_- +\overline{W}\tinyspace \Pnew_+ \right) \epsilon \nn
	& \ \ \ - \frac{\ii}{4\sqrt{2}} \cI_{IJ} F^{J}_{\nu\rho} \Gamma^{\nu\rho} \Gamma_\mu \left( L^I \tinyspace \Pnew_- + \overline{L}^I \tinyspace \Pnew_+ \right) \epsilon \, , \nn
	0 &= \frac{\ii}{2\sqrt{2}} \cI_{IJ} F^{J}_{\mu\nu} \Gamma^{\mu\nu} \left[ {\cG}^{\bar{i} {j}}  \nabla_j L^I \tinyspace  \Pnew_- + \cG^{{i} \overline{j}}  \nabla_{\bar j} \overline{L}^I  \tinyspace \Pnew_+ \right] \epsilon + \Gamma^\mu \left( \partial_\mu z^i \tinyspace\Pnew_- + \partial_\mu \overline{z}^{\bar{i}} \tinyspace \Pnew_+ \right) \epsilon \nn
& \ \ \ - \frac{1}{\sqrt{2}} { \e^{\cK/2} }\left[ \cG^{\bar{i} {j}}  \nabla_j W \tinyspace \Pnew_- + \cG^{{i} \overline{j}}  
\nabla_{\bar j}\overline{W}^I\tinyspace \Pnew_+  \right] \epsilon \, ,
\end{align}
where $\epsilon$ is a Dirac spinor, $\Gamma_\mu$ generate Cliff$(1,3)$, $\Gamma_5 \equiv \ii \Gamma_{0123}$, $\Pnew_\pm \equiv \frac{1}{2}(1 \pm \Gamma_5 )$. In addition the R-symmetry gauge field is defined by the following linear combination of the
gauge fields 
\begin{align}\label{rsymgf}
A^{R}\equiv  \frac{1}{2} \xinew_I A_\mu^I\,,
\end{align}
while the composite K\"ahler connection $\cA$ is defined by
\begin{equation}
\label{eq:Lorentzian_KahlerHodgeConnection}
	\mathcal{A}_\mu = - \frac{\ii}{2}\left(\partial_i \mathcal{K} \partial_\mu z^i - \partial_{\bar{i}} \mathcal{K} \partial_\mu \overline{z}^{\bar{i}}\right) \, .
\end{equation}

After specifying the prepotential $\cF$ and the gaugings $\zeta_I$, there are still two ambiguities that change the parametrization of the solutions. 
First, we  have to fix the sections $X^I(z)$ (see e.g. \cite{Cabo-Bizet:2017xdr} for various choices). 
Second, within the K\"ahler transformations mentioned earlier are those with constant pure imaginary parameter, $h=\ii\alpha$, which act like an Abelian group, with $\cA$ the corresponding connection:\footnote{In the superconformal construction of supergravity, this symmetry appears among the K\"ahler transformations after the gauge-fixing of the chiral $U(1)$ in the $SU(2)\times U(1)$ R-symmetry group of the four-dimensional superconformal group (see for instance \cite{Freedman:2012zz} for a review). }
\begin{align}\label{phaseabelianrots}
\epsilon \to \e^{- \ii \frac{\alpha}{2} \Gamma_5} \epsilon,\quad L^I \to \e^{- \ii \alpha} L^I, \quad \cA_\mu \to \cA_\mu + \partial_\mu \alpha\,.
\end{align} 
This allows us to rotate the phases of the sections in \eqref{eq:Lorentzian_4d_GravitinoVariation}, as explained in e.g. \cite{BenettiGenolini:2024kyy}. In 
\eqref{eq:Lorentzian_4d_GravitinoVariation} we have made the same choice that was used in section 5.3 of \cite{BenettiGenolini:2024kyy}, to facilitate later comparison with the STU model.

\subsection{Euclidean theory}
\label{subsec:EuclideanTheory}

We now define the Euclidean theory using the procedure explained in \cite{Freedman:2013oja}, 
with some further details included in appendix \ref{app:Euclidean_KSE}. 
We take
\begin{align}
{ \ t \to - \ii x^4} \, , \qquad { A_t \to \ii A_{x^4} } \, , \qquad { S \to \ii I } \, .
\end{align}
In general one should double all degrees of freedom and consider complex metric and gauge fields
as well as take $z^i$ and $\zbar^{\bar i}$ to be independent scalar fields. In order to highlight the latter, we write $\zbar^{\bar i} \to \tz^{\tilde i}$, as well
as $\barX^I\to \tX^I$, $\barL^I\to \tL^I$ and $\overline{W}\to\tW$. 
In particular we denote 
$\overline{\mathcal{F}(X^I)}\to \tcalF (\tX^I)$, where for example
the quantity $N_{IJ}$ in  \eqref{cNdef} becomes $N_{IJ}=2\tinyspace \text{Im}\smallspace \mathcal{F}_{IJ}
\to -\ii (\partial_{X^I}\partial_{X^J}\mathcal{F} - \partial_{\tX^I}\partial_{\tX^J}\tcalF)$, 
with similar expressions for other quantities. 
In principle in Euclidean signature one might consider taking 
$\mathcal{F}$, $\tcalF$ to be independent functions, 
but we shall later impose $\tcalF(\tX^I) = -\mathcal{F}(\tX^I)$, 
which is satisfied by the STU model, for example, which has
$\mathcal{F}(X^I) = -2\ii \sqrt{X^0 X^1 X^2 X^3}$.
At a later point we will restrict to real metric and gauge fields, but we continue with the general case for now.

Following \cite{Freedman:2013oja} we obtain the Wick rotated action 
\begin{align}\label{themainaction}
I	&= - \frac{1}{16\pi G_4} \int \Big[ \left( R - 2 \cG_{i\tilde{j}} \partial^\mu z^i \partial_\mu \tz^{\tilde{j}} - \cV (z, \tz)  \right)\vol_4 \nn
	& \qquad \qquad \qquad \qquad + \frac{1}{2} {\cI}_{IJ} F^I \wedge * F^{J} - \frac{\ii}{2} {\cR}_{IJ} F^I \wedge F^J \Big] \, ,
\end{align}
where $\cG_{i\tilde{j}}\equiv \partial_i\partial_{\tilde j}\cK$ and
\begin{align}
\label{eq:EuclideanScalarPotential}
	\cV &= \e^{\cK} \left( \cG^{i\tilde{j}} \nabla_i W \nabla_{\tilde{j}} \tW - 3 W \tW \right) = 2  \cG^{i\tilde{j}} \partial_i \cW \partial_{\tilde{j}} \cW - \frac{3}{2} \cW^2 \, ,
\end{align}
with
\begin{equation}
\label{eq:Euclidean_Superpotentials}
	W \equiv \xinew_I X^I \, , \qquad \tW \equiv \xinew_I \tX^I \, , \qquad\cW \equiv - \sqrt{2}\tinyspace \e^{\cK/2} \sqrt{W \tW} \, .
\end{equation}
The associated equations of motion are given by 
\begin{align}
\label{eq:4d_N2_EinsteinEOM}
	R_{\mu\nu} &= - \frac{1}{2}\mathcal{I}_{IJ}\Big(F^I_{\mu\rho}{F^{J}}^{\ \,  \rho}_\nu - \frac{1}{4} g_{\mu\nu} F^I_{\rho\sigma} F^{J\rho\sigma}\Big) + 2 \cG_{i \tilde{j}} \partial_\mu z^i \partial_\nu \tz^{\tilde{j}} + \frac{1}{2} g_{\mu\nu} \mathcal{V} \, , \nn
	\nabla^\mu ( \cG_{i\tilde{j}} \nabla_\mu \tz^{\tilde{j}}) &= - \frac{1}{8} \partial_i \cI_{IJ} F^I_{\mu\nu}F^{J\mu\nu} + \frac{\ii}{8} \partial_i \cR_{IJ} F^I_{\mu\nu} (* F^J)^{\mu\nu} + \partial_i \cG_{k \tilde{j}} \partial^\mu z^k \partial_\mu \tz^{\tilde{j}} + \frac{1}{2} \partial_i \cV \, ,\nn
	\nabla^\mu ( \cG_{\tilde{i}j} \nabla_\mu z^{j}) &= - \frac{1}{8} \partial_{\tilde{i}} \cI_{IJ} F^I_{\mu\nu}F^{J\mu\nu} + \frac{\ii}{8} \partial_{\tilde{i}} \cR_{IJ} F^I_{\mu\nu} (* F^J)^{\mu\nu} + \partial_{\tilde{i}} \cG_{\tilde{k} j} \partial^\mu \tz^{\tilde{k}} \partial_\mu z^{j} + \frac{1}{2} \partial_{\tilde{i}} \cV \nn
		0 &= \dd \left( \cI_{IJ} * F^J - \ii \cR_{IJ}F^J \right) \, .
\end{align}
For later use we notice that if we take the trace of the Einstein equation we obtain an expression for the 
on-shell action given by 
\begin{align}\label{POSactdef}
I_{\mathrm{OS}}	&=\frac{\pi}{2G_4} \frac{1}{(2\pi)^2} {\int} \Phi_4\,,
\end{align}
where $ \Phi_4$ is the four-form given by  
\begin{align}\label{PHIfourdef}
\Phi_4\equiv - \frac{1}{2}\cV\smallspace \vol_4   - \frac{1}{4} {\cI}_{IJ} F^I \wedge * F^{J} + \frac{\ii}{4}  {\cR}_{IJ} F^I \wedge F^J  \, .
\end{align}

The supersymmetry transformations are parametrized by two Dirac spinors $\epsilon$ and $\tilde \epsilon$ and the corresponding Killing spinor equations are given by 
\begin{align}
\label{eq:Euclidean_KSE_epsilon}
	0 &= \nabla_\mu \epsilon + \frac{\ii}{2} \cA_\mu \gamma_5 \epsilon 
	-  \frac{\ii}{2} A_\mu^R  \epsilon +  \frac{1}{2\sqrt{2}}  \gamma_\mu  \e^{\cK/2}\left( W \tinyspace \Pnew_- + \tW\tinyspace \Pnew_+ \right)\epsilon \nn
	& \ \ \  - \frac{\ii}{4\sqrt{2}} \mathcal{I}_{IJ} F^{J}_{\nu\rho} \gamma^{\nu\rho}\gamma_\mu \left( L^I \tinyspace \Pnew_- + \tL^I\tinyspace \Pnew_+ \right) \epsilon \, ,\nn
	0 &= \frac{\ii}{2\sqrt{2}} \mathcal{I}_{IJ} F^{J}_{\nu\rho} \gamma^{\nu\rho} 
	\left( \cG^{\tilde{i} j} \nabla_j L^I \tinyspace \Pnew_- + \cG^{i\tilde{j}} \nabla_{\tilde j} \tL^I \tinyspace \Pnew_+ \right) \epsilon + \gamma^\mu \left( \partial_\mu z^i \tinyspace \Pnew_- + \partial_\mu \tz^{\tilde{i}} \tinyspace \Pnew_+ \right) \epsilon \nn
	& \ \ \  - \frac{1}{\sqrt{2}} {\e^{\cK/2}} \left( \cG^{\tilde{i} j} \nabla_j W \tinyspace \Pnew_- +  \cG^{i\tilde{j}} \nabla_{\tilde j} \tW \tinyspace \Pnew_+ \right) \epsilon \, ,
\end{align}
and 
\begin{align}
\label{eq:Euclidean_KSE_tepsilon}
	0 &= \nabla_\mu \tepsilon + \frac{\ii}{2} \cA_\mu \gamma_5 \tepsilon 
	+  \frac{\ii}{2} A_\mu^R  \tepsilon + \frac{1}{2\sqrt{2}}  \gamma_\mu \e^{\cK/2} \left( W \tinyspace \Pnew_- + \tW\tinyspace \Pnew_+ \right) \tepsilon \nn
	& \ \ \ + \frac{\ii}{4\sqrt{2}} \mathcal{I}_{IJ} F^{J}_{\nu\rho} \gamma^{\nu\rho}\gamma_\mu \left( L^I \tinyspace \Pnew_- + \tL^I \tinyspace \Pnew_+ \right) \tepsilon \, ,\nn
	0 &= - \frac{\ii}{2\sqrt{2}} \mathcal{I}_{IJ} F^{J}_{\nu\rho} \gamma^{\nu\rho} 
	\left( \cG^{\tilde{i} j} \nabla_j L^I \tinyspace \Pnew_- + \cG^{i\tilde{j}}  \nabla_{\tilde j} \tL^I \tinyspace \Pnew_+ \right) \tepsilon + \gamma^\mu \left( \partial_\mu z^i \tinyspace \Pnew_- + \partial_\mu \tz^{\tilde{i}} \tinyspace \Pnew_+ \right) \tepsilon \nn
	& \ \ \  -  \frac{1}{\sqrt{2}} {\e^{\cK/2}} \left( \cG^{\tilde{i} j} \nabla_j W \tinyspace \Pnew_- + \cG^{i\tilde{j}} \nabla_{\tilde j} \tW \tinyspace \Pnew_+ \right) \tepsilon \, .
\end{align}
In these expressions $\gamma_\mu$,\ which are taken to be Hermitian, 
generate Cliff$(4)$ with $\gamma_5 \equiv \gamma_{1234}$ and
now $\Pnew_\pm \equiv \frac{1}{2}(1 \pm \gamma_5 )$. 
A derivation of these equations from the Lorentzian Killing spinor equations \eqref{eq:Lorentzian_4d_GravitinoVariation} is provided in appendix \ref{app:Euclidean_KSE}.

Corresponding to the doubling of the bosonic degrees of freedom, we note that we should generally take $\epsilon$ and $\tepsilon$ to be independent Dirac spinors.
However, starting in section \ref{realitycondssec},  
we will study a sub-class of solutions of the Euclidean theory where we impose that the bosonic fields (metric, gauge fields and scalars $z^i,\tilde z^i$) are all real.
In addition we will also impose $\tilde \epsilon=\epsilon^c$, with $\epsilon^c$ defined in section \ref{bilinearsubsec},
so that the supersymmetry is parametrised by a single independent Dirac spinor. That is,
the conditions \eqref{eq:Euclidean_KSE_epsilon} and \eqref{eq:Euclidean_KSE_tepsilon} are equivalent with these restrictions.

\subsection{Bilinears}\label{bilinearsubsec}
For an arbitrary spinor $\lambda$ of the Euclidean theory we define the Majorana conjugate via 
$\bar\lambda=\lambda^T\mathcal{C}$, where $\mathcal{C}$ is the Euclidean charge conjugation matrix, which is a unitary matrix satisfying
\begin{equation}
	\mathcal{C} = - \mathcal{C}^T \, , \qquad \gamma_\mu^T = \mathcal{C} \gamma_\mu \mathcal{C}^{-1} \, , \qquad \gamma_\mu^* = \mathcal{C} \gamma_\mu \mathcal{C}^{-1} \, . 
\end{equation}
We can then construct bilinears from the Killing spinors $\epsilon$, $\tepsilon$ via:
\begin{equation}
\label{eq:Bilinears}
	S \equiv \overline{\tepsilon} \epsilon \, , \quad P \equiv \overline{\tepsilon} \gamma_5 \epsilon \, , \quad 
	\xi^\flat \equiv - \ii \overline{\tepsilon} \gamma_{(1)} \gamma_5 \epsilon \, , \quad 
	K \equiv \overline{\tepsilon} \gamma_{(1)} \epsilon \, , \quad 
		U\equiv \ii \overline{\tepsilon} \gamma_{(2)} \epsilon \, ,
\end{equation}
where $S,P$ are scalars, $\xi^\flat, K$ are one-forms and $U$ is a two-form. In general these are all complex quantities.
Notice that in the special case when $\tepsilon=\epsilon^c \equiv - \mathcal{C}^{-1} \epsilon^*$, which we 
restrict to in section \ref{realitycondssec}, we then have
$\overline{\tepsilon}=\epsilon^\dagger$ and 
all the bilinears defined in \eqref{eq:Bilinears} are real (see appendix \ref{app:Euclidean_KSE}).

The Killing spinor $\epsilon$ is necessarily non-vanishing \cite{Ferrero:2021etw}, and is
globally well-defined. It is charged with respect to the R-symmetry gauge field $A^R$ given in \eqref{rsymgf}.\footnote{It is also charged under the K\"ahler gauge connection 
$\mathcal{A}$ in \eqref{eq:Lorentzian_KahlerHodgeConnection}, 
and when the scalars $z^i$, $\tz^{\tilde{i}}$ are global 
functions on $M$ (which we assume), $\mathcal{A}$ is then a global one-form 
on $M$.}
Correspondingly, the bilinears in \eqref{eq:Bilinears}, which 
are uncharged under the R-symmetry, are all globally defined 
forms on $M$.

These bilinears satisfy various algebraic and differential conditions, some of which we discuss in appendix \ref{app:bilinears}. 
Of most significance is that the vector field $\xi$, dual to $\xi^\flat$, is a Killing vector. 
It is immediate from the equations 
in appendix \ref{app:bilinears} that $\mathcal{L}_\xi z^\alpha = 0 
=\mathcal{L}_\xi \tz^{\tilde{\alpha}}$, while 
$\mathcal{L}_\xi F^I=0$ is established in the next 
subsection.
 We also have the important equation
\begin{align}\label{imptbilinearrel}
\diff\xi^\flat = -\sqrt{2}\xinew_I \big(L^I\tinyspace U_{[+]}-\tL^I\tinyspace U_{[-]}\big)-\sqrt{2}\tinyspace \mathcal{I}_{IJ}\big(C^I\tinyspace F^J_{[-]}-\tC^I \tinyspace F^J_{[+]}\big)\, .
\end{align}
Here we have introduced the notation
\begin{align}
U_{[\pm]} \equiv \frac{1}{2}(U\pm *U)\, ,
\end{align}
for the self-dual/anti-self-dual parts of a two-form $U$. 
We have also defined the following combinations of scalar fields and scalar spinor bilinears,
\begin{align}\label{Cdef}
C^I\equiv L^I(S-P)\, , \quad \widetilde{C}^I\equiv \tL^I(S+P)\, .
\end{align}

It is illuminating to identify combinations of the spinor bilinears that transform linearly under the symmetry \eqref{phaseabelianrots}.
If we make the charge assignments that $L^I$ has charge $-1$ and $\tL^I$ has charge $+1$ we find that
bilinears with well-defined charges are given by: 
\begin{align}\label{u1chgeassigns}
	\text{charge 0:} &\quad \xi^\flat \, , \ K \, , \nn
	\text{charge $-1$:} &\quad S + P \, , \ U_{[-]} \, , \nn
	\text{charge $+1$:} &\quad S - P \, , \ U_{[+]} \, .
\end{align}
In particular we see that $C^I$ and $\widetilde{C}^I$ are invariant (i.e. charge $0$). It is useful to note that
\begin{align}
\label{eq:DerivativeVanishingRCharge}
	\dd C^I&=\dd [ L^I (S - P)]= \nabla_i L^I (S - P) \dd z^i + L^I (\dd-\ii\cA)(S - P) \, , \nn
\dd \tC^I&=	\dd [ \tilde{L}^I (S + P)] = \nabla_{\tilde i} \tL^I (S + P) \dd \tz^{\tilde{i}} + \tilde{L}^I (\dd+\ii\cA)(S + P)\,.
\end{align}

\subsection{Equivariantly closed forms}

Using the bilinears introduced in the previous subsection we can construct some polyforms that
are equivariantly closed for supersymmetric solutions\footnote{{A discussion
without assuming supersymmetry is presented in appendix \ref{localnosusy}.}} with respect to the equivariant 
exterior derivative
\begin{align}
\diff_\xi \equiv \diff - \xi \hook\, .
\end{align}
This satisfies $\diff_\xi^2 = -\mathcal{L}_\xi$, and is thus 
nilpotent when acting on polyforms that are invariant under 
$\mathcal{L}_\xi$, the Lie derivative with respect to $\xi$. 

Associated with the Abelian field strengths we define 
\begin{equation}\label{PhiF}
\Phi_{(F)}^{I}\equiv F^I+\Phi_{0 }^I \,,
\end{equation}
where 
\begin{align}\label{defphioiflux}
\Phi_{0 }^I \equiv \sqrt{2}\big(C^I-\tC^I\big)\,,
\end{align}
and recall that $C^I$, $\tC^I$ were defined in \eqref{Cdef}. 
Using the equations in appendix \ref{app:bilinears} 
one can show that $\diff_\xi \Phi_{(F)}^{I}=0$. Generically $\Phi_{(F)}^{I}$ is a complex polyform, but note that with the reality conditions imposed in
\eqref{eq:Reality_Condition}, below, it is real. 
Since $F^I$ is closed by the Bianchi identity we have $\mathcal{L}_\xi F^I = 
\diff(\xi \hook F^I) = 0$, where the latter follows immediately from 
$\Phi_{(F)}^I$ being equivariantly closed. 
Notice that $\Phi_{(F)}^{I}$ in \eqref{PhiF}, \eqref{defphioiflux} is gauge-invariant and globally defined.

If we choose a gauge with $\mathcal{L}_\xi A^I=0$, 
notice that equivariant closure of \eqref{PhiF} 
implies
\begin{align}\label{gaugehookxi}
\xi \hook A^I = -\Phi_{ 0 }^I + c^I  \,,
\end{align}
where $c^I$ are {\it a priori} arbitrary integration constants. 
In particular via a further gauge transformation one can 
set $c^I=0$ if desired.  We can define the \emph{supersymmetric gauge} to be the slightly weaker one for which
$\zeta_I c^I=0$. From \eqref{rsymgf} this is equivalent to the R-symmetry gauge field satisfying
\begin{align}\label{susygaugecond}
\xi \hook A^R_{\mathrm{SUSY}} =-\frac{1}{2}\zeta_I\Phi_{ 0 }^I = -\frac{1}{\sqrt{2}}\zeta_I\big(C^I-\tC^I\big)\, .
\end{align}
At a given fixed point, where $\xi =0$, a regular gauge field would satisfy $\xi \hook A^I=0$. 
So when $\zeta_I\Phi_{ 0 }^I\ne 0$ at a fixed point, which is true in all known examples,
the R-symmetry gauge field $A^R_{\mathrm{SUSY}}$ is \emph{not} in a regular 
gauge at that point. We will not use the supersymmetric gauge in the sequel.

There is also an equivariant completion of the four-form $\Phi_4$ associated with the 
on-shell action in \eqref{POSactdef}. Specifically we find that
\begin{equation}\label{phiecform}
    \Phi=\Phi_4+\Phi_2+\Phi_0\, ,
\end{equation}
is equivariantly closed, $\diff_\xi \Phi=0$, where 
\begin{align}\label{Phidef}
\Phi_4&=-\frac{1}{2}\mathcal{V}\smallspace \vol_{4}-\frac{1}{4}\mathcal{I}_{IJ}F^I\wedge*F^J+\frac{\ii}{4}\mathcal{R}_{IJ}F^I\wedge F^J\, ,\nn
\Phi_2
&=\frac{1}{\sqrt{2}} \xinew_I(L^I \tinyspace U_{[+]} + \tL^I\tinyspace U_{[-]} )- \frac{1}{\sqrt{2}}\mathcal{I}_{IJ}\Big(C^I\tinyspace F^{J}_{[+]}+\tC^I \tinyspace F^{J}_{[-]}\big) +\frac{\ii}{\sqrt{2}}\mathcal{R}_{IJ}F^J\big(C^I-\tC^I\big)\, ,\nn
\Phi_0&=\ii \tinyspace \ex^{\mathcal{K}}\Big[{\mathcal{F}}(X )(S-P)^2+\tcalF(\tX)(S+P)^2\nonumber\\ 
& \qquad\quad 
- \frac{1}{2} (\partial_I \mathcal{F}(X)\tX^I +\partial_I \tcalF(\tX)X^I)(S^2-P^2)\Big]\,.
\end{align}
Since $\mathcal{F}$, $\widetilde{\mathcal{F}}$ are homogeneous of degree two, we can also write 
\begin{equation}\label{Phi0}
\Phi_0=\ii\Big[{\mathcal{F}}(C)+\tcalF(\widetilde{C})- \frac{1}{2} \partial_I \mathcal{F}(C)\widetilde{C}^I- \frac{1}{2} \partial_I\tcalF(\widetilde{C})C^I\Big]\,,
\end{equation}
where  the partial derivative notation $\partial_I \mathcal{F}$ means
derivative with respect to the argument of $\mathcal{F}$ (which 
is $X^I$ in \eqref{Phidef}, but $C^I$ in \eqref{Phi0}).
It is also interesting to point out that 
using \eqref{imptbilinearrel} and \eqref{defphioiflux} in \eqref{Phidef} we can write $\Phi_2$ in the form
\begin{align}\label{phi2xitextexp}
\Phi_2=-\frac{1}{2}*\dd\xi^\flat-\frac{1}{2}(\cI_{IJ}*F^I- \ii \cR_{IJ} F^I)\Phi_0^J\,.
\end{align}
Notice that the polyform $\Phi$ is gauge-invariant and globally defined.
Recalling \eqref{u1chgeassigns} we also observe that the equivariantly closed forms $\Phi_{(F)}^{I}$ and $\Phi$ are all 
invariant under the Abelian transformation \eqref{phaseabelianrots}, as expected.

Finally, in appendix \ref{localnosusy} we show that for a general $D=4$ theory of gravity coupled to gauge fields and scalars, without assuming supersymmetry, 
one can always construct equivariantly closed forms $\Phi_{(F)}^{I}$ and $\Phi$ for solutions to the equations of motion.
In general, these are local expressions and not canonically defined. 
There is a canonical expression for $\Phi_2$ which is in accord with the expression \eqref{phi2xitextexp}, 
but in general $\Phi_0^I$ and hence $\Phi_2$ will not be gauge-invariant.
More generally, here we have shown that for supersymmetric solutions of
the gauged supergravity theory we are considering, canonical expressions for 
$\Phi^I_0$ and $\Phi_2$, $\Phi_0$ 
can be obtained using spinor bilinears
as given in \eqref{defphioiflux} and \eqref{Phidef}, respectively. Notice that we could trivially modify the
expressions for $\Phi_0$ and $\Phi_0^I$ by adding constants 
$\Phi_0^I \to \Phi_0^I + c_0^I$, $\Phi_0  \to \Phi_0 + c_0$ while maintaining equivariant closure 
and similarly we could also modify $\Phi_2 \to \Phi_2 + \zeta_2$ where $\zeta_2$ is a closed globally defined 
basic, two-form (recall a basic form $\alpha$ satisfies $\xi \hook \alpha = 0$ and $\cL_\xi \alpha = 0$).\footnote{Ambiguities in the equivariant forms of the above form are well known and are associated with differing contributions from the fixed point sets arising from Stokes' theorem. For an example
in the compact context, consider the equivariant integration giving the volume of $\mathbb{CP}^2$ written as an $S^3$ fibred over an interval. As described in detail in \cite{BenettiGenolini:2023ndb}, the Reeb vector of $S^3$ has an isolated fixed point and a two-dimensional bolt, and the ambiguity in the volume polyform can be fixed so that the volume is given entirely by each of these or by a combination of the two.}
An interesting and important feature of the canonical expressions \eqref{defphioiflux} and \eqref{Phidef} is that they have the property
that the total contribution to the gravitational free energy from the conformal boundary vanishes, as we show in appendix \ref{app:holrenn}.\footnote{For minimal gauged supergravity, without using localization it was shown by a direct computation that 
the boundary contribution to the on-shell action vanishes if one works in a supersymmetric gauge \cite{BenettiGenolini:2019jdz} and further sets, in their notation, $c_\varphi = c_\sigma = 0$. 
In effect these constants are analogous to the constants $c_0^0$ and $c_0$. The freedom to shift by $\zeta_2$ was not discussed in \cite{BenettiGenolini:2019jdz}, which does lead to boundary contributions to the on-shell action (see \eqref{boundariescancel}).} 

\section{Evaluation of the action}
\label{evalactsec}

In this section we explain how we can use the BVAB theorem to evaluate the on-shell action via localization, as well as obtain important relations between UV and IR quantities. 

\subsection{Reality conditions}\label{realitycondssec}
To do this we will now study a sub-class of solutions of the Euclidean theory where we impose
that the metric and gauge fields are both real, and further just consider solutions which preserve supersymmetry with $\tilde \epsilon=\epsilon^c$: consistency of these assumptions implies additional restrictions which we summarise together as
\begin{equation}
\label{eq:Reality_Condition}
\begin{aligned}
g_{\mu\nu} &\in \mathbb{R}\,,  &\qquad A_\mu^I &\in \mathbb{R}\,, &\qquad z^i, \tz^{\tilde{i}} &\in \mathbb{R}\, , \\
\cA &\in \ii \mathbb{R} \, , &\qquad L^I, \tL^I &\in  \mathbb{R} \, , &\qquad \cG^{\tilde{i} j}, \cG^{i\tilde{j}} &\in \mathbb{R} \, , \\
\tilde\epsilon &=\epsilon^c\equiv - \cC^{-1} \epsilon^* &\Rightarrow \bar{\tilde\epsilon} &=\epsilon^\dagger\,.
\end{aligned}
\end{equation}
The bilinears \eqref{eq:Bilinears} are all real.
Crucially, we will not impose $\tilde z^{\tilde i}=\overline{(z^i)}$, 
which would be $z^i= \tz^{\tilde{i}}$, given \eqref{eq:Reality_Condition}. 
Solutions that satisfy this additional condition will also be solutions of the Lorentzian theory, after Wick rotation. 
For more details, see appendix \ref{app:Euclidean_KSE}.

We also notice that these reality conditions are only compatible with the symplectic constraint given
(in the Lorentzian theory) in \eqref{symconlorentzian}, provided that the prepotential is purely imaginary, and so
in the sequel we shall take
\begin{align}\label{tfcond}
\mathcal{F}\in\ii\R\qquad\Rightarrow \qquad \tcF(\tX)=- \cF(\tX)\,.
\end{align}
For cases where one does not impose the reality condition, it is straightforward to rewrite
the expressions in terms of $\tcF(\tX)$, if needed.

\subsection{Using the BVAB theorem}
\label{subsec:Using_BVAB}

We assume our Euclidean theory admits a
 supersymmetric  $AdS_4$ (i.e. $H^4$) vacuum with vanishing gauge fields and
constant scalar fields satisfying $\tilde z^{\tilde i}=z^i$, and hence it is also a supersymmetric vacuum 
of the Lorentzian theory.\footnote{There could be more than one such
$AdS_4$ solution, and one can treat each one similarly.} This solution defines what we refer to as the vacuum of the dual SCFT$_3$.
We are interested in general Euclidean solutions on a four-dimensional manifold $M$ which asymptotically approach this $AdS_4$ solution 
with conformal boundary $\partial M$. 
As usual, depending on the precise asymptotic behaviour near the boundary, 
such solutions will be dual to deformations of the SCFT, and allow for an arbitrary boundary metric, non-trivial gauge fields on the boundary as well as sources for the operators dual to the bulk scalar fields. 
To evaluate the gravitational free energy of such solutions
we need to evaluate the bulk on-shell Euclidean action \eqref{POSactdef}, and then supplement it with suitable boundary terms to implement holographic
renormalization. We first focus on \eqref{POSactdef}.

Consider, momentarily, equivariantly closed forms on a manifold $M$ in general even dimension $2n$. Denote the fixed point set where $\|\xi\|^2=0$ by
$M_0$, 
which we assume is in the interior of $M$, so $M_0\cap \partial M=\emptyset$. Since $\dd_\xi \xi^\flat=\dd\xi^\flat-\|\xi\|^2$, on $M\setminus M_0$ we can define an inverse $(\dd_\xi \xi^\flat)^{-1}=-\frac{1}{\|\xi\|^2}(1-\frac{\dd\xi^\flat}{\|\xi\|^2})^{-1}=-\frac{1}{\|\xi\|^2}\sum_{k=0}^n(\frac{\dd\xi^\flat}{\|\xi\|^2})^k$, which is equivariantly closed. Using this we can define a polyform $\Phi_\xi=\xi^\flat\wedge
(\dd_\xi \xi^\flat)^{-1}$ satisfying $\dd_\xi\Phi_\xi=1$. Thus, for an arbitrary equivariantly closed form $\Phi$ on $M$, on $M\setminus M_0$ we can write
$\Phi=\Phi\smallspace \dd_\xi\Phi_\xi=\dd_\xi(\Phi\smallspace \Phi_\xi)$ and hence we see it is equivariantly exact on $M\setminus M_0$. Since the top form of an equivariantly exact form
is exact, this shows that the integral of $\Phi$ on $M$, if closed, will only receive contributions from the fixed points 
(as given by the BVAB formula \cite{BV:1982, Atiyah:1984px}). It also shows that if $M$
has a boundary that does not include any fixed points we can obtain an explicit expression for an additional boundary contribution to the integral using Stokes' theorem and the expression for $(\dd_\xi \xi^\flat)^{-1}$ \cite{Couzens:2024vbn}.

We now return to our $D=4$ theory. We want to evaluate the bulk on-shell action \eqref{POSactdef} on a manifold with boundary and we will assume
that the fixed point set does not include the boundary. 
To simplify the exposition we shall assume that $M$ is a manifold, and in particular that the fixed point set does not have orbifold
singularities.\footnote{We will discuss examples where this is relaxed in section \ref{sec:orbifoldexamples}.} 
We write $I_{\mathrm{OS}}=I^{\mathrm{FP}}_{\mathrm{OS}}+I^{\partial M}_{\mathrm{OS}}$
where $I^{\mathrm{FP}}_{\mathrm{OS}}$ is the contribution from the fixed point set in the bulk given by the BVAB formula and
$I^{\partial M}_{\mathrm{OS}}$ is the boundary contribution. Explicitly, on the one hand we have
\begin{align}\label{I4dexpand}
I^{\mathrm{FP}}_{\mathrm{OS}} = \frac{\pi}{2G_4}\bigg\{\sum_{\substack{\mathrm{nuts}}}\frac{ \Phi_0}{b_1b_2} +
\sum_{\substack{\mathrm{bolts}} }\int_\Sigma 
\frac{\Phi_2}{2\pi b}-\frac{\Phi_0\smallspace c_1(L)}{b^2}\bigg\}\, .
\end{align}
Here the sums in \eqref{I4dexpand} are over connected components of the fixed point set.
The $b_i$ and $b$ are weights of the linear action of $\xi$ on the normal bundle to the (connected components of the) fixed point set, 
and
$c_1(L)$ is the first Chern class of the normal bundle to the bolt surface $\Sigma\subset M$. 
The boundary contribution, on the other hand, is given by 
\begin{align}\label{bdybvabterms}
I^{\partial M}_{\mathrm{OS}}&= - \frac{\pi}{2G_4} \frac{1}{(2\pi)^2} \int_{\partial M}\frac{1}{\|\xi\|^2}\xi^\flat\wedge \left(\Phi_2+\Phi_0\frac{\dd\xi^\flat}{\|\xi\|^2}\right) \nn
	&= - \frac{\pi}{2G_4} \frac{1}{(2\pi)^2} \int_{\partial M}\eta\wedge (\Phi_2+\Phi_0\smallspace \dd\eta)\,,
\end{align}
where on $M\setminus M_0$ we have defined $\eta\equiv \frac{1}{\|\xi\|^2}\xi^\flat$, satisfying $\xi\hook\eta =1$.

We next discuss the boundary terms associated with holographic renormalization. As usual these include a Gibbons--Hawking--York term, counterterms to cancel divergences, as well as additional finite boundary terms that implement a supersymmetric renormalization scheme.
The supersymmetric $AdS_4$ vacuum solution with constant scalars and vanishing gauge fields is dual to a $d=3$, $\mathcal{N}=2$ SCFT. In this set-up
each vector multiplet is dual to a conserved current multiplet; the massless gauge fields are dual to the currents with conformal scaling dimension $\Delta=2$ while the scalars are necessarily dual to operators of dimensions $\Delta=1, 2$ to fill out the multiplet. The latter corresponds to bulk scalar fields with $m^2=-2/\ell^2$, where $\ell$ is the radius of the $AdS_4$ vacuum, and in order to obtain scalar operators with $\Delta=1$ we need to implement alternate quantization on half of the scalar fields
via boundary terms that implement a Legendre transform \cite{Klebanov:1999tb}. That this mass spectrum around a given supersymmetric $AdS_4$ vacuum
is indeed realised was proven in appendix A of \cite{Zan:2021ftf}.

Remarkably, it turns out that the boundary terms associated with holographic renormalization exactly cancel the boundary terms arising in \eqref{bdybvabterms}, provided we use the canonical expressions  \eqref{Phidef} for the lower degree forms in the polyform. 
The computation, which extends that of \cite{BenettiGenolini:2019jdz}, who obtained a similar result for minimal gauged supergravity, is
lengthy and presented in appendix \ref{app:holrenn}.
The total on-shell action is thus given by the fixed point contribution in the bulk, $I^{\mathrm{FP}}_{\mathrm{OS}}$,
given by \eqref{I4dexpand}, which we now evaluate. We begin by establishing some preliminary results.

\subsection{Preliminaries}
\label{subsec:Preliminaries}

\subsubsection{Canonical  frame}

With $\tepsilon = \epsilon^c$, the bilinears in \eqref{eq:Bilinears} are real. 
As in \cite{BenettiGenolini:2019jdz} we can construct\footnote{More details can be found in appendix \ref{app:LocalForm_SUSY_Sols}. Note that we have changed the labelling compared to \cite{BenettiGenolini:2019jdz}: $\E^{3,4}_{\rm there} = \E^{4,3}_{\rm here}$.\label{footnote:ChangeOrdering}} a canonical orthonormal frame $\E^a$ and write
the bilinears as
\begin{equation}
\label{eq:IdentityStructure}
	P = S \cos\theta \, , \quad K = - S \sin\theta \, \E^3 \, , \quad \xi^\flat = S \sin\theta \, \E^4 \, , \quad U = - S( \E^{12} - \cos\theta \, \E^{34} ) \, ,
\end{equation}
where we have introduced the angle $\theta$ for convenience.
The volume element is given by $\vol_4=\E^{1234}$, so we have
\begin{equation}
\label{eq:Upm}
	U_{[\pm]}  = - ( 1 \mp \cos\theta) \frac{S}{2} \left( \E^{12} \pm \E^{34} \right) \, .
\end{equation}
As we discuss further below, it is important to note that at a fixed point where $\xi=0$ this frame breaks down. 
Thus, strictly speaking it is a frame on $M\setminus M_0$, rather than on $M$.

\subsubsection{Charge of the Killing spinor}
\label{subsubsec:ChargeKillingSpinor}

We now compute the Lie derivative of the Killing spinor $\epsilon$ with respect to $\xi$, defined as
\begin{equation}
	\cL_\xi \epsilon \equiv \xi^\mu \nabla_\mu \epsilon + \frac{1}{8} \dd\xi^\flat_{\mu\nu} \gamma^{\mu\nu} \epsilon \, .
\end{equation}
To do so, we find it convenient to multiply this equation with $\overline{\tepsilon}$ and consider 
$\overline{\tepsilon} \cL_\xi \epsilon = \xi^\mu \, \overline{\tepsilon} \nabla_\mu \epsilon - \frac{\ii}{8} \dd\xi^\flat_{\mu\nu} U^{\mu\nu}$.
To proceed, we substitute for $\nabla_\mu \epsilon$ using the Killing spinor equation \eqref{eq:Euclidean_KSE_epsilon} and $\dd\xi^\flat$ from \eqref{imptbilinearrel}. Given the reality conditions \eqref{eq:Reality_Condition},
we then utilise the identity structure \eqref{eq:IdentityStructure} as well as $\xi \hook \cA=0$, 
to find
\begin{equation}
	\overline{\tepsilon} \cL_\xi \epsilon 
		= \frac{\ii}{4}  \zeta_I \left( \xi \hook A^I + \Phi_0^{I} \right) S \, ,
\end{equation}
where $\Phi_0^{I}$ is the lowest component of the equivariantly closed  flux polyform \eqref{PhiF}, \eqref{defphioiflux}.
We therefore deduce that if we take $\epsilon$ to have definite charge $Q$ (which we can always do \cite{Ferrero:2021etw}) then
\begin{equation}\label{liederivepeqiqep}
	\cL_\xi \epsilon ={\ii}Q \epsilon\, , \qquad Q=\frac{1}{2}  \xi \hook A^R +  \frac{1}{4}\zeta_I\Phi_0^{I} \,.
\end{equation}
Clearly the charge $Q$ of the Killing spinor depends on the choice of gauge for the R-symmetry gauge field
 $A^R\equiv \frac{1}{2}\zeta_I A^I$ in \eqref{rsymgf}. If we work in a gauge with 
$\cL_\xi A^I = 0$, from \eqref{gaugehookxi} we have $\xi \hook A^I = -\Phi_{ 0 }^I + c^I$, where $c^I$ are constants, and hence 
\begin{equation}
\label{eq:SpinorRcharge}
	Q=\frac{1}{4}  \zeta_I c^I \, .
\end{equation}
Notice that in the supersymmetric gauge \eqref{susygaugecond}, which has $\zeta_Ic^I=0$, we have $Q=0$ and the spinor
is uncharged under the action of $\xi$, and so too is the frame $\E^a$, 
$\mathcal{L}_\xi \E^a=0$.

Having established that the Killing spinor has a definite charge $Q$ under the action of $\xi$, we can now establish a useful
result at a fixed point, where $\xi=0$, by generalizing an argument from \cite{BenettiGenolini:2023ndb}. Start with $\xi^\mu\nabla_\mu\epsilon$
and substitute the expression for $\nabla_\mu\epsilon$ using the Killing spinor equation \eqref{eq:Euclidean_KSE_epsilon}. Using 
$\xi \hook \cA=0$, we then immediately find that at a fixed point $\xi^\mu\nabla_\mu\epsilon=\frac{\ii}{2}\xi^\mu  A^R_\mu$, which may be non-vanishing depending on the choice of gauge. Combining this result with \eqref{liederivepeqiqep} we deduce that on the fixed point set we have
\begin{equation}\label{fptreldxi}
\frac{1}{2} \dd\xi^\flat_{\mu\nu} \gamma^{\mu\nu} \epsilon =\ii\zeta_I\Phi_0^I\smallspace \epsilon \qquad \text{ (at $\xi=0$)}\,,
\end{equation}
and notice that this is a gauge-invariant result.

\subsubsection{Chirality of spinors at fixed point set}
\label{subsubsec:Chirality_FixedPoints}

Recall that the Killing spinor $\epsilon$ is globally well defined and non-vanishing in the bulk.
Given the reality conditions \eqref{eq:Reality_Condition} it is a globally well-defined section
of $\mathcal{S}M\otimes \mathscr{L}^{1/2}$, where 
$\mathcal{S}M$ is the spin bundle of $M$, and $\mathscr{L}$ 
is the R-symmetry line bundle on which $A^R$ is a connection. 
The bilinears in \eqref{eq:Bilinears} are all globally defined and
are uncharged under the R-symmetry.
Given the reality conditions 
\eqref{eq:Reality_Condition}, we also have that $S$ is the square of the norm of the Killing spinor, and hence $S$ is also non-vanishing.
On the other hand $\|\xi\|^2=S^2\sin^2\theta$ and so at a fixed point we necessarily have $\sin\theta=0$ and hence $\cos\theta=\pm1$. Correspondingly, at a fixed point we have $P=\pm S$ and the Killing spinor is necessarily chiral at the fixed point too: 
\begin{align}
\gamma_5\epsilon=\pm\epsilon \qquad \text{ (at $\xi=0$)}\,.
\end{align}
It is convenient to append this chirality data to the fixed points and label them as nuts$_\pm$ and bolts$_\pm$. 
From \eqref{Cdef} we then have, correlated with this chirality,
\begin{align}\label{constrscs}
(C^I,\widetilde{C}^I)|_+=(0,2 S\tL^I)|_+\, ,\qquad \text{or}\qquad
(C^I,\widetilde{C}^I)|_-=(2 SL^I,0)|_-\, ,
\end{align}
and with our reality conditions \eqref{eq:Reality_Condition}, these are real.
It will be convenient to define, at the fixed points,
\begin{align}\label{udefs}
u^I_+&\equiv \frac{\widetilde{X}^I}{\zeta_J \widetilde{X}^J}\Big|_+=\frac{\widetilde{L}^I}{\zeta_J \widetilde{L}^J}\Big|_+=\frac{\widetilde{C}^I}{\zeta_J \widetilde{C}^J}\Big|_+\,,\nn
u^I_-&\equiv \frac{{X}^I}{\zeta_J {X}^J}\Big|_-=\frac{{L}^I}{\zeta_J {L}^J}\Big|_-=\frac{{C}^I}{\zeta_J {C}^J}\Big|_-\,,
\end{align}
where we have assumed $\zeta_J {X}^J\big|_\pm\ne 0$.
Clearly we have the constraints
\begin{align}\label{constsuzeta}
\zeta_Iu^I_+=1\,,\qquad \zeta_Iu^I_-=1\,.
\end{align}
From \eqref{eq:Upm}, at the fixed point we also have
\begin{align}\label{uresults}
\text{$+$ chirality}:&\qquad U=U_{[-]}\,,\nn
\text{$-$ chirality}:&\qquad U=U_{[+]}\,.
\end{align}

As remarked earlier, at a fixed point the canonical frame $\E^a$ is no longer well-defined. Thus, an important point is that we cannot immediately use
\eqref{imptbilinearrel} and \eqref{eq:Upm} to express $\dd\xi^\flat$ in terms of this frame to obtain the weights of the Killing vector at the fixed point, and so we proceed differently below.

\subsection{Action contributions from nuts}
\label{subsec:ActionNut}

A nut is by definition an isolated fixed point. 
We remind the reader that to simplify the exposition we are assuming there are no orbifold singularities on $M$ (examples with orbifolds will be considered
in section \ref{sec:orbifoldexamples}).
Thus, near a nut we can introduce local polar coordinates $\rho_i, \varphi_i$, with $\Delta \varphi_i=2\pi$, 
that parametrize two copies of $\mathbb{R}^2\cong \mathbb{C}$ with the nut located at $\rho_1=0=\rho_2$.
The supersymmetric Killing vector at the nut can be written
\begin{align}\label{xinut}
\xi=b_1\partial_{\varphi_1}+b_2\partial_{\varphi_2}\, ,
\end{align} 
where $b_i$ are the weights of the action at the nut. Indeed,
introducing a local orthonormal frame $\bar e^a$ with $\bar e^a=(\diff\rho_1,\rho_1 \diff\varphi_1, \diff\rho_2,\rho_2\diff\varphi_2)$, at the nut we have
$\diff\xi^\flat=2b_1 \bar e^{12}+2b_2 \bar e^{34}$.
If the orbits of $\xi$ close, which does not have to be the case, then we can write $b_1/b_2=p/q$ for coprime integers $p,q$. 

Notice that in writing \eqref{xinut} we have 
made orientation choices for each $\mathbb{R}^2\cong \mathbb{C}$ factor. 
However,  for a given overall orientation (that we always
fix by supersymmetry to be $\E^{1234}$), we can reverse the orientations of both $\mathbb{R}^2$ factors (equivalently, a simultaneous
complex conjugation of both $\mathbb{C}$ factors) or interchange the two $\mathbb{R}^2$ factors, leaving the overall orientation on $M$ fixed. These transformations act on the 
weights via
\begin{align}\label{beetransfs}
(b_1,b_2)&\to (-b_1,-b_2)\,,\nn
(b_1,b_2)&\to (b_2,b_1)\,,
\end{align}
respectively, and our final answer should respect this. 
They also act on the local frame at the nut via 
$\bar e^{12}, \bar{e}^{34} \to -\bar e^{12},-\bar e^{34}$ and $\bar e^{12}\leftrightarrow \bar{e}^{34}$. 

We can also show that the chiral Killing spinor at the fixed point in this frame must satisfy
\begin{align}\label{kappacond}
\gamma^{\bar 1\bar 2}\epsilon=-\ii\kappa\epsilon\,,\qquad
\gamma^{\bar 3\bar 4}\epsilon=\pm\ii\kappa\epsilon\, ,
\end{align}
where $\kappa=\pm1$ and, moreover, provided that 
$\zeta_I\Phi^I_0\ne 0$ at the fixed point (as we are assuming\footnote{\label{assumpb1eqb2}
This assumption eliminates having $b_1=\pm b_2$ for $\pm$ chirality, respectively. It is worth highlighting, however, that there are certainly examples with
$b_1=\mp b_2$ for $\pm$ chirality.}), we have 
\begin{align}\label{wtscrels}
\text{$+$ chirality}:&\qquad b_1-b_2=-\frac{\kappa}{2}\zeta_I\Phi^I_0=\frac{\kappa}{\sqrt{2}}\zeta_I \tilde C^I\,,\nn
\text{$-$ chirality}:&\qquad b_1+b_2=-\frac{\kappa}{2}\zeta_I\Phi^I_0=-\frac{\kappa}{\sqrt{2}}\zeta_I C^I\,.
\end{align}
To show this we can use \eqref{fptreldxi} to get, at the fixed point,
$2(b_1\gamma^{\bar 1 \bar 2}+b_2\gamma^{\bar 3 \bar 4})\epsilon =\ii\zeta_I\Phi_0^I\epsilon$. Multiplying this expression by
$\gamma^{\bar 1 \bar 2}$ and using the fact that the spinor has a definite chirality at the fixed point, we immediately deduce
that the spinor is an eigenspinor of $\gamma^{\bar 1 \bar 2}$, which has eigenvalues given by plus or minus $\ii$, and the result follows
after using \eqref{defphioiflux} and \eqref{constrscs}.
An alternative derivation utilises \eqref{liederivepeqiqep}: we first note that if we choose a gauge so that the R-symmetry gauge field
is regular at the fixed point, $ \xi \hook A^R=0$, we have 
$\cL_\xi \epsilon ={\ii}Q \epsilon$ with $Q=\frac{1}{4}  \zeta_I \Phi_0^{I}$. On the other hand a spinor of definite chirality that is regular at the fixed point 
will necessarily\footnote{Consider flat 
$\R^2$ in a Cartesian frame $e^1=\diff x$, $e^2=\diff y$. In this frame a constant spinor is regular.  However, since
the Cartesian frame has charge 1 under rotations of $\partial_\varphi$ the 
two different, two-dimensional spinor chiralities will have charges $\pm 1/2$ with respect to $\cL_{\partial_\varphi}$, respectively. One can then
take a tensor product of two copies of this to obtain an analogous result for $\R^4$. This 
can then be applied locally near a nut fixed point.} 
have the property that $\cL_{\partial_{\varphi_1}} \epsilon = -\frac{\ii}{2}\kappa \epsilon$ and
$\cL_{\partial_{\varphi_2}} \epsilon = \pm\frac{\ii}{2}\kappa \epsilon$.

For a given fixed point, we can use the  transformations \eqref{beetransfs} mentioned above to fix the sign of $\kappa$.
However, it is important to emphasize that when there are multiple components of the fixed point set, each comes with its
chirality and $\kappa$, and relative signs and orientation conventions at different 
fixed points need to be determined carefully. 
In some examples studied in this paper we will see how these signs can be constrained, and in \cite{BenettiGenolini:2024hyd} we show in the context of toric examples how these signs can be dealt with in a  systematic way.

We can now obtain an expression for the contribution of an individual nut to the on-shell action in the BVAB formula \eqref{I4dexpand}.
For an individual nut$_+$, from \eqref{constrscs} we have $C^I=0$ and since $\cF$ is homogeneous of degree two, from \eqref{Phi0}
we have $\Phi_0=\ii\tcF(\widetilde C)=-\ii\cF(\widetilde C)$, where the last equality follows from \eqref{tfcond}. Thus, 
the expression \eqref{I4dexpand} can now be written as
\begin{align}\label{I4dexpandnutplus}
I_{\mathrm{OS}}^{\text{nut$_+$}} &= \frac{\pi}{2G_4}
\frac{ \Phi_0}{b_1b_2} \, 
=- \frac{\pi}{2G_4}
\frac{ \ii\cF(\tC)}{b_1b_2} \,.
\end{align}
We next use \eqref{wtscrels} to rewrite this as
\begin{align}\label{I4dexpandnutplus2}
I_{\mathrm{OS}}^{\text{nut$_+$}}
&= -\frac{\pi}{G_4}\frac{(b_1-b_2)^2 }{b_1b_2}\frac{ \ii\cF(\tC)}{(\zeta_I \tC^I)^2}\nn
&= -\frac{\pi}{G_4}\frac{(b_1-b_2)^2 }{b_1b_2} \ii\cF(u_+)\,,
\end{align}
where we used the definition of the $u^I_+$ variables in \eqref{udefs}
and that $\tcF$ is homogeneous of degree two. A similar computation
shows that the contribution for an individual nut$_-$ is given by 
\begin{align}\label{I4dexpandnutminus}
I_{\mathrm{OS}}^{\text{nut$_-$}} = \frac{\pi}{G_4}\frac{(b_1+b_2)^2 }{b_1b_2} \ii\cF(u_-)\, .
\end{align}
Notice that these expressions only depend on the ratio $b_1/b_2$ at the nut, and not the individual values of $b_1,b_2$, and are invariant under 
\eqref{beetransfs}. We also notice that the value of $\kappa\in \{\pm1\}$ does not appear in these expressions.

\subsection{Action contributions from bolts}\label{boltscomp}
A bolt is by definition a connected two-dimensional fixed point set. 
Again, to simplify the exposition we assume there are no orbifold singularities.
In this case we can introduce local coordinates $( x^1,x^2,\rho_2, \varphi_2)$, with $\Delta\varphi_2=2\pi$
and $(x^1,x^2)$ parametrizing the bolt, located at $\rho_2=0$. The supersymmetric Killing vector is given by
\begin{align}
\xi=b_2\partial_{\varphi_2}\,.
\end{align}
We can introduce a frame $\bar e^a$ with $(\bar e^1,\bar e^2)=(\dd x^1,\dd x^2)$ tangent to the bolt and $(\bar e^3,\bar e^4)=(\dd\rho_2, \rho_2\dd\varphi_2)$, so that
$b_2$ is the weight of the action of $\xi$ on the two-dimensional normal space to the bolt:
$\diff\xi^\flat=2b_2 \bar e^{34}$. 
Notice that if we take $\varphi_2\to-\varphi_2$, so $b_2\to -b_2$, 
 and also change the orientation of the bolt $\bar e^{12}\to -\bar e^{12}$ then we keep the overall orientation fixed.

As shown before, the spinor is necessarily chiral at the bolt. 
Using \eqref{fptreldxi} we immediately deduce that $\zeta_I\Phi^I_0\ne 0$ at the bolt and furthermore, 
similar to case of nuts, we can take the chiral Killing spinor at the fixed bolt$_\pm$ in this frame to satisfy
\begin{align}\label{boltprojslf}
\gamma^{\bar 1\bar 2}\epsilon=-\ii\kappa\epsilon\,,\qquad
\gamma^{\bar 3\bar 4}\epsilon=\pm\ii\kappa\epsilon\, ,
\end{align}
where $\kappa=\pm1$ with
\begin{align}\label{wtscrelsbolt}
\text{$+$ chirality}:&\qquad b_2=-\frac{\kappa}{\sqrt{2}}\zeta_I \widetilde C^I\,,\nn
\text{$-$ chirality}:&\qquad b_2=-\frac{\kappa}{\sqrt{2}}\zeta_I C^I\,,
\end{align}
using \eqref{defphioiflux} and \eqref{constrscs}.
 As with the case of nuts, we can also obtain these results using  
\eqref{liederivepeqiqep} and regularity of the spinor at the bolt.
The sign of $\kappa$ can be fixed at a given bolt by appropriate choice of conventions, but when there are multiple bolts and/or nuts, relative signs 
can be important.

Next, using \eqref{constrscs} in \eqref{imptbilinearrel} we can rewrite the expression for
$\Phi_2$ in \eqref{Phidef} at a
bolt with positive and negative chirality as 
\begin{align}\label{Phi2expsbolt}
\Phi_2|_+&=\frac{1}{2}\dd\xi^\flat|_+ -\frac{\ii}{\sqrt{2}}\widetilde \cN_{IJ}\tC^IF^J
=\frac{1}{2}\dd\xi^\flat|_+ +\frac{\ii}{\sqrt{2}}\cF_I(\tC) F^I
\,,\nn
\Phi_2|_-&=-\frac{1}{2}\dd\xi^\flat|_- +\frac{\ii}{\sqrt{2}} \cN_{IJ}C^IF^J=-\frac{1}{2}\dd\xi^\flat|_- +\frac{\ii}{\sqrt{2}}\cF_I(C) F^I\,,
\end{align}
respectively, where to get the first expressions we also used $\cN_{IJ}=\cR_{IJ}+\ii\smallspace\cI_{IJ}$ and $\widetilde\cN_{IJ}=\cR_{IJ}-\ii\smallspace \cI_{IJ}$, while
to get the second we used \eqref{calNrel} and that $\cF$, $\widetilde\cF$ are homogeneous of degree two, with 
$\cF_I(C)\equiv \frac{\partial}{\partial C_I} \cF(C)$
and $\cF_I(\tC)\equiv \frac{\partial}{\partial \tC_I} \cF(\tC)$.
We also note that smoothness of the gauge field at the bolt implies that $\xi\hook F^I=0$ and hence from
\eqref{PhiF} we have $\Phi_0^I$ are constant on the bolt. Then from \eqref{defphioiflux} and \eqref{constrscs} we deduce $\tC^I$ and $C^I$ are constant  on the bolt.\footnote{It is not clear to us whether in addition 
$X^I$, $\widetilde X^I$ and $S$ are also constant on a bolt.} Thus, integrating $\Phi_2$ over a bolt, from \eqref{Phi2expsbolt} we have
\begin{align}\label{intPhi2expsbolt}
\frac{1}{2\pi}\int_{\text{bolt}_+}\Phi_2&= {\ii}{\sqrt{2}}\cF_I(\tC)\smallspace\mathfrak{p}^I\,,\nn
\frac{1}{2\pi}\int_{\text{bolt}_-}\Phi_2&= {\ii}{\sqrt{2}}\cF_I(C)\smallspace \mathfrak{p}^I\,,
\end{align}
where we have defined the magnetic flux through each bolt via 
\begin{align}\label{mathfrakboltdf}
\mathfrak{p}^I \equiv \frac{1}{4\pi}\int_\text{bolt} F^I\,.
\end{align}

With these ingredients, we can evaluate the contribution to the on-shell action for a given
bolt. For a positive chirality bolt, recalling the expression from \eqref{Phi0} and using \eqref{constrscs} 
we have $\Phi_0=-\ii\cF(\tC)$. Then from \eqref{I4dexpand} we find
\begin{align}\label{I4dexpandboltplus}
I_{\mathrm{OS}}^{\text{bolt$_+$}}
&= \frac{\pi}{2G_4}\int_{\text{bolt$_+$}}
\left[\frac{\Phi_2}{2\pi b_2}-\frac{\Phi_0\smallspace c_1(L)}{b_2^2}\right]\nn
&= \frac{\pi}{2G_4}\left[\frac{\ii \sqrt{2}}{b_2}\cF_I(\tC)\smallspace\mathfrak{p}^I  +\frac{\ii}{b_2^2}\cF(\tC)\int_{\text{bolt$_+$}}c_1(L)\right]\,.
\end{align}
We next use \eqref{wtscrelsbolt} and the definition of the $u^I_+$ variables in \eqref{udefs} to rewrite this as
\begin{align}\label{I4dexpandboltplus2}
I_{\mathrm{OS}}^{\text{bolt$_+$}}
&= \frac{\pi}{G_4}\left[-\kappa\ii \cF_I(u_+)\smallspace \mathfrak{p}^I  +\ii\cF(u_+)\int_{\text{bolt$_+$}}c_1(L)\right]\,,
\end{align}
while for a negative chirality bolt we similarly find
\begin{align}\label{I4dexpandboltminus}
I_{\mathrm{OS}}^{\text{bolt$_-$}}
&= \frac{\pi}{G_4}\left[-\kappa \ii \cF_I(u_-)\smallspace \mathfrak{p}^I  -\ii\cF(u_-)\int_{\text{bolt$_-$}}c_1(L)\right]\,,
\end{align}
where $\cF_I(u_\pm)\equiv\frac{\partial}{\partial u_\pm^I}\cF_I(u_\pm)$.

\subsection{Gravitational free energy}
By combining the results for the nuts and bolts we obtain the following general result for the gravitational free energy
for solutions satisfying the reality condition \eqref{eq:Reality_Condition}:
\begin{align}\label{Fgrav}
	\Fgrav &= \frac{\pi}{G_4} \bigg[ \sum_{{\rm nuts}_\pm}  \mp\frac{ (b_1 \mp b_2)^2 }{b_1 b_2} \ii {\cF}( u_\pm)  \nn
	& \qquad \quad + \sum_{{\rm bolts}_\pm} \left( -\kappa\ii \cF_I(u_\pm)\smallspace \mathfrak{p}^I \pm   \ii {\cF}({u_\pm})  \int_{\text{bolt}_\pm} c_1(L) \right)\bigg]\,.
	\end{align}

Some comments are in order. 
First, the result \eqref{Fgrav} is expressed in terms of quantities evaluated at the fixed point set. However, in the next section we explain how one can relate the fixed point IR data
to UV data on the conformal boundary $\partial M$, and in subsequent sections we demonstrate how this works in more detail for various specific examples.
We also notice the expression is written in terms of $u_\pm^I$ defined in \eqref{udefs}. 
This is a very useful set of variables, as they are constant on fixed point sets, and moreover are invariant under $U(1)_R$ rotations (as they are ratios of terms that are linear in $L^I$). Therefore, the final result is independent of the phase chosen for the symplectic sections, as it should be.
We also emphasize that the signs $\kappa$ need to be specified for each individual fixed point component, i.e.\ one $\kappa$ for each nut and for each bolt, 
and moreover these need 
to be determined in a globally consistent way; 
this is carried out for toric examples in \cite{BenettiGenolini:2024hyd}.

Second, it is also worth highlighting an observation made in \cite{BenettiGenolini:2019jdz} (there in the context of minimal gauged supergravity).
The original Killing spinor equation is linear in $\epsilon$, so for any particular solution we can multiply by an arbitrary non-zero complex
number and get another solution. This procedure rescales the Killing vector $\xi$ by an arbitrary positive number and notice that 
\eqref{Fgrav} is invariant under this rescaling. Furthermore,
given the reality conditions \eqref{eq:Reality_Condition}, we notice that any solution with gauge field and Killing spinor $(A,\epsilon)$
can be transformed into another solution with $(-A,\epsilon^c)$, with $\epsilon^c$ having the same chirality as $\epsilon$ at fixed points,
which transforms the Killing vector $\xi\to -\xi$. Correspondingly the final
answer should be invariant under $\mathfrak{p}^I\to -\mathfrak{p}^I$ and $\xi\to -\xi$, which implies that \emph{all} $\kappa\to -\kappa$,
and indeed it is. 

Third, 
if we set all of the vector multiplets to zero we obtain minimal gauged supergravity and, as described in appendix
\ref{mingaugedsugraapp}, we recover the result obtained in \cite{BenettiGenolini:2019jdz}. 

Fourth, it is also possible to include orbifold singularities both in the bulk and on the boundary. 
This requires minor modifications which we discuss in section~\ref{sec:orbifoldexamples}.

Lastly, recall that in obtaining the result \eqref{Fgrav} we used the reality conditions \eqref{eq:Reality_Condition} and the symplectic constraint \eqref{symconlorentzian}
then implied $\tcF(\tX)=- \cF(\tX)$, as in \eqref{tfcond}. In principle, we think it is likely that one can relax the reality conditions, or
impose different reality conditions, allowing one to consider prepotentials that do not satisfy the condition \eqref{tfcond}, but
still be consistent with the symplectic constraint. The result will then be given by
\begin{align}\label{Fgrav_Complex}
	I_{\rm OS}^{\rm FP} &= \frac{\pi}{G_4} \bigg[ \sum_{{\rm nuts}_+}  \frac{ (b_1 - b_2)^2 }{b_1 b_2} \ii \widetilde{\cF}( u_+) + \sum_{{\rm nuts}_-}  \frac{ (b_1 + b_2)^2 }{b_1 b_2} \ii {\cF}( u_-)  \nn
	& \qquad \quad +  \sum_{{\rm bolts}_+} \left( \kappa\ii \widetilde{\cF}_I(u_+)\smallspace \mathfrak{p}^I - \ii \widetilde{\cF}({u_+})  \int_{\text{bolt}_+} c_1(L) \right) \nn
	 & \qquad \quad + \sum_{{\rm bolts}_-} \left( -\kappa\ii \cF_I(u_-)\smallspace \mathfrak{p}^I - \ii {\cF}({u_-})  \int_{\text{bolt}_-} c_1(L) \right) \bigg]\,.
	\end{align}
Notice that if we set $\tcF(\tX)=- \cF(\tX)$ then we recover \eqref{Fgrav}. It is also worth highlighting in the specific examples we consider later, we focus on the STU model with prepotential given in \eqref{STUprepot} which, in particular satisfies 
$\tcF(\tX)=- \cF(\tX)$ independently of any reality conditions being imposed. Furthermore, we also note that in section 
\ref{sec:six} when we consider complex examples, we have explicitly assumed $\tcF(\tX)=- \cF(\tX)$, but it is straightforward to
use \eqref{Fgrav_Complex} to get results for other prepotentials.

\section{Flux integrals, UV-IR relations, and holography}
\label{sec:uvirrels}
In the previous section we showed how equivariant localization can be used to evaluate the on-shell action of a supersymmetric solution of the gauged supergravity theory, obtaining an expression in terms of quantities evaluated at the fixed point set of the R-symmetry. We now show that it is possible to apply equivariant localization to the polyforms of the gauge curvatures \eqref{PhiF}. In particular,
this allows one to relate expressions evaluated at the fixed point sets to boundary data. We also discuss some aspects of the holographic
expansion that is explored in more detail in appendix \ref{app:holrenn}.

\subsection{Flux integrals}
We first consider integrals of the field strengths $F^I$ over two-cycles in $M$. 
For $\Sigma\subset M$ a closed two-dimensional 
submanifold representing a cycle $[\Sigma]\in H_2(M,\mathbb{Z})$, we define the associated magnetic flux,
$\mathfrak{p}^I=\mathfrak{p}^I_{[\Sigma]}$, 
as \begin{align}\label{pfrak}
\mathfrak{p}^I\equiv \frac{1}{2}\int_{\Sigma} c_1(F^I)=\frac{1}{4\pi}\int_{\Sigma} F^I\, . 
\end{align}
For a general gauged supergravity model there is no \emph{a priori} reason for these to be quantized. However,
for gauged supergravity models that can be uplifted to $D=10,11$ supergravity on compact manifolds, the magnetic fluxes are quantized. In section \ref{sec:examples} we discuss various classes of solutions of the STU model, which can be uplifted
on $S^7$ (or a quotient thereof) to obtain solutions of $D=11$ supergravity. We have chosen the normalization 
in \eqref{pfrak} so that $\mathfrak{p}^I\in\mathbb{Z}$ for the STU model with solutions uplifted on $S^7$.

When $\Sigma$ is a bolt, we have already seen how these fluxes directly enter into the localized version of the on-shell action \eqref{Fgrav} via \eqref{intPhi2expsbolt}. If instead $\xi$ is tangent to $\Sigma$, 
hence rotating it, one can also use localization 
to relate these fluxes to the fixed point data at nuts, as illustrated in the examples later
(see sections \ref{rotatingbhs} and \ref{sec:orbifoldexamples}).

We also note that the global existence of the Killing spinor, 
as discussed in section \ref{bilinearsubsec},
 imposes constraints on the flux of the R-symmetry gauge field. As remarked in section \ref{subsubsec:Chirality_FixedPoints}, $\epsilon$ is charged under $A^R$ and at the fixed point set becomes chiral and non-zero. Consider a bolt $\Sigma_\pm$: in this case, the existence of a non-zero section means that one of the line bundles in which $\mathcal{S}M\rvert_{\Sigma_\pm} \otimes \mathscr{L}^{1/2}$ can be decomposed must be a trivial line bundle. It is then straightforward to extend the analysis of \cite{BenettiGenolini:2019jdz} to conclude that for
a bolt of chirality $\pm1$ the R-symmetry flux is given by
\begin{equation}
\label{eq:R_Symmetry_Constraint_Bolt}
	\mathfrak{p}^R_\pm = \frac{1}{2} \zeta_I \mathfrak{p}^I_\pm = \frac{\kappa}{2} \int_{\Sigma_\pm} \left[ \pm c_1(L) - c_1(T\Sigma_\pm) \right] \, .
\end{equation}

\subsection{UV-IR relations}
\label{subsec:UVIR_Noncompact}

We can also obtain important  ``UV-IR relations'' by considering flux integrals over non-compact two-dimensional submanifolds, whose boundaries end on the asymptotic boundary
$\partial M= M_3$. The submanifolds are assumed to have
topology $\mathbb{R}^2$, with a UV boundary circle $S^1_{\mathrm{UV}}\equiv \partial\mathbb{R}^2\subset M_3$, and be left invariant
under the action of $\xi$. 
On such an $\mathbb{R}^2$ we may introduce a local coordinate $\varphi$, with $\Delta\varphi=2\pi$, and
write $\xi\mskip1mu |_{\mathbb{R}^2}=b\smallspace \partial_{\varphi}$. The origin of $\mathbb{R}^2$, which lies on the fixed point set, 
is referred to as the ``IR''. From \eqref{pfrak} we can define the gauge-invariant flux threading the $\mathbb{R}^2$ via
\begin{align}
\label{eq:DefinitionDelta}
 \Delta^I\equiv \frac{1}{4\pi}\int_{\mathbb{R}^2} F^I\,.
\end{align}
Using Stokes' theorem we can then write $\Delta^I
 =\frac{1}{4\pi}( \int_{S^1_{\mathrm{UV}}} A^I - \int_{S^1_{\mathrm{IR}}} A^I)$, where $S^1_{\mathrm{IR}}$ is a small circle surrounding the origin
 of $\mathbb{R}^2$. Since $\mathbb{R}^2$ is topologically trivial we can choose a regular gauge so that $A^I$ is a global one-form on $\mathbb{R}^2$.\footnote{{In the case that there are more than one such $\mathbb{R}^2$ and not homologous to each other, in general it will not be possible 
to choose a gauge that is regular on both of them.} } In this case the IR integral vanishes and $ \Delta^I$ fixes the holonomy of the
gauge field around $S^1_{\rm UV}$ in the boundary theory:
\begin{align}\label{Delholonrel}
 \Delta^I=\frac{1}{4\pi}\int_{S^1_{\mathrm{UV}}} A^I\,.
\end{align}
Note that $\Delta^I$ is not constant, in general; it will be independent of the coordinate on 
$S^1_{\mathrm{UV}}$, which is tangent to $\xi$, but it can depend on the remaining coordinates parametrizing the position of $S^1_{\mathrm{UV}}$.

Continuing, we 
can now consider the lowest components of the equivariantly closed forms for the fluxes in \eqref{PhiF}, \eqref{defphioiflux} both in the UV and
IR:
\begin{align}
\label{eq:Defn_sigma_y}
\sigma^I  \equiv -\left.\frac{\ii }{4\pi}\Phi^I_0\mskip1mu \right|_{S^1_{\rm UV}}\, ,\qquad
y^I \equiv \left.\frac{ 1}{2b}\Phi^I_0 \mskip1mu \right|_{\mathrm{IR}}\, ,
\end{align}
with the factors added for convenience. 
Notice that these quantities arise from evaluating the combinations of spinor bilinears and
scalars appearing in \eqref{defphioiflux} and are gauge-invariant.
Also notice from the argument below \eqref{Phi2expsbolt} that $y^I$ is constant
on a connected component of the fixed point set, with its value dependent on which component
of the fixed point set the $\mathbb{R}^2$ ends on in the bulk.
By contrast, like $\Delta^I$, $\sigma^I$ is not constant in general, but it is independent of the coordinate on $S^1_{\rm UV}$: this follows from the fact that
since $\dd\Phi^I_0 = \xi \hook F^I$ we have $\cL_\xi \Phi_0^I=0$ and that $\xi$ is tangent to $S^1_{\rm UV}$.
Here $\sigma^I$ 
depends on both the UV scalar mass deformations in the dual boundary theory as well as the boundary geometry, as we discuss further in the next subsection.
From the equivariant closure of  $\Phi^{I}_{(F)}$ one then immediately deduces the following relation:
\begin{align}\label{UVIR}
\Delta^I +\ii  \beta\sigma^I = y^I\, ,
\end{align}
where $\beta\equiv 2\pi/b$ so that defining $\psi\equiv \varphi/b$, we can also write $\xi=\partial_\psi$ with $\Delta\psi = \beta$.  
Notice that with our Euclidean reality conditions \eqref{eq:Reality_Condition} the $\sigma^I$ are purely imaginary. 
We also notice that while $\Delta^I$ and $\sigma^I$ are not constant, in general, the linear combination
in \eqref{UVIR} is constant (since $y^I$ is constant). 

We emphasize that equation \eqref{UVIR} is a non-trivial UV-IR relation. On the right-hand side we have the 
IR quantities $y^I$ which depend on
the values of the scalars and bilinear fields at a specific part of the fixed point set. While on the left-hand side we have the
UV boundary quantities $\sigma^I$, characterizing
mass deformations, as well as the $\Delta^I$ which can be thought of as UV holonomies via \eqref{Delholonrel}.
All of the $y^I$, $\sigma^I$ and $\Delta^I$ depend on which $\mathbb{R}^2$ factor one uses to define them.
Now the gravitational free energy \eqref{Fgrav} is expressed in terms of IR quantities evaluated at the fixed point set, 
but for applications to holography it is more natural to write it in terms of the boundary data, and this is what the 
UV-IR relation \eqref{UVIR} allows one to do. A relation analogous to \eqref{UVIR} was conjectured and verified in some examples for which $M\cong \mathbb{R}^2\times \Sigma_g$ with $\Sigma_g$ a Riemann surface, in \cite{Bobev:2020pjk}.

Concretely, we now relate the IR fixed point quantities $y^I$ to the fixed 
point data $u^I_\pm$ defined in \eqref{udefs} that enters the on-shell action in \eqref{Fgrav} as follows.
We first assume that the origin of the $\mathbb{R}^2\subset M$ lies on a bolt. Then, if we identify $\varphi$ and the weight $b$ with 
$\varphi_2$ and $b_2$, using the same local orthonormal frame which is associated with the projections
\eqref{boltprojslf}, and in particular with the same specific $\kappa$ for that bolt, we deduce that
\begin{align}\label{yureln}
y^I_\pm =\pm\kappa u^I_\pm\,,
\end{align}
for positive or negative chirality bolts, respectively.

For a nut$_\pm$ we similarly have
\begin{align}\label{yurelnuts}
y^I_\pm&=-\frac{\kappa}{b}(b_1\mp b_2)u_\pm^I\,,
\end{align}
where $\kappa$ is the same specific $\kappa$ for the nut, as in \eqref{kappacond}.
Notice here that with an appropriate choice of convention 
we could always take $b=b_2$, since we may 
swap $b_1\leftrightarrow b_2$ using the transformations \eqref{beetransfs} 
(noting from \eqref{kappacond} that for positive chirality this then also requires one to take 
$\kappa\rightarrow -\kappa$). However, 
we point out that for general examples 
of toric $M$, studied in \cite{BenettiGenolini:2024hyd}, it is convenient to choose a global ordering 
of $b_1, b_2$ at each fixed point, and correspondingly 
take either $b=b_1$ or $b=b_2$ in \eqref{yurelnuts} according to this choice, for different 
copies of $\R^2\subset M$. 

\subsection{Holography}
\label{subsec:Holography}

We are assuming that the general gauged supergravity theory admits a
supersymmetric $AdS_4$ vacuum with constant scalars and vanishing gauge fields. 
From the equations of motion \eqref{eq:4d_N2_EinsteinEOM} the radius squared of the $AdS_4$
is given by $\ell^2=-6/\cV_*$, where the subscript $*$ denotes evaluation at the vacuum solution. 
To simplify the subsequent formulae it will be convenient to choose units with
$\ell=1$. Supersymmetry requires 
$\nabla_i W_*=\nabla_{\tilde i}\tW_*=0$ and hence we deduce 
\begin{align}\label{vstarexp}
\cV_*=-3(\zeta_I L^I)_*(\zeta_I\tL^I)_* = - 6\,.
\end{align}
In our conventions, the free energy of the dual SCFT on a round $S^3$ in the large $N$ limit,
$F_{S^3}$, is related to the $D=4$ Newton constant via
\begin{align}\label{Gnewtongeneral}
F_{S^3}=\frac{\pi}{2G_4}4\ii\cF(L^I_*)\,.
\end{align}
Note that by a field redefinition we can always assume that the given $AdS_4$ vacuum lies at the origin of the scalar field space,
i.e. $z^i_*=\tz^{\tilde i}_*=0$, and
we will now assume that this is the case.

A supersymmetric solution induces a structure on its conformal boundary $\partial M$, which can be found analysing the asymptotic expansion of the gravity fields near $\partial M$. 
In particular, we find (see also appendix \ref{app:holrenn}) that $\xi$ restricted to the boundary defines a Killing vector $\xi$ with a transversely holomorphic foliation. 
Assuming\footnote{
If $\xi$ has a non-closed orbit a standard argument shows that $\xi$ may be 
approximated by a sequence of vector fields with closed orbits. Note that when $\xi$ has closed orbits the ratios of weights $b_1/b_2$ at the nuts are 
rational numbers. \label{footnote:Orbits}} the orbits of $\xi$ all close, we can write $S^1 \hookrightarrow \partial M \to \Sigma_2$, which is associated to a orbifold bundle $\cL$, 
 i.e. $\Sigma_2$ will in general be an orbifold (and note we are still assuming that $\partial M$ is a manifold). 
The metric on $\partial M$ can be written in terms $\xi$ and a local transverse complex coordinate $u$ as
\begin{equation}
\label{eq:ds2_Bdry}
	\dd s^2_{\rm bdry} = \eta_{(0)}^2 + 4 \e^{V_{(0)}} \dd u \dd\ubar \, ,
\end{equation}
where $V_{(0)}= V_{(0)}(u,\ubar)$ is a local function and $\eta_{(0)}$ is the global almost contact form on $\partial M$ dual to $\xi$ using this boundary metric and $\xi \hook \eta_{(0)} = 1$ with $\|\xi\|^2=1$.\footnote{Note that a conformal transformation may be required to put the boundary metric in this form.} The two-dimensional metric on the leaf space is given by
\begin{equation}
\label{eq:2dMetric}
	\dd s^2_2 = 4 \e^{V_{(0)}} \dd u \dd\ubar \, .
\end{equation}
The R-symmetry gauge field induced on the boundary is determined by the geometry, and its curvature is
\begin{equation}
\label{eq:FR_Leading_Text}
	F^R_{(0)} = \eta_{(0)} \wedge \dd *_2 \dd \eta_{(0)} + \frac{1}{4} \left[  R_{(2d)} - 8 \lVert \dd \eta_{(0)} \rVert^2_2 \right] \vol_2 \, ,
\end{equation}
where $R_{(2d)} = - \e^{-V_{(0)}} \partial_{u\overline{u}}^2 V_{(0)}$ is the Ricci scalar of \eqref{eq:2dMetric}, and $\vol_2$ its volume form.

Deformations of the dual  $d=3$ SCFT are obtained by considering different backgrounds $M_3=\partial M$ and also by turning on background gauge fields as well as allowing suitable boundary conditions for the scalar fields.  
Since we are studying supersymmetric solutions in the bulk, these deformations are necessarily associated with deformations of the SCFT that preserve supersymmetry. 
In particular, as remarked earlier, it is possible to turn on scalar mass deformations
of the boundary SCFT, parametrized by $\sigma^I$, which are related to the lowest component $\Phi^I_0$ of the equivariant form $\Phi^I_{(F)}$
by \eqref{eq:Defn_sigma_y}. 
As discussed in appendix \ref{app:holrenn}, the boundary behaviour of $\Phi^I_0$, defined in \eqref{defphioiflux}, has the following general form
\begin{align}
\label{eq:PhiI0_Text}
	\Phi^I_0 \Big\rvert_{\rm bdry} &= \sqrt{2} \left[ L^I_{*} (*_2 \dd \eta_{(0)}) + \left( \nabla_i L^I \right)_{*} \left( z^i_{(1)} - \tz^{\tilde{i}}_{(1)} \right) \right] \, .
\end{align}
The combination $z^i_{(1)} - \tz^{i}_{(1)}$, appearing at the first subleading order in the expansion of the scalar fields near the conformal boundary, is proportional to the source for the boundary operator of dimension $2$.
From this expression, by restricting to the boundary $S^1$ of the specific $\R^2$ submanifold associated with $\sigma^I$, we immediately obtain
\begin{equation}
\label{eq:sigmaI}
	\ii \sigma^I = \frac{\sqrt{2}}{4\pi} \left[ L^I_* (*_2 \dd\eta_{(0)}) + \left( \nabla_i L^I \right)_* \left( z^i_{(1)} - \tz^{i}_{(1)} \right) \right] {\Big\rvert_{S^1_{\rm UV}}} \, .
\end{equation}
These general results, valid 
for arbitrary supersymmetric backgrounds on $\partial M$, substantially generalize a relation that was conjectured in the context of 
$\partial M \cong S^3$ with vanishing gauge fields, and verified in some examples, in \cite{Zan:2021ftf}.

Notice that there are $n$ independent mass deformations, associated with $( z^i_{(1)} - \tz^{i}_{(1)})$, but there
are $n+1$ components in $\sigma^I$. However, the combination $\zeta_I\sigma^I$ is constrained by the boundary geometry via
\begin{equation}\label{localconssigmaI}
	\ii \zeta_I \sigma^I = \frac{1}{2\pi} *_2 \dd\eta_{(0)}{\Big\rvert_{S^1_{\rm UV}}} \,.
\end{equation}
In the special case that $\sigma^I$ is constant (for any choice of $S^1_{\rm UV}$), 
i.e.
$\Phi^I_0 \big\rvert_{\rm bdry} $ in \eqref{eq:PhiI0_Text} is constant, 
we can integrate over the entire base, obtaining 
\begin{equation}
\label{eq:Rsymm_sigmaI}
	\int_{\Sigma_2} \ii \zeta_I \sigma^I \, \vol_2 = \frac{1}{2\pi} \int_{\Sigma_2} \dd\eta_{(0)} = \frac{1}{b} \int_{\Sigma_2} c_1(\cL) \, ,
\end{equation}
where $c_1(\cL) \in H^2(\Sigma_2, \mathbb{Q})$ is the first Chern class of the orbibundle $\cL$.

The fact that the boundary R-symmetry gauge field is fixed by the geometry as in \eqref{eq:FR_Leading_Text}
allows us to immediately derive some necessary conditions for the possibility of constructing solutions where $F^R$ vanishes. Namely, for \eqref{eq:FR_Leading_Text} to vanish, we need $*_2 \dd\eta_{(0)}$ to be constant on $\Sigma_2$, and its square must be proportional to the Ricci scalar curvature (see \cite{Closset:2019hyt} for a more detailed discussion). This will constrain the examples considered in the next section. 

\section{Real solutions}\label{sec:examples}

We now consider some specific examples of solutions on manifolds $M$, subject to the
reality conditions \eqref{eq:Reality_Condition} and recall that we also imposed 
the condition \eqref{tfcond}. Our formalism is applicable for general gauged supergravity theories with prepotential $\cF$
but we will also focus on the specific case of the STU model, which includes minimal gauged supergravity as a special case. 
Some of these examples were briefly considered in \cite{BenettiGenolini:2024xeo} and here we expand upon the discussion.

\subsection{The STU model}

The STU model has three vector multiplets with prepotential given by
\begin{align}\label{STUprepot}
\cF(X^I)
= -2\ii \sqrt{X^0X^1X^2X^3}\, ,
\end{align}
and we take $\xinew_I=1$ for all $I=0,1,2,3$.
Solutions of the STU model uplift on $S^7$ to give solutions 
of $D=11$ supergravity \cite{Cvetic:1999xp, Azizi:2016noi}.
For example, the
$AdS_4$ vacuum with vanishing scalars and gauge fields uplifts to the $AdS_4\times S^7$ solution dual to the ABJM theory \cite{Aharony:2008ug}.
Minimal gauged supergravity
can be recovered from the general set-up by setting all the vector multiplets to zero, as discussed in appendix \ref{mingaugedsugraapp}.
Alternatively, it can be obtained from the STU model by setting $X^0=X^1=X^2=X^3=\frac{1}{4}$ and
$A^0=A^1=A^2=A^3$. Solutions of minimal gauged supergravity can also be uplifted 
to $D=11$ in other ways \cite{Gauntlett:2007ma}, including on an arbitrary seven-dimensional Sasaki--Einstein space. 

In Lorentzian signature, the complex scalars $z^i$ of the three vector multiplets of the STU model parametrize three copies of the Poincar\'e disc with 
$|z^i|<1$ and K\"ahler potential given by
\begin{equation}
\label{eq:cK_STU}
	\cK = - \sum_{i = 1}^3 \log \left( 1 - z^i \overline{z}^{\overline{i}} \right) \, .
\end{equation}
The associated Hermitian metric and composite connection \eqref{eq:Lorentzian_KahlerHodgeConnection} take the form
\begin{align}
\label{eq:STU_KahlerStuff}
	\cG_{i\overline{j}} &= \frac{\delta_{i\overline{j}}}{(1 - z^i \overline{z}^{\overline{j}})^2} \, , \qquad	\cA = - \frac{\ii}{2} \sum_{i=1}^3  \frac{ \bar{z}^{\bar{i}} \dd z^i - z^i \dd \bar{z}^{\bar{i}} }{1 - z^i \bar{z}^{\bar{i}}} \, .
\end{align}
When needed, we will fix a specific parametrization of the symplectic sections $X^I$ in terms of the physical scalars given by
\begin{equation}
\label{eq:FreedmanPufuScalars}
\begin{aligned}
	X^0 &= \frac{1}{2\sqrt{2}} ( 1 + z^1 )( 1 + z^2 )( 1 + z^3 ) \, , &\qquad X^1 &= \frac{1}{2\sqrt{2}} ( 1 + z^1 )( 1 - z^2 )( 1 - z^3 ) \, , \\
	X^2 &= \frac{1}{2\sqrt{2}} ( 1 - z^1 )( 1 + z^2 )( 1 - z^3 ) \, , &\qquad X^3 &= \frac{1}{2\sqrt{2}} ( 1 - z^1 )( 1 - z^2 )( 1 + z^3 ) \, .
\end{aligned}
\end{equation}
This ``Freedman--Pufu" choice (also used in \cite{Bobev:2020pjk}) will be useful in comparing with the solutions in \cite{Freedman:2013oja}; for a clear discussion
of other choices of parametrizations and how they are related to each other, see \cite{Cabo-Bizet:2017xdr}. In particular, this is helpful
if one wants to uplift the STU solutions to $D=11$ using \cite{Cvetic:1999xp}.
The holomorphic superpotential \eqref{wsuperpots} and the scalar potential \eqref{curlyvpot} take the form
\begin{align}
W= \sqrt{2} (1+z^1 z^2 z^3) \,,\qquad
\cV= 2 \left( 3 - \sum_{i=1}^3 \frac{2}{1- z^i \bar z^{\bar{i}}} \right)\,.
\end{align}
The $AdS_4$ vacuum with unit radius corresponds to $z^i=z^{\bar{i}}=0$.
The explicit expressions for $\cI_{IJ}$ and $\cR_{IJ}$ to give the Lorentzian action \eqref{eq:Lorentzian_I4} or Euclidean action \eqref{themainaction} for the STU model are rather
long and so we don't give them here. For minimal gauged supergravity, the explicit form of the Euclidean action and the supersymmetry transformations are given in appendix \ref{mingaugedsugraapp}.
Regarding the expression for $\sigma^I$ in \eqref{eq:sigmaI}, for the $AdS_4$ vacuum in the parametrization of \eqref{eq:FreedmanPufuScalars} we have
\begin{align}
L^I_* = \frac{1}{2\sqrt{2}}\,,
 \end{align}
 and
 \begin{align}
 (\nabla_i L^0)_*&=\frac{1}{\sqrt{2}}(1,1,1)\,,\qquad\quad\,
  (\nabla_i L^1)_*=\frac{1}{\sqrt{2}}(1,-1,-1)\,,\nn
   (\nabla_i L^2)_*&=\frac{1}{\sqrt{2}}(-1,1,-1)\,,\qquad
  (\nabla_i L^3)_*=\frac{1}{\sqrt{2}}(-1,-1,1)\,.
 \end{align}

For the STU model, we can uplift on $S^7/\mathbb{Z}_k$ to get solutions dual to ABJM theory. In this case, from \eqref{Gnewtongeneral}
the four-dimensional Newton constant is related to the free energy of ABJM theory on $S^3$ in the large $N$ limit via 
\begin{align}\label{abjmGNrel}
F_{S^3}=\frac{\pi }{2G_4} \stackrel{\mathrm{ABJM}}{=}\frac{\sqrt{2k}\pi}{3}N^{3/2}\,.
\end{align}

\subsection{Deformations of Euclidean \texorpdfstring{$AdS_4$}{AdS4}, \texorpdfstring{$A^I=0$}{AI=0}}
\label{subsec:Deformations_AdS4}

For our first class of examples we 
take $M$ to be topologically $\R^4 = \R_1^2\oplus \R_2^2$, with $U(1)^2$ isometry and conformal 
boundary $\partial M\cong S^3$. The boundary metric on $S^3$ has $U(1)^2$ isometry but we make no further assumption
and in general it can be squashed. We don't make any assumption on the gauge fields for the moment. 
We assume that $\xi = \sum_{i=1}^2 b_i\partial_{\varphi_i}$ where $\partial_{\varphi_i}$ rotate 
each $\R_i^2$ and we assume that $b_i\neq 0$ so that there is an isolated fixed point at the origin of $\R^4$. The chirality of the spinor at this nut
can be $\pm 1$ and correspondingly from \eqref{Fgrav} we have an expression for the free energy in terms of the fixed point data given for the two cases, respectively, by
\begin{align}\label{FP}
	\Fgrav = \mp \frac{(b_1\mp b_2)^2}{b_1b_2}\ii\cF(u_\pm )\frac{\pi}{G_4}
\, .
\end{align}
Recall from \eqref{constsuzeta} we have the constraints $\zeta_I u^I_\pm = 1$. 
For a holographic interpretation, we can also write this expression in terms of UV boundary data using the results of section \ref{sec:uvirrels}. 
This example is particularly simple because there is a single nut in the bulk, and moreover, since the topology is $\mathbb{R}^4$, it is possible
to choose a global gauge for the gauge fields when present.
We now focus on various sub-cases and make contact with various known solutions.

We first consider solutions with vanishing gauge fields. This will allow us to make contact with the
solutions explicitly constructed in \cite{Freedman:2013oja,Zan:2021ftf} associated with real mass deformations
of the dual SCFTs on $S^3$.

With vanishing gauge fields, trivially we have $\Delta^I=0$. 
Next, suppose we consider $\R_1^2$ in the decomposition $\R^4 = \R_1^2\oplus \R_2^2$ so that on this submanifold we
have $\xi |_{\R^2_1}=b_1\partial_{\varphi_1}$. Then from
\eqref{UVIR} and \eqref{yurelnuts} we deduce for the nut$_\pm$ cases we have:
\begin{align}\label{yurelnutsex}
	y^I_\pm&=- \frac{\kappa}{{b_1}} (b_1\mp b_2)u_\pm^I= \frac{1}{{b_1}} 2\pi\ii\smallspace \sigma^I_{(1)}\,,
\end{align}
where the subscript in $\sigma^I_{(1)}$ indicates that we are considering $\mathbb{R}^2_1$,
and hence
\begin{align}\label{bupmsigonrel}
(b_1\mp b_2) u^I_\pm=-\kappa2\pi \ii \sigma^I_{(1)}\,.
\end{align}
If instead we had chosen
$\R_2^2$ we would obtain \eqref{bupmsigonrel} but with $\sigma^I_{(2)}$ on the right-hand side. Since the $u^I_\pm$ on the left-hand side are necessary 
equal (evaluated at the origin of $\R^4$), 
we can therefore define the constant deformation parameters 
\begin{align}
\sigma^I\equiv \sigma^I_{(1)}=\sigma^I_{(2)}\,.
\end{align}
In fact, with vanishing gauge fields we can deduce a lot more about the UV deformations. Indeed
the analysis of the bilinear equations shows that outside of the fixed point set, $\dd z^i$ and $\dd \tz^i$ are proportional to the bilinear $K$ (see equation \eqref{algSP}). As a consequence, the scalars $z^i$, $\tz^i$ only depend on the asymptotic radial coordinate (called $y$ in the discussion in appendix \ref{app:LocalForm_SUSY_Sols}), and so the UV mass deformations $z^i_{(1)}$ and $\tz^{\tilde{i}}_{(1)}$  are necessarily constant everywhere on $\partial M$.
Next, from the vanishing of \eqref{eq:FR_Leading_Text} we must have 
$*_2 \dd \eta_{(0)}$ is also a constant on the boundary (and also $R_{(2d)} = 8 \lVert \dd \eta_{(0)}\rVert_2^2$).
Thus, we see that $\Phi^I_0 \rvert_{\rm bdry}$, appearing in \eqref{eq:PhiI0_Text}, is constant and thus
the $\sigma^I$ in \eqref{eq:sigmaI}, for any boundary $S^1$ are all the same constants. Then 
we can invoke \eqref{localconssigmaI}, \eqref{eq:Rsymm_sigmaI}, 
or more immediately \eqref{bupmsigonrel}, to conclude that
the deformation parameters
$\sigma^I$ are constrained to satisfy
\begin{equation}
\label{eq:Rsymmetry_sigmaI_S3}
	2\pi \ii \zeta_I \sigma^I = - \kappa ( b_1 \mp b_2) \, .
\end{equation}
Using the fact that the prepotential is homogeneous of degree two, we can then write \eqref{FP} in terms of UV data as
\begin{equation}\label{pinghere}
	\Fgrav = \mp \frac{b_2}{b_1} \ii \cF \left( \frac{2\pi \ii \sigma }{b_2} \right) \frac{\pi}{G_4} \, .
\end{equation}

A particularly simple case is when $\abs{b_1} = \abs{b_2}$, which is consistent with a conformal boundary that is the round $S^3$ with $SO(4)$ isometry, which we also assume to be the case. 
It is not possible to have a nut$_\pm$ solution with $b_1=\pm b_2$ since then $\sigma^I=0$ (see footnote \ref{assumpb1eqb2}) and there would be no deformations. However, we can consider the special case that $b_1=\mp b_2$ with a nut$_\pm$. 
Normalizing the Killing spinor to have $b_2 = 1$, 
the UV-IR relation simplifies to $u^I_\pm = {\pm} \kappa \pi \ii \sigma^I$, with the mass parameters $\sigma^I$ constrained by
\begin{equation}
\label{eq:Rsymmetry_sigmaI_S3_Simpler}
	2\pi \ii \zeta_I \sigma^I = \pm 2 \kappa  \, ,
\end{equation}
and given in terms of boundary quantities by \eqref{eq:sigmaI}, which for $\partial M \cong S^3$ with the anti-Hopf fibration gives
\begin{equation}\label{eq:Rsymmetry_sigmaI_S3_Simpler2}
	\sqrt{2} \pi \ii \sigma^I = L^I_* + \frac{1}{2}\left( \nabla_i L^I \right)_* \left( z^i_{(1)} - \tilde{z}^i_{(1)} \right) \, ,
\end{equation}
with all terms constant.
Then $\Fgrav$ is given by
\begin{align}\label{FP2}
	\Fgrav = 4\ii\cF(u_\pm )\frac{\pi}{G_4} = \ii \cF \left( \sqrt{2} \pi \ii \sigma \right) \frac{2\pi}{G_4} \, .
\end{align}
In fact this expression\footnote{One should identify
$( z^i_{(1)} - \tilde{z}^i_{(1)})_\text{here}$, which we argued above is constant, with $(\ii e^\alpha_a\mathfrak{m}^a)_\text{there}$, where $e^\alpha_a$ is a frame for the boundary K\"ahler metric on the space of scalar fields.}  was conjectured to be 
the free energy of mass-deformed, three-dimensional $\cN=2$ SCFTs on $S^3$ in the large $N$ limit
in \cite{Zan:2021ftf} (building on \cite{Freedman:2013oja}). Thus, our localization approach provides a general supergravity proof
of this conjecture, assuming that
the solutions actually exist.

For the specific case of the STU model, an explicit solution with $\R^4$ topology, vanishing gauge fields, and $b_1 = \mp b_2$ was constructed in \cite{Freedman:2013oja}, which we review in appendix \ref{subsubsec:FP}. Its conformal boundary is $S^3$ with round metric, and in fact the bulk solution also has $SO(4)$ isometry. The UV mass deformations are constant on the $S^3$.
The free energy was also computed in \cite{Freedman:2013oja}, in agreement with
 \eqref{FP2} after restricting to the STU model using \eqref{STUprepot}, and precisely matches the large $N$ limit of the partition function of ABJM theory on $S^3$ with real mass deformations that was computed in \cite{Jafferis:2011zi}. 
 
More generally, we can also consider $\abs{b_1} \ne \abs{b_2}$, which is naturally associated with having
a non-round metric on the boundary $S^3$. 
The relation between the chirality of the spinor at the nut and the sign of $b_1/b_2$ can only be determined after considering the global structure of the solutions. Such analysis is subtle and not available for a generic supergravity theory, though it has been done for self-dual metrics that are minimal supergravity solutions on $\R^4$ in \cite{Farquet:2014kma}. 
To simplify the following discussion, with no essential loss of generality, we will restrict to the case
\begin{align}
b_1/b_2>0\,,\qquad \hat b^2 = {b_1/b_2}\,,
\end{align}
and then the gravitational free energy can be written
\begin{equation}
\label{eq:Fili}
	\Fgrav = \mp \left( \hat b \mp  \frac{1}{\hat b} \right)^2 \ii \cF(u_\pm) \frac{\pi}{G_4} \, ,
\end{equation}
with $\zeta_I u^I_\pm = 1$. With vanishing gauge fields the UV-IR relation can be written
\begin{align}
	u^I_\pm=-\kappa\frac{2\pi \ii}{b_2(\hat b^2\mp 1)}\sigma^I \,,
\end{align}
with the constant $\sigma^I$ given in \eqref{eq:Rsymmetry_sigmaI_S3_Simpler2}.

A field theory conjecture for the large $N$ limit of the free energy of ABJM theory with real and constant mass deformations on a squashed $S^3$, with squashing parameter $\hat b$, appears in \cite{Bobev:2022eus} (see also the recent \cite{Kubo:2024qhq}). There, the three independent real mass deformations are parametrized in terms
of $\hat \sigma_a$,\footnote{These were labelled $\Delta_a$ in \cite{Bobev:2022eus}.} $a=1,2,3,4$, constrained by $\sum_a \hat\sigma_a = 2$.
In this notation, the large $N$ limit of the ABJM free energy has the form
\begin{equation}
	F = \left( \hat b + \frac{1}{\hat b} \right)^2 \sqrt{\hat\sigma_1 \hat\sigma_2 \hat\sigma_3 \hat \sigma_4} \, F_{S^3} \, .
\end{equation}
Our result \eqref{eq:Fili} specialized to the STU model  
clearly matches the field theory conjecture, provided that we restrict to the nut$_-$ solutions, 
and we identify
\begin{equation}
	u^I_-= -\kappa\frac{2\pi \ii}{b_2(\hat b^2+ 1)}\sigma^I \quad \leftrightarrow \quad \frac{1}{2} \hat \sigma_{I+1} \, .
\end{equation}
It would be interesting to construct such nut$_-$ solutions with vanishing gauge fields
(another possibility, however, is that such solutions require the existence of bulk gauge fields). 
It would also be interesting to know whether or not there are analogous nut$_+$ solutions which would correspond to saddle points in the dual SCFT in addition to those considered in \cite{Bobev:2022eus}.

\subsection{Deformations of Euclidean \texorpdfstring{$AdS_4$}{AdS4}, \texorpdfstring{$A^I\neq 0$}{AI =/= 0}}
\label{subsec:Deformations_AdS42}

We now consider solutions with non-vanishing gauge fields. To simplify the discussion, we again take 
\begin{align}\label{b1b2rat}
b_1/b_2>0\,,\qquad \hat b^2 = {b_1/b_2} \,,
\end{align}
and correspondingly for nut$_\pm$ solutions we have the same fixed point result \eqref{eq:Fili} as in the previous section, that is
\begin{equation}
\label{eq:Fili2}
	\Fgrav = \mp \left( \hat b \mp  \frac{1}{\hat b} \right)^2 \ii \cF(u_\pm) \frac{\pi}{G_4} \, ,
\end{equation}
with $\zeta_I u^I_\pm = 1$. 

The UV-IR relation, though, is now different.
Considering $\R^2_1$ as before, from \eqref{UVIR}, \eqref{yurelnuts} we now have 
\begin{align}
\label{fivenineteen_1}
u^I_\pm = - \kappa\frac{\hat{b}^2}{\hat{b}^2 \mp 1} \left( \Delta^I_{(1)} + \frac{2\pi}{b_1} \ii\sigma^I_{(1)} \right)  \,.
\end{align}
The individual values of $\Delta^I_{(1)}$ and $\sigma^I_{(1)}$, associated with $\R^2_1$, might not be constant, but the
particular combination on the right-hand side of \eqref{fivenineteen_1} is constant (since the left-hand side is). Also, the value of $\sigma^I_{(1)}$ is related to boundary deformations via \eqref{eq:sigmaI} and in particular $\zeta_I\sigma^I_{(1)}$ is constrained by the boundary geometry by \eqref{localconssigmaI}.
Considering instead $\R^2_2$, we similarly obtain 
\begin{align}
\label{fivenineteen_2}
	 u^I_\pm = - \kappa\frac{1}{\hat{b}^2 \mp 1} \left( \Delta^I_{(2)} + \frac{2\pi}{b_2} \ii\sigma^I_{(2)} \right) \,,
\end{align}
which is in fact equal to the right-hand side of \eqref{fivenineteen_1}. We also have
$\zeta_I\sigma^I_{(2)}$ is constrained by \eqref{localconssigmaI} but in general
$\zeta_I\sigma^I_{(1)} \ne \zeta_I\sigma^I_{(2)}$. 
Notice that the UV data is necessarily constrained via
\begin{align}
 b_1 \Delta^I_{(1)} + 2\pi \ii \sigma^I_{(1)}=
 b_2 \Delta^I_{(2)} + 2\pi \ii \sigma^I_{(2)}\,.
 \end{align}
Since the prepotential is homogeneous of degree two, using the UV variables associated with $\mathbb{R}^2_2$, for example, 
we can write
\begin{equation}
\label{eq:Fgrav_S3_UVdataR22_1}
	\Fgrav = \mp \frac{1}{\hat{b}^2} \ii\cF\left( \Delta_{(2)} + \frac{2\pi}{b_2} \ii\sigma_{(2)} \right) \frac{\pi}{G_4} \, .
\end{equation}
It is striking that $\Fgrav$ depends on a specific combination of UV deformation data that is constant, 
even if the full UV deformation data is not constant; this
applies more generally to subsequent examples we shall consider.
We can now compare this result with known solutions and the large $N$ limit of holographic field theories. 

\subsubsection{\texorpdfstring{$AdS_4$}{AdS4} instantons}\label{ads4instantons}

We can make a direct comparison with the supergravity solutions of minimal gauged supergravity constructed in \cite{Martelli:2011fu, Farquet:2014kma}.
These have a non-vanishing gauge field with boundary given by $S^3$ with a squashed, cohomogeneity-one metric that is only $U(1)^2$ invariant.
In fact the latter is conformally related to the round $S^3$, and the solution is simply Euclidean $AdS_4$ with a non-vanishing instanton gauge field and a Killing spinor with negative chirality at the origin.
Considering the case of nut$_-$ solutions, the gravitational free energy predicted by \eqref{eq:Fili2} for these solutions is
\begin{equation}
\label{eq:Fgrav_S3_Squashed_Masses_Minimal}
	\Fgrav = \frac{1}{4}\left( \hat b + \frac{1}{\hat b} \right)^2 {F_{S^3}} \, ,
\end{equation}
where we have used the results of appendix \ref{mingaugedsugraapp} to restrict to minimal gauged supergravity.
This is in exact agreement with the result of \cite{Martelli:2011fu, Farquet:2014kma} and moreover
matches the large $N$ limit of the partition function of a large class of $3d$ $\cN=2$ Chern--Simons-matter theories computed on a squashed $S^3$ \cite{Martelli:2011fu}.
Presumably there are no similar supergravity solutions in the nut$_+$ class for minimal gauged supergravity.
It is straightforward to generalize the solutions of \cite{Martelli:2011fu, Farquet:2014kma} to $\cN=2$ gauged supergravity models
with vector multiplets: for nut$_+$ solutions (if they exist) and nut$_-$ solutions the free energy will be given by
\eqref{eq:Fgrav_S3_UVdataR22_1}. 

\subsubsection{Supersymmetric defects and R\'enyi entropy}
\label{subsec:Defects}

We now discuss how our localization results can be used to obtain the Euclidean on-shell action for explicitly known solutions 
in the STU model that are associated with supersymmetric monodromy defects of $d=3$ SCFTs \cite{Arav:2024wyg} 
(see also \cite{Gutperle:2018fea,Chen:2020mtv}). 
As we shall recall below, the same solutions can also be used to compute the supersymmetric R\'enyi entropy \cite{Nishioka:2013haa} for circular entangling regions of the SCFTs \cite{Hosseini:2019and,Arav:2024wyg}.
Moreover, we will see that the localization viewpoint clarifies the standard prescription, coming from \cite{Casini:2011kv,Hung:2011nu}, for computing the supersymmetric R\'enyi entropy.
 In addition to recovering results for the STU model, we can also straightforwardly generalize them to more general gauged supergravity theories.

Considering the weights for the supersymmetric Killing vector again as in \eqref{b1b2rat}, and now using the notation
$b_1/b_2=\hat{b}^2 \equiv n$, the gravitational free energy \eqref{eq:Fili2} can be written 
\begin{equation}
\label{defectfestpt}
	\Fgrav = \mp \frac{(n \mp 1)^2}{n^2} \ii \cF(u_\pm)\frac{\pi}{G_4} \, .
\end{equation}
For holographic applications, we need to introduce boundary conditions on $\partial M$ and use
the UV-IR relations to express this in terms of holographic data. One choice of boundary is the $U(1)^2$-preserving squashed $3$-sphere just discussed in section \ref{ads4instantons}, but this is not the only possibility, and in fact it is not the conformal representative useful for analysing the solutions in \cite{Arav:2024wyg}. 

To be concrete, we start by noticing that a natural choice of boundary metric consistent with the choice of Killing vector is
\begin{align}
\label{eq:Defect_Bdry_v1}
	\dd s^2_{\rm bdry} &= \dd\theta^2 + \cos^2\theta \, \dd\phi_1^2 + n^2 \sin^2\theta \, \dd\phi_2^2 \nn
	&= \eta_{(0)}^2 + \dd\theta^2 + \sin^2\theta \cos^2\theta ( \dd\phi_1 - n \, \dd\phi_2)^2 \, , 
\end{align}
where $\theta \in [0,\pi/2]$, $\phi_1, \phi_2\in [0,2\pi]$, and for definiteness, we are considering $\xi = \partial_{\phi_1} + \frac{1}{n} \partial_{\phi_2}$. As clear from the first line, for integer $n$ this is the metric on a round $S^3$ branched $n$ times over the $S^1$ parametrized by $\phi_2$, with branch locus at $\theta=0$. 
In the second line, we have isolated the one-form $\eta_{(0)}$ dual to $\xi$, showing that the metric has the form \eqref{eq:ds2_Bdry} with
\begin{equation}
	\eta_{(0)} = \cos^2\theta \, \dd\phi_1 + n^2 \sin^2\theta \, \dd\phi_2 \, ,
\end{equation}
and $\xi\hook \eta_{(0)}=1$ (i.e. we have $\|\xi\|^2=1$ and are also consistent with \eqref{eq:ConformalBoundary_Metric0}, \eqref{eq:ConformalBoundary_Metric}).
This boundary metric is the natural one to use in order to compute the supersymmetric R\'enyi entropy.
Importantly, while it does have a conical singularity along the branch locus, it has finite volume and the conclusion of appendix \ref{app:holrenn} that there will be a vanishing boundary contribution to the free energy is still valid.
Thus, we can use \eqref{defectfestpt} to compute the on-shell action for this boundary metric.

Introducing $\sinh\rho = \cot\theta$, with $\rho\in [0,\infty)$, we find 
\begin{equation}
\label{eq:Defect_Bdry_v2}
	\dd s^2_{\rm bdry} = \frac{1}{\cosh^2\rho} \left( \dd\rho^2 + \sinh^2\rho \, \dd\phi_1^2 + n^2 \, \dd\phi_2^2 \right) \equiv \frac{1}{\cosh^2\rho} \widetilde{\dd s^2}_{\rm bdry} \, .
\end{equation}
Thus, the $n$-fold branched $3$-sphere is conformally related to the direct product of $H^2\times S^1$ where the ratio of the radius of the $S^1$ to the $H^2$ factor is given by $n$. The branch locus is now at $\rho = \infty$, so it has been moved infinitely far away. The conformal factor is always positive, vanishing only at $\rho = \infty$, which is then removed from the space, hence decompactifying the $3$-sphere.
Choosing the conformal boundary to be $H^2\times S^1$, with $n$ an arbitrary real number, is naturally associated with the boundary metric for the supersymmetric monodromy defects in \cite{Arav:2024wyg} and is also used to compute the supersymmetric R\'enyi entropy \cite{Casini:2011kv}.
As we recall below, the infinite volume of the boundary $H^2$ factor needs to be appropriately regulated in computing the on-shell action for a specific solution. 
We also highlight that the analysis of appendix \ref{app:holrenn} is not directly applicable with this boundary.
Thus, we cannot directly use \eqref{defectfestpt} to compute the action of a solution with $H^2\times S^1$ bundary (the 
conformal transformation relating it to \eqref{eq:Defect_Bdry_v1} is not smooth as it changes the topology).
 Nonetheless, we shall see that the results agree, provided we regulate the volume of $H^2$.

We now consider a solution with topology $\R^4$ but boundary $H^2\times S^1$ and we assume that the boundary gauge fields and scalars are trivial on $H^2$. To use our localization results, our prescription is to carry out the computations using the conformally scaled metric \eqref{eq:Defect_Bdry_v1}.
We can obtain UV-IR relations using the $\R^2_2$ factor (bounded by the $S^1$ factor) and \eqref{fivenineteen_2} 
takes the form
\begin{equation}
	u^I_\pm = - \kappa \frac{1}{n \mp 1} \left( \Delta^I_{(2)} + 2\pi n \ii \sigma^I_{(2)} \right) \, .
\end{equation}
As noted above, using the boundary metric \eqref{eq:Defect_Bdry_v1}, we get no boundary contribution to the on-shell action and so we again obtain the result \eqref{defectfestpt}, which can be expressed as
\begin{equation}
\label{eq:Defect_Localization}
	\Fgrav = \mp \frac{1}{n} \ii \cF \left( \Delta_{(2)} + 2\pi n \ii \sigma_{(2)} \right)\frac{\pi}{G_4}  \, .
\end{equation}
Here $\ii\sigma^I_{(2)}$ are determined by \eqref{eq:sigmaI} and after using $*_2\dd\eta_{(0)} = -2$, which follows from \eqref{eq:Defect_Bdry_v1}, we obtain
\begin{equation}
\label{eq:sigmaI2}
	\ii \sigma^I_{(2)} = \frac{\sqrt{2}}{4\pi} \left[ -2L^I_*  + \left( \nabla_i L^I \right)_* \left( z^i_{(1)} - \tz^{i}_{(1)} \right) \right] {\Big\rvert_{S^1_{\rm UV}}} \, .
\end{equation}
In the absence of mass sources, as in the known solutions that we presently discuss, we have $z^i_{(1)} - \tz^{i}_{(1)} =0$.

We now compare this localization result to the result obtained for the supersymmetry monodromy defect solutions of the STU model in \cite{Arav:2024wyg}, which we first briefly review. The Lorentzian solutions have the form $AdS_2\times \mathbb{R}^2_2$ with the scalars and gauge fields depending on the $\mathbb{R}^2_2$ factor.\footnote{They are obtained from a double Wick rotation of
the supersymmetric, electrically charged hyperbolic black holes with non-compact $H^2$ horizons of \cite{Cvetic:1999xp}. 
In other words, from the same Euclidean $H^2\times \mathbb{R}^2_2$ solution, with $H^2\times S^1$ boundary, we can Wick rotate on the $H^2$ factor to get 
defect type solutions with $AdS_2\times S^1$ boundary and real gauge fields, or on the polar angle in the $\R^2_2$ factor to get a hyperbolic black hole with imaginary gauge fields.} The conformal boundary of these solutions is given by $AdS_2\times S^1$, with the ratio of the radius of the $S^1$ to the $AdS_2$ factor given by $n>0$. Working with regular gauge fields in the bulk solution (i.e. the gauge fields vanish at the origin of $\mathbb{R}^2_2$), the gauge fields have non-trivial holonomy around the boundary $S^1$, and are denoted by
\begin{equation}
\label{eq:Defects_Holonomies}
	\frac{1}{4\pi}\int_{S^1} A^I_{(0)} = g \mu^I \, .
\end{equation}
The interpretation as monodromy defects is obtained by using a conformal transformation to map the $AdS_2\times S^1$ boundary to flat space $\mathbb{R}^{1,2}$ with a conical singularity at the origin of the spatial slice with deficit angle $2\pi(1-{n})$, and 
observing that the gauge fields have non-trivial holonomy, or ``monodromy", as one circles the point defect on a constant time slice. 
After Wick rotating $AdS_2 \to H^2$, the resulting Euclidean solutions are smooth, with real metric, gauge field and scalars $z^i = \tz^i \in \R$ (but vanishing mass sources). Thus, they satisfy the reality requirements \eqref{eq:Reality_Condition} we have assumed in using localization.

Two branches of solutions were found in \cite{Arav:2024wyg}, which are parametrized by  a sign $s=\pm 1$ appearing in the 
phase of the Killing spinor, which is related to another sign $\hat{\kappa} = \pm 1$ denoting the chirality of the Killing spinor on $AdS_2$ (and we have added the hat notation here). The first branch of solutions have $s=-\hat\kappa$ and exist for all $n>0$. The second branch of solutions have $s=+\hat\kappa$ and only exist for $0<n<1$.
After regulating the volume of $H^2$ via $\Vol(H^2)=-2\pi$
(\textit{cf}. \cite{Hung:2011nu}),
the Euclidean defect free energy can be written, in the notation of \cite{Arav:2024wyg},  as
\begin{align}
\label{resultdefect}
	I&=\mp\frac{4}{n}\sqrt{\left({  g\mu^0}+\frac{\hat\kappa n}{2}\right) 
		\left({  g\mu^1}+\frac{\hat \kappa n}{2}\right) 
		\left({ g\mu^2}+\frac{\hat \kappa n}{2}\right) 
		\left({  g\mu^3}+\frac{\hat \kappa n}{2}\right) }
		\, F_{S^3}\, ,
\end{align}
where the overall sign refers to the two branches of solutions.
Moreover, supersymmetry of the two branches imposes that the holonomies \eqref{eq:Defects_Holonomies} are constrained by $\sum_I g\mu^I = -\hat{\kappa} (n\mp 1)$, or equivalently
\begin{equation}
\label{resultdefect2}
	\sum_{I=0}^3 \left( g\mu^I + \frac{\hat\kappa n}{2} \right) = \hat{\kappa} (n\pm 1) \,.
\end{equation}

In order to match \eqref{resultdefect} with \eqref{eq:Defect_Localization} restricted to the STU model, we first notice that in the chosen regular gauge, the $\Delta^I_{(2)}$ defined in \eqref{eq:DefinitionDelta} are precisely the holonomies \eqref{eq:Defects_Holonomies}. Furthermore, for the STU model we have $L^I_* = 1/2\sqrt{2}$ so that
$2\pi\ii\sigma_{(2)}^I = - \frac{1}{2}$. Thus, we find agreement and consistency with \eqref{resultdefect2} 
provided that
\begin{equation}
\label{eq:Defects_Identifications}
	\Delta_{(2)}^I \leftrightarrow g\mu^I \, , \qquad {2\pi} n\ii \sigma_{(2)}^I \leftrightarrow \frac{\hat\kappa n}{2}  \, , 
\end{equation}
and we set $\hat\kappa=-1$.
We highlight that the canonical prescription of regulating\footnote{
Another connection with localization is as follows. Consider the metric on a unit radius $H^2$: $\dd s^2= \dd\rho^2 + \sinh^2\rho\, \dd\phi^2$, with $\phi \sim \phi + 2\pi$, and the Killing vector $\xi=\partial_\phi$. 
Then, 
$\Phi = \vol - \frac{1}{2} * \dd \xi^\flat=\sinh\rho \, \dd\rho\wedge \dd\phi-\cosh\rho$ is equivariantly closed.
Using the BVAB theorem to get the volume we obtain $\Vol(H^2)= (\mathrm{boundary}) -2\pi$, where the boundary contribution is divergent and $-2\pi$ is the contribution from the fixed point.  In the spirit of holographic renormalization, we can add boundary terms to cancel the boundary divergence, finding a regulated volume $\Vol_{\mathrm{reg}}(H^2)=-2\pi$, as in (2.11) of \cite{Hung:2011nu}.}  the volume of $H^2$ in solutions with $H^2\times S^1$ boundary, via $\Vol(H^2)=-2\pi$,
gives a gravitational free energy that is the same as using ordinary holographic renormalization on the conformally rescaled metric \eqref{eq:Defect_Bdry_v1}.

\subsection{Taub-bolt saddle solutions}
\label{subsec:TaubBolt}

The second class of examples we consider consists of a spacetime $M$ with topology
\begin{equation}
\label{eq:TaubBolt_Topology}
	\mathbb{C} \hookrightarrow \mathcal{O}(-p) \to \Sigma_g \, , 
\end{equation}
where $\Sigma_g$ is a Riemann surface of genus $g$ and $-p$ is the degree of the complex line bundle constructed over it. We assume that $\xi = b \partial_\varphi$ rotates the $\mathbb{C}$ fibre, where $\varphi \sim \varphi + 2\pi$. If $g>1$ this is necessarily the case, but if $g=0$, $1$ it is also possible that $\xi$ rotates both fibre and base, in which case the symmetry is at least toric and is discussed in section \ref{subsec:BlackHoles} (and in more detail and generality in \cite{BenettiGenolini:2024hyd}). The conformal boundary $\partial M$ is the three-manifold $\cM_{g,p}$, which is by definition a circle bundle over $\Sigma_g$ with first Chern number $-p$.

By construction, the R-symmetry Killing vector vanishes at the origin of the $\mathbb{C}\cong \R^2$ fibre, and thus fixes the base $\Sigma_g$, which is a bolt with normal bundle $\cO(-p)$. The gravitational free energy can be immediately obtained from \eqref{Fgrav}: there is only one bolt, so we can choose $\kappa = +1$, and the only remaining sign characterizing the solution is the chirality of the spinor at the bolt
\begin{equation}
\label{eq:Fgrav_TaubBolt_Saddles}
	F_{\rm grav} = - \left[ \pm p\, \ii \cF(u_\pm) + \ii \cF_I (u_\pm) \mathfrak{p}^I_\pm \right] \frac{\pi}{G_4} \, .
\end{equation}
Here we recall that $ \zeta_I u^I_\pm = 1 $ and because of supersymmetry (see \eqref{eq:R_Symmetry_Constraint_Bolt}), we have
\begin{equation}
\label{eq:zetaIpI_TaubBolt_Saddles}
	\zeta_I \mathfrak{p}^I_\pm = \mp p -2(1-g) \, .
\end{equation}
For the purposes of holography, it is more convenient to express the result in terms of the ``UV'' data discussed in section \ref{sec:uvirrels}. 
There is a natural family of $\R^2$ non-compact submanifolds, namely  copies of the fibre $\mathbb{C}\cong \R^2$ of 
\eqref{eq:TaubBolt_Topology} at a chosen point on the 
Riemann surface base $\Sigma_g$. 
Equations
 \eqref{UVIR} and \eqref{yureln} relate $u^I_\pm$, the constant values of $u^I$ at the chosen point on the bolt $\Sigma_g$ of chirality $\pm$, and the holonomy and flux of the gauge field at the boundary
\begin{equation}\label{udelsigmagain}
	u^I_\pm = \pm \left( \Delta^I + \ii \beta \sigma^I \right) \, .
\end{equation} 
Localization in the bulk hence allows us to  deduce that the right-hand side, which 
are {\it a priori} functions on the conformal boundary $\partial M \cong 
\mathcal{M}_{g,p}$, are independent of the Riemann surface 
directions, and are therefore constant, although note that the individual elements $\Delta^I $ and $\sigma^I$ may not be constant. 

If we furthermore assume that the $\sigma^I$ are constant, {i.e.
$\Phi^I_0 \big\rvert_{\rm bdry} $ in \eqref{eq:PhiI0_Text} is constant,}
in this case
the constraint \eqref{eq:Rsymm_sigmaI} simplifies and we can write
\begin{equation}
\label{eq:zetaIsigmaI_TaubBolt_Saddles}
	2\pi \ii \zeta_I \sigma^I = - \frac{p}{\Vol(\Sigma_g)} \frac{2\pi}{b} \, .
\end{equation}
Together with the constraint on $u^I_\pm$, this also implies that the holonomy of the R-symmetry gauge field should satisfy
\begin{equation}
\label{eq:zetaIDeltaI_TaubBolt_Saddles}
	\zeta_I \Delta^I = \pm 1 + \frac{p}{\Vol(\Sigma_g)} \frac{2\pi}{b^2} \, .
\end{equation}
For the STU model, we note that the result \eqref{eq:Fgrav_TaubBolt_Saddles} can  be written as 
\begin{equation}
\label{eq:Fgrav_TaubBolt_Saddles_STU}
	F_{\rm grav} = - 2\sqrt{u_\pm^0 u_\pm^1 u_\pm^2 u_\pm^3} \left[ \pm 2 p  + \sum_{I=0}^3\frac{\mathfrak{p}^I_\pm}{u_\pm^I}  \right]  F_{S^3}\, .
\end{equation}

Solutions with the topology \eqref{eq:TaubBolt_Topology} and $p\neq 0$ are not known in the STU model. Within the STU model, solutions with non-vanishing scalars are only known via analytical and numerical analysis when $p=0$ \cite{Bobev:2020pjk}. The topology in this case is that of a direct product $\mathbb{C} \times \Sigma_g$. The resulting gravitational free energy obtained from \eqref{eq:Fgrav_TaubBolt_Saddles_STU} is  
\begin{equation}
\label{eq:Fgrav_TTI}
	\Fgrav = - 2\sqrt{u^0_\pm u^1_\pm u^2_\pm u^3_\pm} \sum_{I=0}^3 \frac{\mathfrak{p}^I_\pm}{u_\pm^I} F_{S^3}\, ,
\end{equation}
and it is a function of the fluxes $\mathfrak{p}^I_\pm$ and the gauge field variables $\Delta^I$, $\sigma^I$ that are constrained to satisfy \eqref{eq:zetaIpI_TaubBolt_Saddles} and, if we also assume that $\sigma^I$ are constant,
\eqref{eq:zetaIsigmaI_TaubBolt_Saddles} and \eqref{eq:zetaIDeltaI_TaubBolt_Saddles}, namely
\begin{equation}
\label{eq:TTI_Constraints}
	\sum_{I=0}^3 \mathfrak{p}^I_\pm = 2(g-1) \, , \qquad \sum_{I=0}^3 \Delta^I = \pm 1 \, , \qquad \sum_{I=0}^3 \sigma^I = 0 \, .
\end{equation}
Formula \eqref{eq:Fgrav_TTI} and the constraints \eqref{eq:TTI_Constraints} match the expressions found by explicit solution of the equations of motion in \cite{Bobev:2020pjk}.\footnote{The map between \cite{Bobev:2020pjk} and \eqref{eq:Fgrav_TTI} is given by 
$\left(\xi^2 g^2 \right)_{\rm there} = \frac{1}{2}$ and $\hat{u}^I_{\rm there} = 2\pi u^I$.
}
The dual field theory is ABJM on $S^1\times \Sigma_g$ deformed by an R-symmetry flux through $\Sigma_g$ (a ``topological twist"), as required in order to preserve supersymmetry, and the free energy has a Hilbert space interpretation as a topologically twisted index \cite{Benini:2015noa, Benini:2016hjo, Closset:2016arn}. As expected, \eqref{eq:Fgrav_TTI} matches the large $N$ limit of the topologically twisted index computed in \cite{Benini:2015eyy}, and the constraints \eqref{eq:TTI_Constraints} match those required to have the large $N$ limit scaling like $N^{\frac{3}{2}}$ and thus a bulk gravity dual.

Explicit solutions with the topology \eqref{eq:TaubBolt_Topology} and $p\neq 0$ are known in minimal supergravity, both for $g=0$ \cite{Martelli:2012sz} and for $g>0$ \cite{Toldo:2017qsh}.\footnote{In the case of \cite{Martelli:2012sz}, only the non-self-dual $\frac{1}{4}$-BPS solutions satisfy the assumptions in this section, as the R-symmetry Killing vector rotates only the fibre of the fibration. Whilst the $\frac{1}{2}$-BPS solutions have the same topology \eqref{eq:TaubBolt_Topology}, the R-symmetry Killing vector rotates the sphere as well. This case is then captured more effectively using methods from toric geometry discussed in \cite{BenettiGenolini:2024hyd}.} Applying the results of appendix \ref{mingaugedsugraapp} to \eqref{eq:Fgrav_TaubBolt_Saddles}, we find that the gravitational free energy of these solutions of minimal gauged supergravity is given by\footnote{{When comparing with \cite[(4.35)]{BenettiGenolini:2019jdz}, recall that a $\pm$ solution there corresponds to a $\mp$ solution here.}}
\begin{equation}
	\Fgrav = \frac{\pi}{2G_4} \left(\pm \frac{p}{4} + 1 - g \right) \, ,
\end{equation}
where from \eqref{eq:zetaIpI_TaubBolt_Saddles} the flux through the bolt is given by
\begin{equation}
\label{eq:TaubBolt_Constraints_Minimal1}
	\mathfrak{p}_\pm^0 = \mp \frac{p}{4} - \frac{1-g}{2}\,.
	\end{equation}
If we also assume that $\sigma^0$ is constant (which would follow from the spinor bilinear $P$ being constant on the boundary or, equivalently, the
right-hand side of \eqref{localconssigmaI} being constant on the boundary) then from 
\eqref{eq:zetaIsigmaI_TaubBolt_Saddles}, \eqref{eq:zetaIDeltaI_TaubBolt_Saddles}
we have the UV data
\begin{equation}
\label{eq:TaubBolt_Constraints_Minimal}
	\mathfrak{p}_\pm^0 = \mp \frac{p}{4} - \frac{1-g}{2} \, , \qquad 2\pi\ii \sigma^0 = - \frac{p}{\Vol(\Sigma_g)} \frac{2\pi}{4b} \, , \qquad \Delta^0 = \pm \frac{1}{4} + \frac{p}{\Vol(\Sigma_g)} \frac{2\pi}{4b^2} \, .
\end{equation}
It is immediate to check that $\Fgrav$ matches the on-shell action computed for the explicit solutions
(see (3.73), (3.74) of \cite{Toldo:2017qsh}), and that the flux through the bolt is also in agreement (see (3.48) of \cite{Toldo:2017qsh}, accounting for the factor of 2 difference in the normalization used in \eqref{pfrak}).
Moreover, it is also straightforward to check that the expression for $\sigma^0$ given in \eqref{eq:TaubBolt_Constraints_Minimal} is in agreement with that constructed out of the Killing spinors of the solution (taking care that the choice of radial coordinate in
\cite{Toldo:2017qsh} effectively fixes the otherwise arbitrary norm of the Killing spinor, and thus in turn the weights of the Killing vector and $\sigma^0$).

More interestingly, the expressions obtained from application of the localization theorem provide predictions for the large $N$ limit of the free energy on $\partial M\cong \cM_{g,p}$ even in cases when the gravity solution is not known, and they match the field theory result, when this is known (see also \cite{Hong:2024uns}). For instance, \eqref{eq:Fgrav_TaubBolt_Saddles_STU} gives a prediction for the large $N$ limit of the free energy of ABJM theory on $\cM_{g,p}$ without the need of an explicit solution of the STU model on \eqref{eq:TaubBolt_Topology}. Remarkably, the two possibilities corresponding to the two chiralities of the spinor at the bolt match the two solutions of the extremization problem in the large $N$ limit of ABJM found in 
\cite{Toldo:2017qsh}, which read 
\begin{align}\label{twftfe}
	-\log Z^\text{ABJM}_{\mathcal{M}_{g,p}}&=-\frac{2\pi N^{3/2}}{3}\sqrt{2 k [m_1][m_2][m_3][m_4]}\left(-2p+\sum_{i=1}^4\frac{\mathfrak{n}_i}{[m_i]}\right)\,, \nn
	-\log Z^\text{ABJM}_{\mathcal{M}_{g,p}}&=-\frac{2\pi N^{3/2}}{3}\sqrt{2 k [\hat m_1][\hat m_2][\hat m_3][\hat m_4]}\left(+2p+\sum_{i=1}^4\frac{\mathfrak{\hat n}_i}{[\hat m_i]}\right)\,.
\end{align}
The variables $[m_i]$ and $[\hat m_i]\equiv 1-[m_i]$ are interpreted as the (fractional part of the) masses and $\mathfrak{n}_i$, $\mathfrak{\hat n}_i$ as the magnetic charges of the chiral fields, and they satisfy the constraints 
\begin{equation}\label{constraintstw}
	\sum_{i=1}^4 [m_i]=\sum_{i=1}^4 [\hat m_i]= 1 \,, \quad
	 \sum_{i=1}^4 \mathfrak{n}_i = p-2(1-g)\,, \quad \sum_{i=1}^4 \mathfrak{\hat n}_i = -p-2(1-g)\,.
\end{equation} 
If we identify the sets of variables via
\begin{equation}
\label{eq:TaubBolt_HolographicMap}
	\mathfrak{u}^I_+ \leftrightarrow [\hat{m}_i]\, , \quad \mathfrak{u}^I_- \leftrightarrow [{m}_i]\, , \quad \mathfrak{p}^I_+ \leftrightarrow \hat{\mathfrak{n}}_i\, , \quad \mathfrak{p}^I_- \leftrightarrow {\mathfrak{n}}_i\, ,
\end{equation}
then after using \eqref{abjmGNrel} we immediately see that our expressions for the free energy and constraints \eqref{eq:Fgrav_TaubBolt_Saddles_STU} and
\eqref{eq:zetaIpI_TaubBolt_Saddles} correctly reproduce \eqref{twftfe} and \eqref{constraintstw}.
{Note that on the supergravity side the UV/IR relation \eqref{udelsigmagain} allows for both $\sigma^I$ (parametrizing mass deformations/boundary geometry) and $\Delta^I$ (parametrizing holonomies), which are not necessarily constant in general,
and still giving the same free energy.}

Even more generally, the authors of \cite{Toldo:2017qsh} computed the large $N$ limit of the free energy on $\cM_{g,p}$ for a large class of quiver gauge theories with $U(N)$ gauge factors, bifundamental and (anti-)fundamental chiral multiplets, and Chern--Simons levels with some additional
conditions imposed which ensures there is a well-behaved M-theory description at large $N$, with degrees of freedom scaling like $N^{3/2}$. The theory is assumed to   have a polynomial superpotential
\begin{equation}
	W = \sum_\alpha \prod_i \Phi_i^{q^\alpha_i} \, ,
\end{equation}
where $\Phi_i$ are the chiral multiplets, labelled by 
$i$, while $\alpha$ labels monomial terms in the superpotential. 
The $q_i^\alpha\in\Z_{\geq 0}$ hence specify $W$.   
The large $N$ limit is computed by extremizing an effective twisted superpotential $\cW$ (also referred to as Bethe potential) \cite{Benini:2015eyy}.\footnote{This is the superpotential of a $(2,2)$ two-dimensional theory obtained reducing the theory to two-dimensional $\R^2$. The fact that $\cW$ is relevant also for three-dimensional quantum field theories on $\cM_{g,p}$ comes from the realization that their partition function on $\cM_{g,p}$ is in fact determined by the low-energy description on the Coulomb branch of the theory formulated on $\R^2\times S^1$ \cite{Closset:2017zgf}.} The authors of \cite{Toldo:2017qsh}, building on the work of \cite{Hosseini:2016tor, Hosseini:2016ume}, find two solutions to the extremization problem, namely\footnote{
Here we have corrected the expression (5.16) in \cite{Toldo:2017qsh}, 
taking the variable of $\cW_{\rm ext}$ to be $[\hat{m}_i] \equiv 1 - [m_i]$, and changing the sign of the first term.
}
\begin{align}
	- \log Z_{\cM_{g,p}} &= p\, 2\pi \ii \cW([ m_i]) - \sum_i \mathfrak{n}_i \partial_i \left[ 2\pi\ii \cW([m_i]) \right] \, , \nn
\label{eq:TaubBolt_Z_GenQuiver}
	- \log Z_{\cM_{g,p}} &= - p\, 2\pi \ii \cW([\hat{m}_i]) - \sum_i \hat{\mathfrak{n}}_i \partial_i \left[ 2\pi\ii \cW([\hat{m}_i]) \right] \, ,
\end{align}
where the variables are constrained to satisfy
\begin{align}
	\sum_i q_i^\alpha [m_i] &= 1 = \sum_i q^\alpha_i [\hat{m}_i] \, , \nn
\label{eq:TaubBolt_Constraints_GenQuiver}
	\sum_i q^\alpha_i \mathfrak{n}_i &= p - 2 (1-g) \, , \qquad	\sum_i q^\alpha_i \hat{\mathfrak{n}}_i = - p - 2(1-g) \, .
\end{align}
Note there is one constraint for each monomial term in the superpotential $W$. 
In this case we should not use the simple dictionary \eqref{eq:TaubBolt_HolographicMap}, but it is clear that if we write
\begin{align}
	\frac{\pi}{G_4} \ii \cF( \mathfrak{u}_+) \leftrightarrow 2\pi\ii \cW_{\rm ext}([\hat{m}_i]) \, , \quad \frac{\pi}{G_4} \ii \cF( \mathfrak{u}_-) \leftrightarrow 2\pi\ii \cW_{\rm ext}([{m}_i]) \, , \nn 
	\zeta_I \mathfrak{u}^I_+ \leftrightarrow \sum_i q^\alpha_i [\hat{m}_i] \, , \quad \zeta_I \mathfrak{u}^I_- \leftrightarrow \sum_i q^\alpha_i [m_i] \, , \quad \zeta_I \mathfrak{p}^I_+ \leftrightarrow q^\alpha_i \hat{\mathfrak{n}}_i \, , \quad \zeta_I \mathfrak{p}^I_- \leftrightarrow q^\alpha_i \mathfrak{n}_i\, ,
\end{align}
then \eqref{eq:Fgrav_TaubBolt_Saddles} and \eqref{eq:zetaIpI_TaubBolt_Saddles} match \eqref{eq:TaubBolt_Z_GenQuiver} and \eqref{eq:TaubBolt_Constraints_GenQuiver}. We remark that the two sets of conditions \eqref{eq:zetaIpI_TaubBolt_Saddles} and \eqref{eq:TaubBolt_Constraints_GenQuiver} have been derived in different frameworks: the first set purely by analysis of supersymmetric solutions in gauged supergravity, the second set by a large $N$ limit of the free energy of quiver gauge theories. A crucial step in the match is the identification of the gauged supergravity prepotential evaluated at the fixed sets with the extremum of the effective twisted superpotential. This relation had  first been observed for the topology $p=0$ (that is, when the dual observable is the topologically twisted index) in \cite{Hosseini:2016tor}. Here we have shown that
the conjectural holographic relation is valid for any $p$. In turn, the gauged supergravity prepotential is also related to the large $N$ limit of the partition function on $S^3$ by the conjecture proved in section \ref{subsec:Deformations_AdS4} \cite{Hosseini:2016tor, Zan:2021ftf}. 

Building on these results, we can further conjecture that quite generally for this class of gauge theories,
the on-shell action \eqref{Fgrav} on $M$ gives the large $N$ limit of a holographic theory on $\partial M$ provided $\ii\cF$ is identified with the extremum of the effective twisted superpotential $2\pi\ii \cW_{\rm ext}$. 

\section{Complex solutions and black holes}\label{sec:six}

The examples considered in the previous section are solutions satisfying the reality conditions \eqref{eq:Reality_Condition}. However, the equations of motion \eqref{eq:4d_N2_EinsteinEOM} and the Killing spinor equations \eqref{eq:Euclidean_KSE_epsilon} are analytic in the supergravity fields, so we can also consider analytically continued \textit{complex} solutions. We shall now see that, despite the fact that the derivation in section \ref{evalactsec} relies on the reality assumptions \eqref{eq:Reality_Condition}, the final result\footnote{In this section we assume that $\tcF(\tX)=- \cF(\tX)$, which is satisfied, for example, in the STU model. For prepotentials not satisfying
this condition, it is straightforward to get analogous results using \eqref{Fgrav_Complex}.}
\eqref{Fgrav} still holds even for the complexified solutions, for some specific examples.

As a first example, we return to the solution of the STU model on $\R^4$ with vanishing gauge fields found in \cite{Freedman:2013oja}, which is reviewed in some detail in appendix \ref{subsubsec:FP}. In order to derive the gravitational free energy \eqref{FP2}, we assumed that $z^i, \tz^i \in \R$. In particular, these scalar fields depend on three constants $c_i$, which we took to be real. However, in the construction of the explicit solutions in \cite{Freedman:2013oja} one can analytically continue $c_i$ and take them to be complex. On the other hand, the result \eqref{FP2}, which here in particular takes the form \eqref{eq:FP_Final_Fgrav}, reproduces the results in \cite{Freedman:2013oja} even if $c_i\in\mathbb{C}$, which is suggestive that the localization approach can be extended when the reality constraints are relaxed.

\subsection{Black holes and indices}
\label{subsec:BlackHoles}

Another interesting class of complex supersymmetric solutions appears when considering non-extremal deformations of Wick-rotated supersymmetric black holes \cite{Cabo-Bizet:2018ehj}. 
By supersymmetric black holes, we mean (real) Lorentzian and supersymmetric black hole solutions. Typically the supersymmetric Lorentzian solutions have pathologies, such as naked singularities or closed time-like
curves, unless one imposes that they are also extremal with vanishing temperature. In this case the metric includes a near-horizon $AdS_2$ factor with an infinite throat \cite{Kunduri:2007vf}, which makes it challenging to compute the on-shell Euclidean action after Wick rotation. 
To resolve this issue, it has been proposed that one should consider a broader class of Euclidean solutions, with complex metrics in general, that are 
supersymmetric but non-extremal \cite{Cabo-Bizet:2018ehj,Cassani:2019mms, Bobev:2019zmz}.\footnote{For a discussion of the relation with the extremization advocated in \cite{Benini:2015eyy, Benini:2016rke} see \cite{Bobev:2020pjk}.} 
In this setting it is possible to compute the on-shell action to obtain the free energy and
develop a thermodynamic picture, including a quantum statistical relation. 
Importantly, the on-shell action computed on the supersymmetric non-extremal solutions can be expressed in terms of ``reduced'' chemical potentials which remain finite (and complex) when taking the extremal limit. The reduced chemical potentials are the natural variables used to express the dual supersymmetric observables, and we shall find them appearing below when considering the UV-IR relations. Furthermore, by taking a Legendre transform of the 
on-shell action one can recover the entropy of the extremal supersymmetric Lorentzian solutions.

The field theory observable dual to the on-shell action can be identified as a supersymmetric index on $\Sigma \times S^1$. That is, it can be interpreted as a trace over the Hilbert space of the dual SCFT quantized on $\Sigma$, with supersymmetric boundary conditions that guarantee that the resulting observable is independent of the size of $S^1$ \cite{Witten:1982df}. This explains why one can use supersymmetric non-extremal solutions to extract information about the extremal solutions.

We are therefore interested in complex supersymmetric and non-extremal solutions of the Euclidean gauged supergravity theory.
Such solutions do not satisfy, in general, the reality conditions that we imposed in \eqref{eq:Reality_Condition} in order to derive the localization formula.
Nevertheless, proceeding formally we obtain formulae consistent with known results,  which strongly
indicates that the localization results can be extended to this larger class of Euclidean solution.

The relevant supersymmetric, non-extremal solutions have $M$ with topology $\Sigma \times \R^2$. 
Topologically the conformal boundary is $\partial M=\Sigma \times S^1$, but the metric may not in general be a direct product. Different choices of $\Sigma$ and Killing vector $\xi$ correspond to different black hole solutions after performing a Wick rotation and taking the extremal limit. Assuming
that the solutions exist we can formally compute the gravitational free energy and compare with some known solutions.

\subsection{Topologically twisted black holes}
\label{topbhsec}

We first consider the case of a direct product $M={\Sigma_g \times \R^2}$ where $\Sigma_g$ is a compact Riemann surface of genus $g$ and $\xi$ rotates only the $\R^2$ factor. This case was also considered in section \ref{subsec:TaubBolt} as the subcase of a bolt with trivial normal bundle $\cO(0)$: the resulting gravitational free energy for a generic prepotential is
\begin{equation}
	\Fgrav = - \frac{\pi}{G_4} \ii \cF_I(u_\pm) \mathfrak{p}^I_\pm\, ,
\end{equation}
with the constraint on the flux given by
\begin{equation}\label{topbhfluxcons}
	\zeta_I \mathfrak{p}^I_\pm = 2(g-1) \, ,
	\end{equation}
and, assuming $\sigma^I$ are constant, the UV constraints
\begin{equation}
	 \zeta_I \Delta^I = \pm 1 \, , \qquad \zeta_I \sigma^I = 0 \, .
\end{equation}
Recall that the $\sigma^I$ are given as in \eqref{eq:sigmaI}. 
If we further set the mass deformations to zero, we can take 
\begin{align}
\sigma^I=0\,,
\end{align} (since the bolt has trivial normal bundle) 
and we then have
\begin{equation}
\label{Fgavtoptwbh}
	\Fgrav = - \frac{\pi}{G_4} \ii \cF_I(\Delta) \mathfrak{p}^I_\pm \, .
\end{equation}
As already mentioned, the dual observable is the topologically twisted index on $\Sigma_g\times S^1$ \cite{Benini:2015noa, Benini:2016hjo, Closset:2016arn}.

Solutions of this topology, with $\sigma^I = 0$,  
are supersymmetric deformations of static dyonic black holes with supersymmetry preserved via a topological twist on $\Sigma_g$ (see e.g. \cite{Romans:1991nq, Caldarelli:1998hg, Gauntlett:2001qs, Cacciatori:2009iz} for some early references). Equation \eqref{Fgavtoptwbh} with $ \zeta_I \Delta^I = \pm 1$ provides an expression for the grand-canonical free energy of these solutions in terms of the magnetic charges and the reduced electrostatic potentials.\footnote{The latter follows from the fact that the electrostatic potential is defined as the difference of $\xi \hook A^I$ between the conformal boundary and the horizon, and this agrees with the definition of $\Delta^I$, which are necessarily constant in this topology {when $\sigma^I=0$} (see below \eqref{eq:DefinitionDelta}).} As suggested in \cite{Cabo-Bizet:2018ehj}, extending the canonical prescriptions of \cite{Gibbons:1976ue} to complex solutions, the entropy of the Lorentzian extremal supersymmetric static dyonic black hole can be obtained by taking the real part of the (constrained) Legendre transform of $\Fgrav$ with respect to the $\Delta^I$, obtaining a function of the conjugate variable (the electric charges). Concretely, one extremizes
\begin{equation}
\label{eq:S_top_BH}
	S = -\Fgrav - \ii q_I \Delta^I\, ,
\end{equation}
with the constraint $\zeta_I \Delta^I = \pm 1$, and 
then imposes that one obtains a real positive solution $S_*(q, \mathfrak{p})$. The entropy of static extremal supersymmetric black hole solutions in a generic $\cN=2$ gauged supergravity is given in terms of a (quartic) invariant of the special geometry describing the scalar manifold \cite{Halmagyi:2014qza}. It would be interesting to show that this result can be recovered using the extremization of \eqref{eq:S_top_BH} described above.

Analytic solutions of the above type are known for the special case of the $X^0X^1$ model, which is a truncation of the STU model obtained by setting pairwise scalars and gauge fields to be equal \cite{BenettiGenolini:2023ucp}. For these solutions,  
the expression \eqref{Fgavtoptwbh} matches the gravitational free energy. Specifically, for the $X^0X^1$ model we find
\begin{equation}
	\Fgrav = \mp \frac{2\pi}{G_4} \left( \Delta^1_\pm\mathfrak{p}^0_\pm  + \Delta^0_\pm\mathfrak{p}^1_\pm  \right) \, , \qquad 2\Delta^0_\pm + 2\Delta^1_\pm = \pm 1 \,, \quad \mathfrak{p}^0_\pm + \mathfrak{p}^1_\pm = g-1 \, ,
\end{equation}
which matches eq. (4.15) of \cite{BenettiGenolini:2023ucp} upon identifying 
\begin{equation}
	\left( 2\Delta^I \right)_{\rm here} \leftrightarrow \left( \varphi_I \right)_{\rm there} \, , \qquad \left( \mathfrak{p}^I_\pm \right)_{\rm here} \leftrightarrow \left( 2G_4 P_I \right)_{\rm there} \, .
\end{equation}
For this case, the constrained extremization of \eqref{eq:S_top_BH} gives $q_0 = q_1$ and $\mathfrak{p}^0_\pm = \mathfrak{p}^1_\pm = (g-1)/2$, from which
$S = \frac{\pi}{G_4} \mathfrak{p}^0_\pm \mp \frac{\ii}{2} q_0 $. Demanding that this is real then imposes $q_0=0$ and the entropy is given by
\begin{equation}
S_* = \frac{\pi}{G_4} \mathfrak{p}^0_\pm \, ,
\end{equation}
with $\mathfrak{p}^0_\pm = (g-1)/2$.

\subsection{Rotating black holes }
\label{rotatingbhs}

Next we consider solutions with topology $S^2\times \R^2$ with $\xi = \varepsilon \partial_{\varphi_1} + \partial_{\varphi_2}$, where $\varphi_{1,2}\sim \varphi_{1,2} + 2\pi$ are angular coordinates on $S^2$ and $\R^2$, respectively. 
For $\varepsilon\ne 0$, there are two isolated fixed points, corresponding to the origin of $\R^2$ at the North and South poles of $S^2$, each of which is labelled with a chirality $\chi_{N,S}$ and a sign $\kappa_{N,S}$ (as introduced in section~\ref{subsec:ActionNut}); when $\varepsilon =0$ there is a bolt fixed point set as just discussed in section \ref{topbhsec}.
The weights of the $U(1)$ action at the poles (in the conventions of
\cite{BenettiGenolini:2024hyd}) are given by
\begin{equation}
\label{eq:Weights_RotatingBlackHole}
	(b^1_N, b^2_N) = (-1,\varepsilon) \, , \qquad (b^1_S, b^2_S) = (-\varepsilon, -1) \, .
\end{equation}
A first expression for the gravitational free energy, expressed in terms of fixed point data, can be obtained from \eqref{Fgrav}:
\begin{equation}
\label{eq:Fgrav_RotatingBlackHole_1}
		\Fgrav = \frac{\pi}{G_4} \frac{1}{\varepsilon} \left[ \chi_N \left( 1 + \chi_N \varepsilon \right)^2 \ii \cF(u_N) - \chi_S \left( 1 - \chi_S\varepsilon \right)^2 \ii \cF(u_S) \right] \, .
\end{equation}
An expression more useful for applications to holography can be found following the discussion in section \ref{sec:uvirrels}. We first consider the flux of the gauge fields through the $S^2$: applying the BVAB theorem on $\Phi^I_{(F)}$ in \eqref{PhiF} gives
\begin{align}
	\mathfrak{p}^I &= \frac{1}{2\varepsilon} \left( \Phi^I_0\rvert_N - \Phi^I_0\rvert_S \right) \, .
\end{align}
Moreover, using \eqref{wtscrels} we find a condition for the R-symmetry gauge field flux
\begin{equation}\label{barbara}
	\zeta_I \mathfrak{p}^I = \frac{\kappa_N + \chi_S\kappa_S }{\varepsilon} + \chi_N \kappa_N - \kappa_S \, .
\end{equation}

The fixed point set here contains disconnected components (namely two nuts), and thus we need to impose a global regularity condition in order to relate $\kappa_N$ and $\kappa_S$, as mentioned in section \ref{subsec:ActionNut}. Namely, the spinor must be regular on the $S^2$ submanifold. This implies that it should have a definite fixed charge under the $U(1)$ generated by $\partial_{\varphi_2}$, which 
recall rotates the $\R^2$ direction and hence 
by construction fixes the $S^2$ submanifold. In the notation of the local frame at the nut introduced in section \ref{subsec:ActionNut}, we see from the weights in \eqref{eq:Weights_RotatingBlackHole} that the normal directions to $S^2$ at the North pole are $\overline{e}^{12}$, and thus, from \eqref{kappacond}, we find that the charge of the Killing spinor at the North pole is $-\kappa_N/2$. On the other hand, again from the weights in \eqref{eq:Weights_RotatingBlackHole}, we see that the normal directions to $S^2$ at the South pole are $\overline{e}^{34}$, and from \eqref{kappacond} we find that the charge of the Killing spinor at the South pole is $\chi_S \kappa_S/2$. Requiring that the spinor is regular everywhere on $S^2$ then implies that we should choose $\kappa_N = - \chi_S \kappa_S$ (see also \cite{BenettiGenolini:2024hyd}). 
The R-symmetry flux \eqref{barbara} hence simplifies to 
\begin{align}\label{fluxconstrS2}
\zeta_I \mathfrak{p}^I = - \kappa_S (1 + \sigma) \, ,
\end{align}
where $\sigma \equiv \chi_N \chi_S$.\footnote{{The sign $\sigma$ should not be confused with the scalar variables $\sigma^I$.  We also note that if we want to further compare with the notation of \cite{BenettiGenolini:2024hyd} we should also relate $\kappa_N\chi_N=(\sigma_1)_\text{there}$ and $\kappa_S=(-\sigma_0)_\text{there}$.}}  
The case $\sigma=+1$, when the chiralities are the same at each pole, corresponds to a topological twist with $\zeta_I \mathfrak{p}^I = - 2\kappa_S $, which is the case that allows us to take the limit $\varepsilon\to 0$ recovering
\eqref{topbhfluxcons} (with $\kappa_S = +1$); while the case $\sigma=-1$, when the chiralities are opposite at each pole, corresponds to an anti-twist
\cite{Ferrero:2021etw}, has $\zeta_I \mathfrak{p}^I=0$
(these two cases are referred to as the A-twist and the identity  twist, respectively, in \cite{Hristov:2021qsw}).

For the fluxes through the non-compact $\R^2$ submanifolds with origin at the North and South poles, from the discussion in section \ref{subsec:UVIR_Noncompact} we introduce
\begin{equation}
\label{eq:ys_RotatingBlackHole}
	y^I_N \equiv \frac{1}{2b^1_N} \Phi^I_0 \rvert_N = - \frac{1}{2} \Phi^I_0 \rvert_N \, , \qquad y^I_S \equiv \frac{1}{2b^2_S}\Phi^I_0 \rvert_S = - \frac{1}{2} \Phi^I_0 \rvert_S \, , \qquad y^I_N = y^I_S - \varepsilon \mathfrak{p}^I \, ,
\end{equation}
which are then related to $u^I_{N,S}$ by \eqref{yurelnuts}. This allows us to write $\Fgrav$ as
\begin{equation}
\label{eq:Fgrav_RotatingBlackHole_2}
	\Fgrav = \frac{\pi}{G_4} \frac{1}{\varepsilon} \left[ \chi_N \ii \cF(y_S - \varepsilon \mathfrak{p}) - \chi_S \ii \cF(y_S) \right] \, .
\end{equation}
In terms of holographic boundary data we have
\begin{equation}
\label{eq:HolographicData_RotatingBlackHole}
	y^I_N = \Delta^I_N - 2\pi \ii \sigma^I_N \, , \qquad y^I_S = \Delta^I_S - 2\pi\ii \sigma^I_S \, ,
\end{equation}
where $\Delta^I_{N,S}$, $\sigma^I_{N,S}$ are constants (since they are independent of the boundary $S^1$ coordinate bounding the associated $\R^2$); in general they are not equal to each other. From 
\eqref{yurelnuts} we have the constraints
\begin{equation}
\label{eq:HolographicData_RotatingBlackHole_2}
	\zeta_I y^I_N = \chi_S \kappa_S \left( 1 + \sigma \chi_S \varepsilon \right) \, , \qquad \zeta_I y^I_S  = \chi_S \kappa_S \left( 1 - \chi_S\varepsilon \right) \, ,
\end{equation}
which gives rise to corresponding constraints on $\Delta^I_{N,S}$, $\sigma^I_{N,S}$ in addition to the constraint \eqref{localconssigmaI}.
While the holonomies and mass deformations on the boundary are not constant in general, it is only this
constant UV data $\Delta^I_{N,S}$, $\sigma^I_{N,S}$ that appears in the free energy.
Recall from \eqref{eq:sigmaI} that this data includes both the source for 
 boundary dimension-2 mass operator as well as $*_2 \dd\eta_{(0)}=0$, evaluated along the $S^1$ at the $N,S$ poles, respectively.

The solutions described in this section would be supersymmetric, non-extremal (complex) deformations of Wick-rotated supersymmetric rotating dyonic black holes. The expression \eqref{eq:Fgrav_RotatingBlackHole_2} agrees with the known supergravity solutions, and also with the conjectures based on the study of the large $N$ behavior of the dual field theory observable.

For example, if we further assume that we are preserving supersymmetry with the anti-twist
(i.e. $\sigma=-1$ or equivalently $\chi_N = - \chi_S$), and restrict to the STU model we find
\begin{align}
	\Fgrav &= \frac{2\pi}{G_4}\frac{1}{\varepsilon} \chi_N \left[ \sqrt{ ( y^0 - \varepsilon \mathfrak{p}^0) ( y^1 - \varepsilon \mathfrak{p}^1) ( y^2 - \varepsilon \mathfrak{p}^2) ( y^3 - \varepsilon \mathfrak{p}^3 ) } + \sqrt{y^0y^1y^2y^3} \right] \,, \nn
\label{eq:Fgrav_RotatingBlackHole_STU}
	\sum_{I=0}^3 y^I &= - \kappa_S \left( \chi_N + \varepsilon \right) \, , \qquad \sum_{I=0}^3 \mathfrak{p}^I = 0 \, .
\end{align}
There are no known supersymmetric non-extremal deformations of rotating black holes in the STU model 
of such a general form. However, we notice that
when $\mathfrak{p}^I=0$, \eqref{eq:Fgrav_RotatingBlackHole_STU} reproduces the large $N$ limit of the leading singularity of the Cardy limit of the superconformal index of ABJM theory \cite{Choi:2019zpz, Bobev:2022wem}. 
Also, supersymmetric non-extremal solutions obtained by deforming a rotating electrically charged black hole are known in the $X^0 X^1$ model \cite{Cassani:2019mms}.
Restricting to the $X^0X^1$ model by setting $y^0= y^1$, $y^2 = y^3$
and also setting $\mathfrak{p}^I = 0$, \eqref{eq:Fgrav_RotatingBlackHole_STU} matches the on-shell action computed using holographic renormalization in \cite{Cassani:2019mms}. Interestingly, \eqref{eq:Fgrav_RotatingBlackHole_STU} also allows the possibility of adding magnetic charges through the $S^2$, provided they sum to zero: this is consistent with the known results about supersymmetric rotating black holes in the STU model \cite{Hristov:2019mqp}. We also note that the limit $\varepsilon\to 0$ is not defined for this expression.

Analogously, if we instead assume that $\sigma=+1$, so that we are preserving supersymmetry with the topological twist, and restrict to the STU model, then
\begin{align}
	\Fgrav &= \frac{2\pi}{G_4}\frac{1}{\varepsilon} \chi_N \left[ \sqrt{ ( y^0 - \varepsilon \mathfrak{p}^0) ( y^1 - \varepsilon \mathfrak{p}^1) ( y^2 - \varepsilon \mathfrak{p}^2) ( y^3 - \varepsilon \mathfrak{p}^3 ) } - \sqrt{y^0y^1y^2y^3} \right] \,, \nn
\label{eq:Fgrav_TwistedRotatingBlackHole_STU}
	\sum_{I=0}^3 y^I &= \kappa_S \left( \chi_N - \varepsilon \right) \, , \qquad \sum_{I=0}^3 \mathfrak{p}^I = -2\kappa_S \, .
\end{align}
Supersymmetric non-extremal rotating solutions of the STU model with supersymmetry on $S^2$ preserved via a topological twist are not known, but if they exist, \eqref{eq:Fgrav_TwistedRotatingBlackHole_STU} would be their action. Moreover, \eqref{eq:Fgrav_TwistedRotatingBlackHole_STU} reproduces the large $N$ limit of the refined topologically twisted index \cite{Benini:2015noa} as computed in \cite{Hosseini:2022vho}.
We also note that the limit $\varepsilon\to 0$ is well defined for this expression, in contrast to the previous case.

As already discussed, one use of the supersymmetry-preserving non-extremal solutions discussed in this section is to obtain the entropy of the Lorentzian extremal supersymmetric black holes. To do so, one should take the real part of the Legendre transform of $\Fgrav$ with respect to $\Delta^I$ (in terms of which $y^I$ are expressed) and $\varepsilon$, subject to the constraint in \eqref{eq:Fgrav_TwistedRotatingBlackHole_STU}, obtaining the micro-canonical entropy as a function of the electric charges and the angular momentum. Indeed, starting from \eqref{eq:Fgrav_RotatingBlackHole_STU} and \eqref{eq:Fgrav_TwistedRotatingBlackHole_STU}, one obtains the entropy of the known supersymmetric black holes in the STU model, respectively found in \cite{Hristov:2019mqp} and \cite{Hristov:2018spe}.

\section{Orbifold solutions}
\label{sec:orbifoldexamples}

Until now we have assumed that $M$ is a manifold and that the fixed point sets do not have orbifold singularities. However, this is not necessary, and the results can be straightforwardly  generalized to include the case that a fixed point set $F\subset M$ is an orbifold with orbifold structure group with order $d_F\in \mathbb{N}$, by extension of the argument in section \ref{subsec:Using_BVAB}. Under these assumptions, the contribution from the fixed points to the gravitational free energy written in \eqref{I4dexpand} takes the form
\begin{equation}
\label{eq:BVAB_Orbifold}
	I^{\rm FP}_{\rm OS} = \frac{\pi}{2G_4} \left\{ \sum_{F_0} \frac{1}{d_{F_0}} \frac{\Phi_0}{b_1 b_2} + \sum_{F_2} \frac{1}{d_{F_2}} \int_{F_2} \frac{\Phi_2}{2\pi b} - \frac{\Phi_0 c_1(L)}{b^2} \right\} \, ,
\end{equation}
denoting by $F_0$ the nuts and $F_2$ the bolts.
As explained in section \ref{subsec:Using_BVAB}, the boundary contribution consists of a term obtained from integration of the bulk on-shell action, namely \eqref{bdybvabterms}, the Gibbons--Hawking--York term, holographic renormalization counterterms, and finally the finite terms implementing the Legendre transform needed to match the functional dependence of the boundary field theory free energy. 
Using the canonically defined expressions for the equivariant forms constructed from the bilinears, 
we have shown that the gravitational free energy of a supersymmetric solution on a manifold $M$ does not receive any contribution from the boundary, and $\Fgrav = I^{\rm FP}_{\rm OS}$. This result is shown in appendix \ref{app:holrenn}, and it is straightforward to follow the same steps and conclude that the computations extend also to the case that $M$ (including $\partial M$) contains orbifold points.  Similarly, the computations of the contributions of nuts and bolts carry over without modification, and we find the final result extending \eqref{Fgrav} is
\begin{align}\label{Fgravorbi}
	\Fgrav &= \frac{\pi}{G_4} \bigg[ \sum_{F_{0,\pm}} \mp \frac{1}{d_{F_0}} \frac{ (b_1 \mp b_2)^2 }{b_1 b_2} \ii {\cF}( u_\pm) \nn
	& \qquad \quad + \sum_{F_{2,\pm}} \frac{1}{d_{F_2}}\left( -\kappa\ii \cF_I(u_\pm)\smallspace \mathfrak{p}^I \pm \ii {\cF}({u_\pm})  \int_{F_{2,\pm}} c_1(L) \right) \bigg] \, .
\end{align}
This formula is expressed in terms of the bulk fixed point values $u_\pm$, and the relation with the UV boundary data is discussed in section \ref{sec:uvirrels}. In the presence of non-trivial orbifold structure, the relation \eqref{UVIR} is modified to include the order of the orbifold group
\begin{equation}
	d_F \Delta^I + \ii \beta \sigma^I = y^I \, ,
\end{equation}
where $d_F$ is the order of the orbifold group of the fixed set at the origin of the chosen $\R^2$ submanifold (whether a bolt or a nut).

\subsection{Spindle solutions}
\label{sec:spindleexamples}

As a concrete example, we can generalize the Euclidean supersymmetric black hole spacetime with topology $S^2 \times \R^2$ discussed in section \ref{rotatingbhs}, by replacing the $S^2$ horizon with a spindle $\Sigma=\mathbb{WCP}^1_{[n_N,n_S]}$.
We consider $\xi = \varepsilon \partial_{\varphi_1} + \partial_{\varphi_2}$, with $\partial_{\varphi_1}$ rotating $\Sigma$ and $\partial_{\varphi_2}$, rotating $\R^2$, 
which has two isolated fixed points at the North and the South of the sphere with the order of the orbifold groups given by $d_{N/S}=n_{N/S}$.
The weights of the $U(1)$ action at the poles generalize those in \eqref{eq:Weights_RotatingBlackHole}  (in conventions\footnote{Note that one should also relate $\kappa_N\chi_N=(\sigma_1)_\text{there}$ and $\kappa_S=(-\sigma_0)_\text{there}$.} 
taken from \cite{BenettiGenolini:2024hyd}) and are given by
\begin{equation}
	(b^1_N, b^2_N) = (-1,\varepsilon/n_N) \, , \qquad (b^1_S, b^2_S) = (-\varepsilon/n_S, -1) \, ,
\end{equation}
with the associated gravitational free energy given by
\begin{equation}
\label{eq:Fgrav_RotatingBlackHole_Spindle_1}
	\Fgrav = \frac{\pi}{G_4} \frac{1}{\varepsilon} \left[ \chi_N \left( 1 + \chi_N \frac{\varepsilon}{n_N} \right)^2 \ii \cF(u_N) - \chi_S \left( 1 - \chi_S\frac{\varepsilon}{n_S} \right)^2 \ii \cF(u_S) \right] \, .
\end{equation}

In order to relate this to boundary data and constrain the fluxes, we follow the same procedure as was done after \eqref{eq:Fgrav_RotatingBlackHole_1}. 
Again ensuring that the spinor is globally defined requires
 $\kappa_N = - \chi_S\kappa_S $,  and we deduce  
\begin{equation}\label{constspndfluxsum}
	\zeta_I \mathfrak{p}^I = - \kappa_S \sigma \frac{n_S + \sigma n_N}{n_N n_S} \, ,
\end{equation}
where we have introduced $\sigma\equiv \chi_N\chi_S=\pm 1$ which parametrizes whether supersymmetry is being
preserved by a twist or anti-twist, respectively \cite{Ferrero:2021etw}.
We next introduce the variables $y^I$ associated with the North and the South poles of the spindle, as in \eqref{eq:Defn_sigma_y}, finding the same relation between the two as in \eqref{eq:ys_RotatingBlackHole}, namely 
\begin{align}
y^I_N = y^I_S - \varepsilon  \mathfrak{p}^I\,.
\end{align}
Since the relation between $y^I$ and $u^I$ is again \eqref{yurelnuts}, the expression for the gravitational free energy on the spindle in terms of $y^I$ and $\mathfrak{p}^I$ is also the same as that on the two-sphere \eqref{eq:Fgrav_RotatingBlackHole_2}, namely
\begin{equation}
\label{eq:Fgrav_RotatingBlackHole_3}
	\Fgrav = \frac{\pi}{G_4} \frac{1}{\varepsilon} \left[ \chi_N \ii \cF(y_S - \varepsilon \mathfrak{p}) - \chi_S \ii \cF(y_S) \right] \, .
\end{equation}
However, the constraint on the gauge field fluxes \eqref{constspndfluxsum} differs from the $S^2$ case, as do
the relationships between $y^I$ and the boundary holographic data which now read
\begin{equation}
	y^I_N = n_N \Delta^I_N - 2\pi \ii \sigma^I_N \, , \qquad y^I_S = n_S \Delta^I_S - 2\pi\ii \sigma^I_S \, ,
\end{equation}
and again $ \Delta^I_{N,S}$, $\sigma^I_{N,S}$ are constants.  
Furthermore, the constraints
$\zeta_Iu^I_\pm =1$ with \eqref{yurelnuts} imply
\begin{equation}
	\zeta_I y^I_N 
	= -\kappa_N \left( 1 + \chi_N \frac{\varepsilon}{n_N} \right) \, , \qquad \zeta_I y^I_S
	 = \kappa_S \left( \chi_S - \frac{\varepsilon}{n_S} \right) \, .
\end{equation}
If we assume that $*_2 \dd\eta_{(0)}=0$, associated with a direct product $\Sigma\times \R^2$ geometry,
then from \eqref{localconssigmaI} we have $\zeta_I \sigma^I_{N,S}=0$, so then
$\zeta_I y^I_{N,S}=\zeta_I\Delta^I_{N,S}$,
 and also from 
\eqref{eq:sigmaI} the $\sigma^I$ are directly fixed by the mass deformations. 

Remarkably, these results for the gravitational free energy of Euclidean 
supersymmetric spindle solutions  precisely agree
with minus the off-shell entropy function introduced in \cite{BenettiGenolini:2024kyy}. 
The change of variable is simply $y^I_{N,S} = - x^I_\pm/2$, $\varepsilon = b_0$, 
and then we obtain agreement with (5.36)--(5.39) of \cite{BenettiGenolini:2024kyy} after making the choice\footnote{To compare with \cite{BenettiGenolini:2024xeo} one should also set $\chi_N=-1$.}
$\kappa_N=1$, and $\kappa_S = -\chi_S$. 
In \cite{BenettiGenolini:2024kyy} we had considered the equivariant localization of a two-form obtained by reduction of the four-dimensional action on a geometry $AdS_2 \times \Sigma$ and, generalizing Sen's approach \cite{Sen:2005wa}, we argued that extremizing the resulting function over the  unconstrained variables would give the entropy of the relevant black hole with near-horizon $AdS_2\times \Sigma$.
This indeed reproduces  the entropy of various accelerating black hole solutions, with
different subfamilies (and/or near-horizon limits) studied in 
\cite{Ferrero:2020twa, Ferrero:2021ovq, Couzens:2021cpk, Cassani:2021dwa, Boido:2022iye}. 
Here we strengthen this interpretation of the result, as we are proving that, assuming the existence of a family of supersymmetric deformations of the extremal black holes, the on-shell action of the family agrees with the ``entropy function'' computed from the near-horizon geometry, and is thus independent of the deformation parameters . Indeed the on-shell action only depends on the constant UV data $\sigma^I_{N,S}$ and $\Delta^I_{N,S}$.  
This is not unexpected from the holographic interpretation, as the dual observable is a supersymmetric index
and so will be independent of most UV deformations 
\cite{Witten:1982df}. In fact, the class of spaces considered here is larger than that of \cite{BenettiGenolini:2024kyy}, since it also includes the possibility of metrics that are not just the direct product $\R^2 \times \Sigma$, such as dyonic black holes. From the perspective in this paper the extremization 
simply performs a Legendre transform,  passing from 
canonical to micro-canonical ensemble, hence reproducing 
the entropy. 

In \cite{Cassani:2021dwa} instead the on-shell action 
of a complex locus of supersymmetric but non-extremal accelerating black holes in minimal gauged supergravity was computed. These black holes 
are rotating, accelerating, and carry dyonic charge. 
Formally  our gravitational free energy \eqref{eq:Fgrav_RotatingBlackHole_3} 
(in the  minimal gauged supergravity case)
agrees with the on-shell action computed in \cite{Cassani:2021dwa}, 
identifying $\varepsilon = \mp \omega/2\pi \ii$ and setting $\sigma=-1$ (for the 
``anti-twist''), where 
$\omega$ is a complex chemical potential associated with rotation, and the sign
here is associated to a choice of branch of the complex solutions. 
This agreement is only formal due to the reality conditions 
we have imposed in the present paper, but as commented 
already
we expect it to be possible to relax these conditions, allowing complex 
solutions.
 Importantly, the formula \eqref{eq:Fgrav_RotatingBlackHole_3} for the gravitational 
free energy
is also in precise agreement with the large
$N$ field theory result for the ``spindle index'' \cite{Inglese:2023wky, Inglese:2023tyc}, obtained
recently in \cite{Colombo:2024mts}, providing a microscopic 
``counting'' of the entropy of accelerating black holes. 

Additional topologies including spindles with toric symmetry are further considered in the companion paper \cite{BenettiGenolini:2024hyd}. Moreover, infinite classes of explicit supersymmetric solutions of minimal gauged supergravity on manifolds that are total spaces of orbifold line bundles over spindles have been recently constructed in \cite{Crisafio:2024fyc}. As shown there, the direct computation of the on-shell action using holographic renormalization agrees with the result of the application of the equivariant localization theorem \eqref{Fgravorbi}.

\subsection{Seifert three-manifolds}
\label{sec:Seifertexamples}

We conclude this section with one final class of examples, which generalizes the examples in section
\ref{subsec:TaubBolt}, by considering four-dimensional solutions which have a three-dimensional boundary 
that is a Seifert manifold. 
A Seifert manifold can be viewed 
as a circle orbibundle over a two-dimensional orbifold $\Sigma_{g,k}$ (see e.g. \cite{Closset:2018ghr,Closset:2019hyt} for a review). We want to consider 
the four-dimensional solution\footnote{There may be other four-dimensional manifolds with the
same Seifert manifold boundary, but we do not consider them here.}
 to have the same topology as the complex line orbibundle that is associated with the $S^1$ orbibundle:
\begin{equation}
\label{eq:seifertcase}
	\mathbb{C} \hookrightarrow L \to \Sigma_{g,k} \, .
\end{equation}
The orbifold $\Sigma_{g,k}$ has genus $g$ and $k$ orbifold points, where
in a given patch containing the $i$th orbifold singularity the coordinate system is modelled on $\mathbb{C}/\mathbb{Z}_{n_i}$
with $n_i$ a positive integer.
That is, if $z_i$ is a local coordinate on $\mathbb{C}$ then we identify $z_i\sim \omega_i z_i$
with $\omega_i=\ex^{2\pi \ii/n_i}$. To specify the line orbibundle near that point, each patch is modelled on $(\mathbb{C}\times \mathbb{C})/\mathbb{Z}_{n_i}$ with the quotient specified by $(z_i, s_i)\sim (\omega_i z_i, \omega_i^{m_i}s_i)$ 
with ``charges"
$m_i\in\mathbb{Z}_{n_i}$. The integrated orbifold
first Chern class is then given by
\begin{align}\label{chernintl}
\int_{\Sigma_{g,k}}c_1(L)=\text{deg}(L)+\sum_{i=1}^k\frac{m_i}{n_i}\,,
\end{align}
which in general is a rational number.
Here\footnote{This can be compared with 
(2.17) of \cite{Ferrero:2021etw} with the sign difference in the $m_i$ accounted for by
the fact that a single azimuthal angular coordinate $\varphi$ was used at both poles in \cite{Ferrero:2021etw}.} 
we have taken $0\leq m_i<n_i$, and this formula then 
defines the \emph{degree} $\text{deg}(L)\in\mathbb{Z}$. 
 We will also need
the rational-valued orbifold Euler number for $\Sigma_{g,k}$, which is given by
\begin{align}\label{chernintl2}
\int_{\Sigma_{g,k}} c_1(T\Sigma_{g,k})=2(1-g)+\left(\sum_{i=1}^{k}\frac{1}{n_i}\right)-k\,.
\end{align}

Now, by construction, the R-symmetry Killing vector vanishes at the origin of the $\mathbb{C}\cong \R^2$ fibre in
\eqref{eq:seifertcase}, and hence fixes the base $\Sigma_{g,k}$, which is a bolt with normal bundle $L$. The gravitational free energy can be obtained from \eqref{Fgravorbi}. There is only one bolt, so we choose $\kappa = +1$, and the only remaining sign characterizing the solution is the chirality of the spinor at the bolt
\begin{align}\label{eq:F_Grav_Seifert}
	F_{\rm grav} = \left( -\ii \cF_I(u_\pm)\smallspace \mathfrak{p}^I_\pm  \pm \ii {\cF}({u_\pm})  \int_{\Sigma_{g,k}} c_1(L) \right)  \frac{\pi}{G_4} \, .
\end{align}
As usual we have $ \zeta_I u^I_\pm = 1 $ and because of supersymmetry (see \eqref{eq:R_Symmetry_Constraint_Bolt}), we have
\begin{equation}
\label{eq:R_Symmetry_Constraint_Boltseifert}
	\mathfrak{p}^R_\pm = \frac{1}{2} \zeta_I \mathfrak{p}^I_\pm = \frac{1}{2} \int_{\Sigma_{g,k}} \left[ \pm c_1(L) - c_1(T\Sigma_{g,k}) \right] \, .
\end{equation}
where the integrated Chern classes are given in \eqref{chernintl}, \eqref{chernintl2}.

These results generalize those of \eqref{eq:Fgrav_TaubBolt_Saddles}. 
We can relate the free energy to UV data following the discussion in section \ref{subsec:TaubBolt}. 
Assuming that the corresponding supergravity solutions exist, this then provides a prediction for
the large $N$ limit of the partition function of $\mathcal{N}=2$ SCFTs on Seifert manifolds
obtained in \cite{Closset:2018ghr}.

 It is also possible to make a connection with the toric results of 
\cite{BenettiGenolini:2024hyd} by considering the special case of a spindle with $g=0$ and $k=2$. With normal bundle $\mathcal{O}(-p)\rightarrow
\mathbb{WCP}^1_{[n_1,n_2]}$, by definition we have $\int_{\Sigma_{g,k}}c_1(L)=-\frac{p}{n_1 n_2}$ with $p\in \mathbb{Z}$, and
$\int c_1(T\Sigma_{g=0,k=2}) =\frac{1}{n_1}+\frac{1}{n_2}$ to get
\begin{align}\label{GGI}
F_{\rm grav} = \left( -\ii \cF_I(u_\pm)\smallspace \mathfrak{p}^I \mp \ii {\cF}({u_\pm})  \frac{p}{n_1 n_2} \right)  \frac{\pi}{G_4} \, ,
\end{align}
and \eqref{eq:R_Symmetry_Constraint_Boltseifert} becomes
\begin{equation}
\label{eq:R_Symmetry_Constraint_Boltseifert2}
\zeta_I \mathfrak{p}^I_\pm =	\mp \frac{p}{n_1 n_2} - \left(\frac{1}{n_1}+\frac{1}{n_2}\right)\, .
\end{equation}
This result can now be compared with \cite{BenettiGenolini:2024hyd}. 
In the limit that the R-symmetry Killing vector just rotates the fibre, in \cite{BenettiGenolini:2024hyd} we need to set the signs $\sigma_0=\sigma_2$. Also identifying $\sigma_1 y^I_{N,S}$ with $u_\pm^I$, 
we then see that setting $\sigma_0=-1$ one has $\chi_S=\sigma_1=\pm$ in the above notation, 
and the formulae \eqref{GGI}, \eqref{eq:R_Symmetry_Constraint_Boltseifert2}
precisely agree with (43), (44) in \cite{BenettiGenolini:2024hyd}, respectively. 

Furthermore, for the case of trivial fibration over a spindle, obtained by setting $p=0$,
we can also see these results are consistent with those of section \ref{sec:spindleexamples} in the following sense. In section \ref{sec:spindleexamples} we should assume that the 
R-symmetry Killing vector does not rotate the spindle and set $\varepsilon=0$. Further restricting to the twist
class $\sigma=+1$, which can be achieved by taking $\chi_N=\chi_S=-1$, 
and also setting $\kappa_S=+1$, we find precise agreement with \eqref{GGI}, \eqref{eq:R_Symmetry_Constraint_Boltseifert2}.

\section{Discussion}\label{secdisc}

 We have shown that supersymmetric solutions of $D=4$, $\mathcal{N}=2$ Euclidean gauged supergravity coupled
 to vector multiplets have a set of equivariantly closed forms. These forms, which are globally defined and gauge-invariant, 
 are expressed in terms of spinor bilinears and the supergravity fields. 
We have explained how 
this structure allows one to compute various flux integrals and also the on-shell action of a supersymmetric solution using the BVAB fixed point formula. The on-shell action is naturally expressed in terms of bulk fixed point data but we also explained
how it can be expressed in terms of holographic boundary data.
We illustrated the formalism with various concrete examples. 
 
In developing the formalism we focussed on supersymmetric solutions that satisfy a particular reality condition
given in \eqref{eq:Reality_Condition}.  In particular, we assumed that the metric, the gauge fields and
the scalars $z^i$ and $\tilde z^i$ are real, and also imposed a corresponding reality condition on the Killing spinors. 
While this still covers rich classes of solutions, there is good evidence that these assumptions can be relaxed.
For example, explicit solutions were constructed in \cite{Freedman:2013oja} that are dual to ABJM theory on $S^3$, with 
real metric and gauge fields but with complex $z^i,\tilde z^i$, as reviewed in appendix \ref{subsubsec:FP}. For real $z^i,\tilde z^i$ we precisely recover
the on-shell action computed in \cite{Freedman:2013oja} and we also do if we formally assume that $z^i,\tilde z^i$ are complex.
Similarly, in section \ref{sec:six}, again proceeding formally, we were also able to obtain the on-shell
action and hence the entropy of classes of black hole solutions which have complex metrics.
Thus, it is of interest to systematically analyze what happens when the reality conditions \eqref{eq:Reality_Condition} are relaxed. 

It is also of interest to generalize the supergravity models to include hypermultiplets in addition to vector multiplets.
Based on the results of \cite{BenettiGenolini:2024kyy} we anticipate that there will be additional equivariantly closed forms which, moreover,
will lead to the imposition of specific constraints on the variables appearing in the on-shell action.
Another interesting generalization is to consider supergravity theories with higher derivatives, as this will lead to additional
general predictions for the dual SCFTs beyond the leading large $N$ contributions that we have obtained here.
It seems likely that the results for the supersymmetric index of supersymmetric black holes in ungauged $\mathcal{N}=2$ supergravity, recently presented in \cite{Hegde:2024bmb},
can be obtained in this way.

Some of the $\mathcal{N}=2$ gauged supergravity theories that we have considered can be obtained
from a consistent KK truncation of $D=10,11$ supergravity. For example, solutions of the STU model can be uplifted on
$S^7$ \cite{Cvetic:1999xp} to obtain solutions dual to ABJM theory, while
solutions of minimal gauged supergravity can also be uplifted on $SE_7$ (seven-dimensional Sasaki--Einstein manifolds)
and other manifolds to $D= 11$ \cite{Gauntlett:2007ma} giving solutions dual to other SCFTs. For such theories, our expressions for the $D=4$ on-shell action immediately lead to expressions
for the on-shell action of the corresponding supersymmetric solutions of $D=10,11$ supergravity.
However, we believe that even when such a consistent KK truncation does not exist, the corresponding gauged supergravity sector that we study is sufficient to be able to evaluate the gravitational free energy of the associated $D=10,11$ solutions, at least
in some cases. For example, in section \ref{subsec:TaubBolt} we obtained results for the on-shell action that are in agreement
with field theory results, in the large $N$ limit, for general classes of $\cN= 2$ SCFTs \cite{Toldo:2017qsh} that have $D=11$
holographic duals involving $SE_7$ but are not associated with consistent KK truncations (aside from the connection with minimal gauged supergravity). This comparison
requires one to identify the prepotential of the gauged supergravity with the twisted superpotential or Bethe potential arising in the field theory computation (\textit{cf}. \cite{Hosseini:2016tor}). It would be interesting to study this further.

\section*{Acknowledgements}

\noindent 
We thank Nikolay Bobev, Cyril Closset, Chris Couzens, Sameer Murthy, Jaeha Park, Julian Sonner, Chiara Toldo
and Gustavo Joaquin Turiaci for helpful discussions.
This work was supported by STFC grants  ST/X000575/1 and
ST/X000761/1, EPSRC grant EP/R014604/1 and SNSF Ambizione grant PZ00P2\_208666.
PBG would like to thank the UCSB Physics Department for hospitality during the final stages of this work. 
JPG is a Visiting Fellow at the Perimeter Institute. 
AL is supported by a Palmer Scholarship.

\appendix
\section{Localization without supersymmetry}
\label{localnosusy}

Here we discuss the existence of equivariantly closed forms in a general context, without invoking supersymmetry.
We start by presenting some general results regarding equivariantly closed forms. We assume that we have a $k$-dimensional
manifold with a metric and Killing vector $\xi$. We also consider gauge fields with $F=\dd A$ and $\mathcal{L}_\xi F=0$.

For the latter, we can always construct, locally, an equivariantly closed completion of $F$ by working in a gauge satisfying
$\mathcal{L}_\xi A=0$. We then find that $\Phi=F-\xi\hook A +c$, with $c$ an arbitrary constant, is equivariantly closed.
In principle, one can then use the BVAB formula to compute flux integrals, but some care is required
to ensure that one gets a globally well-defined answer. By contrast, in the supersymmetric
setting in our expressions \eqref{PhiF}, \eqref{defphioiflux}
the zero-form completion is expressed in terms of globally well-defined spinor bilinears and scalar fields.

Similarly, consider the volume form defined by the metric, $\vol_{k}$, which is invariant under the action of $\xi$, and define $\Phi_{k}=f\, \vol_{k}$ where $f$
is a globally defined function satisfying $\mathcal{L}_\xi f=0$. We may always construct an equivariantly 
closed extension of this, \emph{locally}, as follows.  
Introduce local coordinates $\psi, x^i$ with $\xi=\partial_\psi$. Consider (for notational simplicity) even dimensions, $k=2p$. Then we may consider the polyform given by
\begin{align}
\Phi=\diff \psi\wedge \beta_{2p-1}+(\alpha_{2p-2}+\diff\psi\wedge \beta_{2p-3})+\dots (\alpha_{2}+\diff\psi\wedge \beta_{1}) +\alpha_{0}\, ,
\end{align}
with $\xi\hook \alpha=\xi\hook\beta =0$. On dimensional grounds we have $\diff\beta_{2p-1}=0$.
The condition $\cL_\xi\Phi=0$ implies that $\cL_\xi \alpha=\cL_\xi \beta=0$.
 It is straightforward to check that 
the polyform $\Phi$ will be equivariantly closed provided that
\begin{align}
\diff\alpha_{2p-2}=\beta_{2p-1}\, ,\quad
\diff \alpha_{2p-4}=\beta_{2p-3}\, ,\quad
\ldots\quad\,  , \quad 
\diff\alpha_{2}=\beta_{3}\, ,\quad
\diff \alpha_{0}=\beta_{1}\, .\quad
\end{align}
With $\diff\beta_{2p-1}=0$, as noted, we can use the Poincar\'e Lemma to construct a local $\alpha_{2p-2}$ satisfying
$\diff\alpha_{2p-2}=\beta_{2p-1}$. The remaining forms in $\Phi$ can then be obtained by freely choosing
the remaining basic forms $\alpha$ and setting $\beta_{2p-3}=\diff\alpha_{2p-4}$, $\ldots$, $\beta_1=\diff\alpha_{0}$. 
This argument shows the importance of defining $\Phi$ as a \emph{global} 
polyform, with this in turn of course being crucial when applying 
integral formulae such as the BVAB theorem: it is the global constraint 
on $\Phi$ that effectively relates its components of different degrees.

We also note that if we are given $\Phi_{2p}$ with $\mathcal{L}_\xi\Phi_{2p}=0$ and also that there exists a $\Phi_{2p-2}$ with
$\mathcal{L}_\xi\Phi_{2p-2}=0$ and $\dd\Phi_{2p-2}=\xi\hook \Phi_{2p}$ then we have $\dd(\xi\hook\Phi_{2p-2})=0$. This 
immediately follows from $0=\mathcal{L}_\xi\Phi_{2p-2}=(\xi\hook \dd+\dd\, \xi\hook\, )\Phi_{2p-2}$. In the above coordinates
this says that $\beta_{2p-3}$ is necessarily closed. 
Then using the Poincar\'e Lemma we can construct a local
$\alpha_{2p-4}$ with $\diff \alpha_{2p-4}=\beta_{2p-3}$ and moreover $\mathcal{L}_\xi\alpha_{2p-4}=0$.
This structure will appear below.

With these comments in mind, we note that in theories of gravity in arbitrary dimensions we can always find a globally defined and
canonical expression for $\Phi_{k-2}$, but in theories with gauge fields it is generally gauge-dependent. 
This construction arises in the ``Wald formalism" for conserved quantities \cite{Wald:1993nt}.
To illustrate, we consider a $D=4$ Euclidean theory
of a similar form to that considered in the main text. The theory
depends on a metric $g_{\mu\nu}$, gauge fields $A^I$, with field strengths $F^I=\diff A^I$, and also real scalar fields $z^\alpha$ 
with action 
given by 
\begin{align}
I &= - \frac{1}{16\pi G_4} \int \Big[ \left( R - 2 g_{\alpha{\beta}}\partial_\mu z^\alpha \partial^\mu{z}^{\beta}  -   \mathcal{V} \right) \vol_4
+ \frac{1}{2}\mathcal{I}_{IJ}F^I \wedge * F^J \nonumber\\ 
& \qquad \qquad \quad \qquad - \frac{\ii}{2} \mathcal{R}_{IJ} F^I \wedge F^J \Big] \,,
\end{align}
where $ \mathcal{V}$, $g_{\alpha{\beta}}$, $\mathcal{I}_{IJ}$ and $\mathcal{R}_{IJ}$ (both symmetric in $I,J$), are all functions of the scalar fields.
We will utilise the Einstein equations and the gauge equations of motion which are given by
\begin{align}
\label{eq:4dN2_EinsteinEOM}
	R_{\mu\nu} &= - \frac{1}{2}\mathcal{I}_{IJ}\Big(F^I_{\mu\rho}F^J_\nu{}^\rho - \frac{1}{4} g_{\mu\nu} F^I_{\rho\sigma} F^{J\rho\sigma}\Big) + 2 g_{\alpha {\beta}} \partial_\mu z^\alpha \partial_\nu {z}^{{\beta}} + \frac{1}{2}g_{\mu\nu}  \mathcal{V}\,,\nn 
	0 &= \dd \left( \mathcal{I}_{IJ} * F^J - \ii \mathcal{R}_{IJ}F^J \right) \, .
\end{align}

We first consider configurations, not necessarily solving the equations of motion, with a Killing vector $\xi$ that also preserves the field strengths: $\mathcal{L}_\xi g_{\mu\nu}=0$, $\mathcal{L}_\xi F^I=0$.
As above, we immediately deduce that $\dd(\xi\hook F^I)=0$ and hence we can define equivariantly closed forms via
\begin{align}\label{genthyfluxeqclosed}
\Phi^I=F^I+\Phi_0^I\,,\qquad \dd_\xi\Phi^I=0\,,
\end{align}
where $\Phi_0^I$ are functions, locally defined in general, satisfying\footnote{The $\Phi_0^I$ will 
exist as global functions precisely when $\xi\hook F^I = 0 \in H^1(M,\R)$. A non-example is $M=T^2\times \Sigma$ 
with $\Sigma$ any two-dimensional space, $T^2$ parametrized by 
periodic coordinates $\psi_1,\psi_2$, and taking  
$F^I=\diff\psi_1\wedge \diff\psi_2$ with $\xi=\partial_{\psi_1}$. 
In this case $\xi$ has no fixed points, and $\Phi_0^I=-\psi_2$ is only 
defined as a local function, since $\psi_2$ is periodic.}
\begin{align}\label{defphiziapp}
\dd\Phi^I_0=\xi\hook F^I\,.
\end{align}
For example, if we work in a gauge satisfying $\mathcal{L}_\xi A^I=0$, then we can choose
\begin{align}
\Phi^I_0=-\xi\hook A^I+c^I\, ,
\end{align}
where $c^I$ are constants. Note that $\mathcal{L}_\xi \Phi^I_0=0$. 

We next consider solutions to the equations of motion and we further assume that the Killing vector also leaves the scalars invariant:
$\mathcal{L}_\xi z^\alpha=0$. 
By taking the trace of the Einstein equations in \eqref{eq:4dN2_EinsteinEOM} we deduce that the on-shell
action can be written 
\begin{equation}
\label{eq:IbulkOS}
	I_{\mathrm{OS}} =  \frac{\pi}{2G_4} \bigg[\frac{1}{(2\pi)^2} \int \Phi_4 \bigg]\,,
\end{equation}
with
\begin{equation}
\label{eq:NonMinimalPhi4}
	\Phi_4 \equiv -  \frac{1}{2} \mathcal{V}\smallspace \vol_4- \frac{1}{4}\mathcal{I}_{IJ}F^I \wedge * F^J + \frac{\ii}{4}\mathcal{R}_{IJ} F^I \wedge F^J \, .
\end{equation}
Our goal is to construct an equivariantly closed completion of $\Phi_4$ of the form
\begin{align}
\Phi=\Phi_4+\Phi_2+\Phi_0\,,\qquad \dd_\xi\Phi=0\,,
\end{align}
and $\mathcal{L}_\xi\Phi=0$. 
Using the equations of motion in \eqref{eq:4dN2_EinsteinEOM} we prove below that we can take
\begin{equation}
\label{eq:NonMinimalPhi2}
	\Phi_2 =  - \frac{1}{2} *\dd\xi^\flat - \frac{1}{2}\left( \mathcal{I}_{IJ} *F^I - \ii \mathcal{R}_{IJ}  F^I \right) \Phi_0^{J}  + \zeta_2 \, ,
\end{equation}
where $\Phi^I_0$ was defined in \eqref{defphiziapp}, $\zeta_2$ is an arbitrary two-form satisfying $\diff\zeta_2=0$ and 
$\mathcal{L}_\xi\zeta_2=0$. 
Furthermore, using the equations of motion we can show that $\dd ( \xi \hook \Phi_2) = 0$
and hence, locally, we can write
\begin{align}
\label{eq:NonMinimal_dPhi0}
	\dd\Phi_0&\equiv \xi \hook \Phi_2\nn
	 &= \frac{1}{2} * ( \xi^\flat \wedge \dd \xi^\flat) - \frac{1}{2}\xi \hook \left( \mathcal{I}_{IJ} *F^I - \ii \mathcal{R}_{IJ} F^I \right) \Phi_0^{J} + \xi \hook \zeta_2 \, ,
\end{align}
where we used $\xi\hook (*\dd\xi^\flat)=- * ( \xi^\flat \wedge \dd \xi^\flat)$. 
As emphasized earlier, in order to deploy the BVAB fixed point theorem 
to compute the bulk on-shell action \eqref{eq:IbulkOS}
one needs to ensure that the various forms we have introduced 
are globally well-defined. In particular, if $\Phi_0^I$ 
are global then so is $\Phi_2$ in \eqref{eq:NonMinimalPhi2}, 
and $\Phi_0$ will exist as a global 
function satisfying \eqref{eq:NonMinimal_dPhi0} precisely 
when the closed one-form on the right-hand side is globally exact. 
A sufficient condition for all of this to hold is simply $H^1(M,\R)=0$.\footnote{In particular this is true for the general  toric class 
of solutions studied in \cite{BenettiGenolini:2024hyd}.} 

We now prove these results. To do so we utilise the following
identity which is satisfied for a Killing vector $\xi^\mu$:
\begin{align}
\nabla_\nu\nabla_\mu\xi_\rho=R_{\rho\mu\nu\sigma}\smallspace \xi^\sigma\,.
\end{align}
In Euclidean signature we then deduce\footnote{The Hodge star definition we use is such that for a three-form $\omega$, for example, we have $*\omega_\mu=\frac{1}{3!}\epsilon_\mu{}^{\rho\sigma\delta}\omega_{\rho\sigma\delta}$, with $*^2\omega=-\omega$.}
\begin{align}\label{kvidsapp}
(*\dd*\dd\xi^\flat)_\mu&=-2\nabla^2\xi_\mu=2R_{\mu\nu}\smallspace \xi^\nu\,,\nn
(*\dd*[\xi^\flat\wedge \dd\xi^\flat])_{\nu\rho}&=-4\xi_{[\nu}R_{\rho]\sigma}\xi^\sigma\,.
\end{align}

We first show $\dd\Phi_2=\xi\hook\smallspace\Phi_4$. Substituting the Einstein equations into the first identity in \eqref{kvidsapp}
we deduce
\begin{align}
\frac{1}{2}(*\dd*\dd\xi^\flat)_\mu=-\frac{1}{2}\mathcal{I}_{IJ}[F^I_{\mu\nu}(\xi\hook F^J)^\nu-\frac{1}{4}F^I_{\nu\rho}F^{J\nu\rho}\xi_\mu]+
\frac{1}{2}\mathcal{V}\smallspace\xi_\mu\,.
\end{align}
Noting that $F^I_{\mu\nu}(\xi\hook F^J)^\nu=*(\xi\hook F^I\wedge *F^J)_\mu$ as well as $\xi^\flat=*(\xi\hook \vol_4)$, we conclude that
\begin{align}
-\frac{1}{2}\dd*\dd\xi^\flat =\frac{1}{2}\mathcal{I}_{IJ}(\xi\hook F^I\wedge *F^J)
-\frac{1}{2}\big(\frac{1}{4}\mathcal{I}_{IJ}F^I_{\nu\rho}F^{J\nu\rho}+\mathcal{V}\big)\xi\hook\vol_4\,.
\end{align}
We now use this result to compute $\dd\Phi_2$:  after utilising the gauge equation of motion \eqref{eq:4dN2_EinsteinEOM} and also \eqref{defphiziapp} we find
\begin{align}
\dd\Phi_2=-\frac{1}{2}\left(\frac{1}{4}\mathcal{I}_{IJ}F^I_{\nu\rho}F^{J\nu\rho}+\mathcal{V}\right)\xi\hook\vol_4+
\frac{\ii}{2}\mathcal{R}_{IJ}(F^I\wedge \xi\hook F^J)\,,
\end{align}
and hence $\dd\Phi_2=\xi\hook\Phi_4$.
 
We now show that $\dd ( \xi \hook \Phi_2) = 0$.
Substituting the Einstein equations into the second identity in \eqref{kvidsapp}, after using $\mathcal{L}_\xi z^\alpha=0$
we deduce
\begin{align}
(*\dd*[\xi^\flat\wedge \dd\xi^\flat])_{\nu\rho}&=2\mathcal{I}_{IJ}\xi_{[\nu}F^I_{\rho]\gamma}(\xi\hook F^J)^\gamma\,.
\end{align}
Then using $2\mathcal{I}_{IJ}\xi_{[\nu}F^I_{\rho]\gamma}(\xi\hook F^J)^\gamma=*[\mathcal{I}_{IJ}(\xi\hook F^I)(\xi\hook *F^J)]_{\nu\rho}$ we can conclude
\begin{align}
\dd*[\xi^\flat\wedge \dd\xi^\flat]=\mathcal{I}_{IJ}(\xi\hook F^I)(\xi\hook *F^J)\,.
\end{align}
Using this and the expression for $\xi \hook \Phi_2$ in \eqref{eq:NonMinimal_dPhi0} we obtain
\begin{align}
	\dd(\xi \hook \Phi_2)
	 &= \frac{1}{2}\mathcal{I}_{IJ}(\xi\hook F^I)(\xi\hook *F^J)- \frac{1}{2}\dd\Big[\xi \hook \left( \mathcal{I}_{IJ} *F^I - \ii \mathcal{R}_{IJ} F^I \right) \Phi_0^{J}\Big]  \, ,\nn
	 &=  \frac{1}{2}\mathcal{I}_{IJ}(\xi\hook F^I)(\xi\hook *F^J)+  \frac{1}{2}\xi \hook \left( \mathcal{I}_{IJ} *F^I - \ii \mathcal{R}_{IJ} F^I \right) \wedge \dd\Phi_0^{J}  \, ,\nn
	 &=0\,.
\end{align}
In the first line we used $\diff\zeta_2=0$ and $\mathcal{L}_\xi\zeta_2=0$. In the second we used $\mathcal{L}_\xi=\dd\, \xi\hook+\, \xi\hook\dd$, the gauge equation of motion and the invariance of the scalars and $F^I$ under the action of $\xi$. Finally the third line follows from \eqref{defphiziapp} and the fact that $(\xi\hook F^I)\wedge (\xi\hook F^J)$ is anti-symmetric in $I,J$.

\section{Euclidean supersymmetry equations}
\label{app:Euclidean_KSE}

A Lorentzian solution to the supergravity theory \eqref{eq:Lorentzian_I4} is supersymmetric if there exist a global Dirac spinor $\epsilon$ satisfying the Killing spinor equations \eqref{eq:Lorentzian_4d_GravitinoVariation}. We now want to construct the supersymmetry conditions for the Euclidean theory, taking into account the possibility of doubling the degrees of freedom.

We first construct the equations satisfied by the charge conjugate spinor $\epsilon^c$ in the Lorentzian theory. To define it, we introduce the unitary matrices $\cA_L$ and $\cC_L$ satisfying
\begin{equation}
\label{eq:Intertwining_Lorentzian_1}
	\Gamma_\mu^\dagger = - \cA_L \Gamma_\mu \cA_L^{-1} \, , \qquad \Gamma_\mu^T = - \cC_L \Gamma_\mu \cC_L^{-1} \, , \qquad \cC_L^T = - \cC_L \, ,
\end{equation}
which in turn allow us to define a unitary $\cB_L$ with
\begin{equation}
\label{eq:Intertwining_Lorentzian_2}
	\Gamma_\mu^* = \cB_L \Gamma_\mu \cB^{-1}_L \, , \qquad \cB_L \equiv  - \ii \cC_L \cA_L \, .
\end{equation}
Notice that $\cB_L \Gamma_5 \cB^{-1}_L=-\Gamma_5^*$.
Concretely, we can take 
\begin{align}\label{aldefchoice}
\cA_L = \Gamma_0\,,
\end{align}
so that $\Gamma_0^\dagger=-\Gamma_0$ and $\Gamma_i=\Gamma_i^\dagger$). For any spinor $\lambda$, the charge conjugate spinor is defined by
\begin{equation}
	\lambda^c \equiv \cB_L^{-1} \lambda^* = \ii \Gamma^0 \cC_L^{-1} \lambda^* \, .
\end{equation}
If $\epsilon$ satisfies the supersymmetry equations \eqref{eq:Lorentzian_4d_GravitinoVariation}, then its charge conjugate satisfies the following differential equations
\begin{align}
	0 &= \nabla_\mu \epsilon^c + \frac{\ii}{2} \cA_\mu  \Gamma_5 \epsilon^c + \frac{\ii}{2} A^{R} \epsilon^c + \frac{1}{2\sqrt{2}} \Gamma_\mu \e^{\cK/2} \left( W \tinyspace \Pnew_{-} + \overline{W} \tinyspace\Pnew_{+} \right) \epsilon^c \nn
	& \ \ \ + \frac{\ii}{4\sqrt{2}} \cI_{IJ} F^{J}_{\nu\rho} \Gamma^{\nu\rho} \Gamma_\mu \left( L^I \tinyspace \Pnew_{-} + \overline{L}^{I} \tinyspace \Pnew_{+} \right) \epsilon^c \, , \nn[5pt]
\label{eq:Lorentzian_Charge_Conj_KSE}
	0 &= - \frac{\ii}{2\sqrt{2}} \cI_{IJ} F^{J}_{\mu\nu} \Gamma^{\mu\nu} \left[ \cG^{\bar{i} {j}} \nabla_{j} {L}^{I}  \tinyspace \Pnew_{-} + {\cG}^{{i} \bar{j}}  \nabla_{\bar{j}} \overline{L}^{I} \tinyspace  \Pnew_{+} \right] \epsilon^c + \Gamma^\mu \left( \partial_\mu z^i \tinyspace\Pnew_- + \partial_\mu \overline{z}^{\bar{i}} \tinyspace \Pnew_+ \right) \epsilon^c \nn
& \ \ \ - \frac{1}{\sqrt{2}} {\e^{\cK/2}} \left[ \cG^{\bar{i} {j}}  \nabla_j W \tinyspace \Pnew_- + \cG^{{i} \bar{j}}  
\nabla_{\tilde j}\overline{W}^I\tinyspace \Pnew_+  \right] \epsilon^c \, .
\end{align}

We now introduce the Euclidean theory by taking the Wick rotation
\begin{equation}
\label{eq:Wick_Rotation_App}
	x^0 \to - \ii x^4 \,, \qquad A_0^I \to \ii A_4^I \, , \qquad \Gamma_0 \to \ii \gamma_4 \, , \qquad \Gamma_i \to \gamma_i \, , \qquad \Gamma_5 \to \gamma_5 \, ,
\end{equation}
where $\gamma_\mu$ generate Cliff$(4)$ and $\gamma_5 \equiv \gamma_{1234}$.
We should also double all degrees of freedom, taking complex metric and gauge fields and letting $z^i$ and $\overline{z}^{\bar{i}}$ be independent, obtaining the bosonic Euclidean action \eqref{themainaction}, as discussed in section \ref{subsec:EuclideanTheory}. In terms of the spinors, we analogously declare that $\epsilon$ and $\epsilon^c$ are independent spinors, and to stress this, we denote the latter by $\tepsilon$. Since the spinor equations are analytic in the fields, we find that the two independent Dirac spinors satisfy in Euclidean signature the equations \eqref{eq:Euclidean_KSE_epsilon} and \eqref{eq:Euclidean_KSE_tepsilon}, where the latter follows from \eqref{eq:Lorentzian_Charge_Conj_KSE}. 

It will be important for our analysis to introduce a ``real'' contour in Euclidean signature. 
To do so, we introduce unitary matrices $\cA_E$ and $\cC_E$ satisfying
\begin{equation}
\label{eq:Intertwining_Euclidean_1}
	\gamma_\mu^\dagger = \cA_E \gamma_\mu \cA_E^{-1} \, \qquad \gamma_\mu^T = \cC_E \gamma_\mu \cC_E^{-1} \, , \qquad \cC_E^T = - \cC_E \, , 
\end{equation}
which in turn lead to a unitary $\cB_E$ with
\begin{equation}
\label{eq:Intertwining_Euclidean_2}
	\gamma_\mu^* = \cB_E \gamma_\mu \cB_E^{-1} \, , \qquad \cB_E \equiv - \cC_E \cA_E \, ,
\end{equation}
Notice that $\cB_E \gamma_5 \cB^{-1}_E=+\gamma_5^*$.
Concretely, associated with \eqref{aldefchoice}, we take
\begin{align}
\cA_E = \mathbb{I}\,,
\end{align}
so that the Euclidean $\gamma$ matrices are Hermitian.
Notice that in the intertwining relation with the transpose representation we are using a different sign compared to 
the Lorentzian case in \eqref{eq:Intertwining_Lorentzian_1}. Therefore, in order to be consistent with the Wick rotation \eqref{eq:Wick_Rotation_App}, the relation between the two intertwiners is 
\begin{align}
\cC_L \to \cC_E \gamma_5\,.
\end{align}

Similar to the Lorentz case, the charge conjugate spinor in Euclidean signature is defined by
\begin{equation}
	\lambda^c \equiv \cB_E^{-1} \lambda^* = - \cC_E^{-1} \lambda^* \, .
\end{equation}
Tensor bilinears in Euclidean signature are also defined using the Majorana conjugate, now defined using $\cC_E$ as
\begin{equation}
	\overline{\lambda} \equiv \lambda^T \cC_E \, .
\end{equation}
A special family of bilinears will be particularly useful for our analysis in the main text, and they are constructed using both spinors: for any element $\mathbb{A}$ in the Clifford algebra, we write
\begin{equation}
\label{eq:Dirac_Bilinears_Euclidean_1}
	\overline{\tepsilon} \mathbb{A} \epsilon \, .
\end{equation} 
They are generically complex.

Requiring that $\tepsilon = \epsilon^c$, as we do in section \ref{realitycondssec},
is consistent with \eqref{eq:Euclidean_KSE_tepsilon} 
and the complex conjugate of \eqref{eq:Euclidean_KSE_epsilon},
provided that additional reality conditions are imposed on the bosonic fields, namely
\begin{equation}
\label{eq:Reality_Euclidean_App}
\begin{aligned}
g_{\mu\nu} &\in \mathbb{R}\,,  &\qquad A_\mu^I &\in \mathbb{R}\,, &\qquad z^i, \tz^{\tilde{i}} &\in \mathbb{R}\, , \\
\cA &\in \ii \mathbb{R} \, , &\qquad L^I, \tL^I &\in  \mathbb{R} \, , &\qquad \cG^{\tilde{i} j}, \cG^{i\tilde{j}} &\in \mathbb{R} \, ,
\end{aligned}
\end{equation}
which are also summarized in \eqref{eq:Reality_Condition}. If $\tepsilon = \epsilon^c$, the Euclidean bilinears \eqref{eq:Dirac_Bilinears_Euclidean_1} simply read
\begin{equation}
	\overline{\epsilon^c} \mathbb{A} \epsilon = \epsilon^\dagger \mathbb{A} \epsilon \, , 
\end{equation}
which are real when $\mathbb{A}$ is chosen among $\{ \mathbb{I}, \gamma_5, \gamma_\mu, \ii \gamma_\mu \gamma_5, \ii \gamma_{\mu\nu} \}$, corresponding to the bilinears introduced in \eqref{eq:Bilinears}.

\section{Bilinear equations}
\label{app:bilinears}

Our starting point is the Killing spinor equations of the Euclidean theory for $\epsilon$ and $\tepsilon$ given in
\eqref{eq:Euclidean_KSE_epsilon} and \eqref{eq:Euclidean_KSE_tepsilon}. From these we can obtain the following
useful differential and algebraic relations that hold for an arbitrary element of the Clifford algebra $\mathbb{A}$:
\begin{align}\label{eq:KSEdiffrel}
	\nabla_\mu \left( \overline{\tepsilon} \mathbbm{A} \epsilon \right)&= - \frac{\ii}{2} \cA_\mu \, \overline{\tepsilon} \{ \gamma_5, \mathbbm{A} \} \epsilon  -  \frac{1}{4\sqrt{2}} \xinew_I \overline{\tepsilon} \left[ ( L^I + \widetilde{L}^I ) \{ \gamma_\mu , \mathbbm{A} \} - ( L^I - \widetilde{L}^I) [ \gamma_5 \gamma_\mu, \mathbbm{A} ] \right] \epsilon \nn
	& \ \ \ + \frac{\ii}{8\sqrt{2}} \cI_{IJ} F^J_{ab} \overline{\tepsilon} \left[ ( L^I + \widetilde{L}^I) \left( \gamma_\mu \gamma^{ab} \mathbb{A} + \mathbb{A} \gamma^{ab} \gamma_\mu \right) \right.  \nn
	& \left. \qquad \qquad \qquad \quad - ( L^I - \widetilde{L}^I) \left( \gamma_5 \gamma_\mu \gamma^{ab} \mathbb{A} + \mathbb{A} \gamma^{ab} \gamma_\mu \gamma_5 \right) \right] \epsilon \, ,
\end{align}
and
\begin{align}\label{eq:KSEalgrel}
	 & 0 = \overline{\tepsilon} \left[\frac{1}{2} \partial_a (z^i + \tz^{\tilde{i}}) [ \mathbbm{A}, \gamma^a ]_\pm 
	+ \frac{1}{2}\partial_a ({z^i - \tz^{\tilde{i}}}) [ \mathbbm{A}, \gamma_5 \gamma^a ]_\mp \right] \epsilon \nn
&  + \frac{\ii}{4\sqrt{2}}\cI_{IJ} F^J_{ab} \overline{\tepsilon} \left[ \left( \cG^{\tilde{i} {j}} \nabla_j L^I + \cG^{i \tilde{j}} \nabla_{\tilde{j}}\tL^I \right) [ \mathbbm{A}, \gamma^{ab} ]_\pm - \left( \cG^{\tilde{i} {j}} \nabla_j L^I - \cG^{i \tilde{j}} \nabla_{\tilde j} \widetilde{L}^I \right) [ \mathbbm{A}, \gamma_5\gamma^{ab} ]_\pm \right] \epsilon \nn
& - \frac{1}{2\sqrt{2}} \xinew_I \overline{\tepsilon} \left[ \left( \cG^{\tilde{i} {j}} \nabla_j L^I + \cG^{i \tilde{j}} \nabla_j\widetilde{L}^I \right) \mathbbm{A} ( 1 \pm 1) - \left( \cG^{\tilde{i} {j}} \nabla_j L^I - \cG^{i \tilde{j}} \nabla_{\tilde j}\widetilde{L}^I \right) [ \mathbbm{A}, \gamma_5 ]_\pm \right] \epsilon \, ,
\end{align}
where $[\cdot\,, \cdot]_\pm$ denote anti-commutator and commutator, respectively.

\subsection{Bilinear relations}
Recall that the spinor bilinears were defined in \eqref{eq:Bilinears}. Using 
appropriate choices of $\mathbb{A}$ in \eqref{eq:KSEdiffrel} and \eqref{eq:KSEalgrel} allows us to obtain various relations on 
the bilinears. In particular, from \eqref{eq:KSEdiffrel} we obtain the following differential constraints:
\begin{align}
		\label{diffSP}
		\dd K&=0\,,\nn
		( \dd + \ii \cA) ( S + P) &= -  \frac{\xinew_I}{\sqrt{2}} L^I \, K + \sqrt{2} \cI_{IJ} L^I \, \xi \hook F^{J}_{[-]}  \, , \nn
		( \dd - \ii \cA) ( S - P) &=  -  \frac{\xinew_I}{\sqrt{2}} \widetilde{L}^I \, K - \sqrt{2} \cI_{IJ} \widetilde{L}^I \, \xi \hook F^{J}_{[+]} \, , \nn
		(\dd + \ii \cA)U_{[-]} &=  \frac{3}{2\sqrt{2}} \xinew_I {L}^I(\xi\hook\vol_4)  - \frac{1}{\sqrt{2}} \cI_{IJ} L^I \, K \wedge F^{J}_{[+]}  \, ,\nn
		(\dd - \ii \cA)U_{[+]} &= \frac{3}{2\sqrt{2}} \xinew_I \widetilde{L}^I  (\xi\hook\vol_4) - \frac{1}{\sqrt{2}} \cI_{IJ} \widetilde{L}^I \, K \wedge F^{J}_{[-]} \, ,
\end{align}
together with 
\begin{align}\label{dxiflatfirst}
	\dd \xi^\flat &= -   \sqrt{2}\xinew_I \left[ L^I U_{[+]} - \widetilde{L}^I U_{[-]} \right] - \sqrt{2} \cI_{IJ} \left[ C^I F^{J}_{[-]} - \widetilde{C}^I F^{J}_{[+]}\right] \, ,
\end{align}
and hence
\begin{align}\label{stardxiapp}
	*\dd \xi^\flat &= -  \sqrt{2} \xinew_I \left[ L^I U_{[+]} + \widetilde{L}^I U_{[-]} \right] + \sqrt{2} \cI_{IJ} \left[ C^IF^{J}_{[-]} + \widetilde{C}^I F^{J}_{[+]} \right] \, .
\end{align}

On the other hand from \eqref{eq:KSEalgrel} we obtain the following set of useful algebraic relations for the bilinears:
\begin{align}
	\label{algSP}
	0&=\xi\hook\dd z^i=\xi\hook\dd \tilde z^{\tilde i}\,,\nn
	0 &= \dd z^i ( S - P)  + \sqrt{2} \cI_{IJ} \cG^{i \tilde{j}} \nabla_{\tilde j}\widetilde{L}^I \, \xi \hook F^{J}_{[-]} -  \frac{\xinew_I}{\sqrt{2}} 
	\cG^{i \tilde{j}} \nabla_{\tilde j}\widetilde{L}^I  K  \, , \nn
	0 &= \dd \tz^{\tilde{i}} ( S + P) - \sqrt{2} \cI_{IJ} \cG^{\tilde{i} {j}} \nabla_{j} L^I  \, \xi \hook F^{J}_{[+]} -  \frac{\xinew_I}{\sqrt{2}} \cG^{\tilde{i} {j}} \nabla_{j} L^I K \, , \nn
	0 &= \dd z^i\wedge U_{[+]}  + \frac{1}{\sqrt{2}} \cI_{IJ} \cG^{i \tilde{j}} \nabla_{\tilde j}\widetilde{L}^I K\wedge F^{J}_{[+]}+ \frac{\xinew_I}{2\sqrt{2}} \cG^{i \tilde{j}} \nabla_{\tilde j}\widetilde{L}^I (\xi\hook\vol_4)  \, , \nn
	0 &=\dd \tz^{\tilde{i}} \wedge U_{[-]}  + \frac{1}{\sqrt{2}} \cI_{IJ} \cG^{\tilde{i} {j}} \nabla_{j} L^I K\wedge F^{J}_{[-]} + \frac{\xinew_I}{2\sqrt{2}} 
	\cG^{\tilde{i} {j}} \nabla_{j} L^I (\xi\hook\vol_4) \, .
\end{align}
From \eqref{diffSP} and \eqref{algSP} we can then derive
\begin{equation}
\label{eq:dWU}
	\dd \left[ ( \zeta_IL^I) U^+ + (\zeta_I\widetilde{L}^I) U^-  \right] = - \frac{1}{\sqrt 2}  \cV \, \xi \hook \vol_4 + \frac{1}{2\sqrt 2}  \zeta_I F^I \wedge K \, ,
\end{equation}
and notice that on the right-hand side we have the R-symmetry field strength $F^R\equiv \dd (\frac{1}{2} \xinew_I A_\mu^I)$ (see 
\eqref{rsymgf}).

Finally, we also record the result
\begin{align}
	\xi\hook U_{[\pm]}=\mp\frac{1}{2}(S\mp P)K\,,
\end{align}
which follows from Fierz identities.

\subsection{Some useful supergravity relations}
We now record some useful relations for the Euclidean supergravity theory we are considering (obtained from the Lorentzian theory; see, for example, \cite{Andrianopoli:1996vr,Craps:1997gp,Freedman:2012zz, Lauria:2020rhc, Bobev:2020pjk}).

One can show that
\begin{align}\label{altKexp}
\e^{-\cK}=-X^I N_{IJ}\tX^J\,.
\end{align}
We also have
\begin{align}\label{cNXFrelapp}
\cN_{IJ}X^J=\cF_I\,,
\end{align}
and hence
\begin{align}\label{NRXFrels}
\cN_{IJ}X^I X^J=2\cF\,,\qquad
\cR_{IJ}X^I \widetilde{X}^J=\frac{1}{2}\left(\cF_I\widetilde{X}^I+\widetilde{\cF}_I X^I\right)\, ,
\end{align}
where we used that $\cF$ is homogeneous of degree two to get the first equation.

From \eqref{altKexp} we can deduce that
\begin{align}
\nabla_i X^I N_{IJ}\tX^J=0\,,
\end{align} 
and then using \eqref{cNXFrelapp} we have
\begin{align}\label{covdereffeye}
\nabla_i \cF_I=\widetilde{\cN}_{IJ}\nabla_i X^J\,.
\end{align}
Expressions for $\nabla_i X^I$ and $\nabla_i L^I$ are given in 
\eqref{nablaXdef}, \eqref{nablaLdef} and we recall $\nabla_{\tilde{ i} }L^I=0$. We then compute
\begin{align}
\label{eq:dLI}
\dd L^I&=\nabla_i L^I \diff z^i-\ii\cA L^I\,.
\end{align}
A standard relation (see above references) which is important for our purposes is
\begin{align}\label{gff+L=I}
	\cG^{i \tilde{j}} \nabla_i L^I \nabla_{\tilde{ j} }\widetilde{L}^J+  \widetilde{L}^I {L}^J = - \frac{1}{2} ( \cI^{-1})^{IJ} \, .
	\end{align}
By taking the holomorphic covariant derivative of \eqref{cNXFrelapp} and using \eqref{covdereffeye} we deduce
\begin{align}
\nabla_i(\mathcal{N}_{IJ}L^J)=\widetilde{\cN}_{IJ}\nabla_i L^J\,.
\end{align}
Since $\cN_{IJ}$ does not transform under K\"ahler transformations we deduce $(\partial_i\cN_{IJ})L^J=-2\ii\cI_{IJ}\nabla_iL^J$. 
From \eqref{cNXFrelapp} we have $0=\partial_{\tilde{ j}}\cF_I=(\partial_{\tilde{ j}}\cN_{IJ})X^J$. After multiplying the latter by $\ex^{\cK/2}$ we
obtain 
\begin{equation}\label{dNL}
	(\dd \cN_{IJ})L^J=-2\ii\cI_{IJ}\nabla_i L^J \dd z^i\,, \quad (\dd \widetilde{\cN}_{IJ})\widetilde{L}^J
	=2\ii\cI_{IJ}\nabla_{\tilde{ i} } \widetilde{L}^J \dd \tilde z^{\tilde i}\,,
\end{equation}
from which we derive the further relations 
\begin{align}\label{dILL}
\dd(\cN_{IJ}L^I L^J)&=2\cR_{IJ}L^I\nabla_i L^J \dd z^i-2\ii\cA(\cN_{IJ} L^IL^J)\,,\nn
\dd(\cR_{IJ}L^I \widetilde{L}^J)&=\widetilde{\cN}_{IJ}\widetilde{L}^J\nabla_i L^J \dd z^i
+\cN_{IJ}L^J\nabla_{\tilde{ i}}\widetilde{L}^I \dd \tilde z^{\tilde i}\,.
\end{align}

\subsection{Equivariantly closed forms}
We now obtain expressions for equivariantly closed forms associated with the fluxes and the on-shell action. 
From the results of appendix \ref{localnosusy} we know these equivariantly closed forms must exist;
here we show that for supersymmetric configurations
they can be canonically expressed in terms of supergravity fields and the bilinears.

We first consider the equivariantly closed forms for the fluxes. 
We first recall \eqref{eq:DerivativeVanishingRCharge}:
\begin{align}
	\dd C^I\equiv 
	\dd[L^I(S-P)]=\nabla_i L^I\dd z^i(S-P)+L^I(\dd-\ii\cA)(S-P)\,, \nn
	\dd \widetilde{C}^I\equiv 
	\dd[\widetilde{L}^I(S+P)]=\nabla_{\tilde i}\widetilde{L}^I\dd z^{\tilde i}(S+P)+\widetilde{L}^I(\dd+\ii\cA)(S+P)\,.
\end{align}
Then using \eqref{diffSP}, \eqref{algSP} and \eqref{gff+L=I} we find
\begin{align}
\dd \left[ \sqrt{2} \left( C^I - \widetilde{C}^I  \right) \right] =  \xi \hook F^I \, .
\end{align}
Thus, we immediately deduce that $\Phi^{I}_{(F)} $, given by
\begin{align}\label{appbfcexp}
		\Phi^{I}_{(F)} = F^I + \sqrt{2} \left( C^I- \widetilde{C}^I \right) \, ,
\end{align}
is equivariantly closed. In particular, we have derived a canonical expression for $\Phi_0^I$ in~\eqref{genthyfluxeqclosed}.

We now turn to the on-shell action. We want to show that following polyform is equivariantly closed
\begin{equation}\label{actappsimple}
    \Phi=\Phi_4+\Phi_2+\Phi_0\,,
    \end{equation}
where
\begin{align}\label{phisappact}
    \Phi_4&=-\frac{1}{2}\mathcal{V}\smallspace \vol_{4}-\frac{1}{4}\mathcal{I}_{IJ}F^I\wedge*F^J+\frac{\ii}{4}\mathcal{R}_{IJ}F^I\wedge F^J\, ,\\
    \Phi_2 &=\frac{1}{\sqrt{2}} \xinew_I(L^I \tinyspace U_{[+]} + \tL^I\tinyspace U_{[-]} )- \frac{1}{\sqrt{2}}\mathcal{I}_{IJ}\big(C^I\tinyspace F^{J}_{[+]}+\tC^I \tinyspace F^{J}_{[-]}\big) +\frac{\ii}{\sqrt{2}}\mathcal{R}_{IJ}F^J\big(C^I-\tC^I\big)\, ,\nn
    \Phi_0&=\ii \tinyspace \ex^{\mathcal{K}}\Big[{\mathcal{F}}(X )(S-P)^2+\tcalF(\tX)(S+P)^2
    - \frac{1}{2} (\partial_I \mathcal{F}(X)\tX^I +\partial_I \tcalF(\tX)X^I)(S^2-P^2)\Big]\nonumber\,.
\end{align}
The expression for $\Phi_4$ is associated with the on-shell action and is the same as in the general context in \eqref{eq:NonMinimalPhi4}.
The expression for $ \Phi_2$ can be obtained from the general result in \eqref{eq:NonMinimalPhi2} (with $\zeta_2=0$) and then using
the expression for $*\dd\xi^\flat$ in \eqref{stardxiapp}. 
Specifically, using \eqref{stardxiapp} and \eqref{appbfcexp} in \eqref{phisappact} we have
\begin{align}\label{phi2xiappbexp}
\Phi_2=-\frac{1}{2}*\dd\xi^\flat-\frac{1}{2}(\cI_{IJ}*F^I-\ii\cR_{IJ} F^I)\Phi_0^J\,,
\end{align}
in agreement with \eqref{eq:NonMinimalPhi2}.
That $\dd\Phi_2=\xi\hook\Phi_4$ has already been shown in appendix \ref{localnosusy}; alternatively it can also be shown directly
using the bilinear relations above, together with the Bianchi identity, $\dd F^I=0$, and \eqref{dNL}.

In appendix \ref{localnosusy} we showed that $\Phi_0$ must exist, but not in  any canonical form. 
We now show that the expression
for $ \Phi_0$ given in \eqref{phisappact} does indeed satisfy $\dd\Phi_0=\xi\hook\Phi_2$. On the one hand, using Fierz identities
we have
\begin{align}\label{xiinphi2}
	\xi\hook\Phi_2=&-\frac{1}{2\sqrt{2}}\zeta_I(C^I-\widetilde{C}^I)K-\frac{1}{\sqrt{2}}\cI_{IJ}\big[(\xi\hook F^J_{[+]})C^I+(\xi\hook F^J_{[-]})\widetilde{C}^I \big]\nn &+\frac{\ii}{\sqrt{2}}\cR_{IJ}(\xi\hook F^J)(C^I-\widetilde{C}^I)\,.
\end{align}
On the other hand, using \eqref{NRXFrels} we see that $\Phi_0$ can be written as
\begin{equation}
	\Phi_0=\ii\Big[\frac{1}{2}\cN_{IJ}L^IL^J(S-P)^2+\frac{1}{2}\widetilde{\cN}_{IJ}\widetilde{L}^I\widetilde{L}^J(S+P)^2
	-\cR_{IJ}L^I\widetilde{L}^J(S+P)(S-P)\Big]\,.
\end{equation}
Taking the exterior derivative and using \eqref{dILL} we have 
\begin{align}
	\dd \Phi_0 =\ii\Big[
	&\cR_{IJ}L^I\nabla_i L^J \dd z^i(S-P)^2
	+\cR_{IJ}\widetilde{L}^I\nabla_{\tilde i} \widetilde{L}^J \dd \tilde z^{\tilde i}(S+P)^2\nn
	&-( \widetilde{\cN}_{IJ} \widetilde{L}^ J\nabla_iL^I \dd z^i     +\cN_{IJ}L^J\nabla_{\tilde i} \widetilde{L}^ I \dd \tilde z^{\tilde i} )(S+P)(S-P)
	\Big] \nn +\ii\Big[&\cN_{IJ}L^IL^J(S-P)(\dd-\ii\cA)(S-P)+\widetilde{\cN}_{IJ}\widetilde{L}^I\widetilde{L}^J(S+P)(\dd+\ii\cA)(S+P) \nn
	&-\cR_{IJ}L^I\widetilde{L}^J((S+P)(\dd-\ii\cA)(S-P)+(S-P)(\dd+\ii\cA)(S+P))\Big]\,.
\end{align}
The differential terms can be replaced using \eqref{dILL}, \eqref{diffSP}, and subsequently \eqref{algSP}. 
Further using \eqref{gff+L=I}, this reduces to 
\begin{align}
	\dd \Phi_0 &=\frac{\ii}{2\sqrt{2}}(\cI^{JK})^{-1}\Big[(\cN_{IJ}-\cR_{IJ})L^I(S-P)+(\widetilde{\cN}_{IJ}-\cR_{IJ})\widetilde{L}^I(S+P)\Big]\zeta_K K\nn
	&+\frac{\ii}{\sqrt{2}}\Big[\big(\cN_{IJ}(\xi\hook F^J_{[+]})+\cR_{IJ}(\xi\hook F^J_{[-]})\big)L^I(S-P)\nn
	&\qquad-\big(\widetilde{\cN}_{IJ}(\xi\hook F^J_{[-]})+\cR_{IJ}(\xi\hook F^J_{[+]})\big)\widetilde{L}^I(S+P)\Big]\,,
\end{align}
which recalling \eqref{Cdef} is equal to \eqref{xiinphi2}.
This completes the proof that $\Phi$ 
given in \eqref{actappsimple} and \eqref{phisappact}, canonically expressed in terms of the supergravity fields and bilinears,
 is equivariantly closed for supersymmetric configurations.

\subsection{Local form of the supersymmetric solutions}
\label{app:LocalForm_SUSY_Sols}

Here we use the bilinears \eqref{eq:Bilinears} to introduce coordinates and construct a local form for supersymmetric solutions, following the analogous analysis in \cite{BenettiGenolini:2019jdz}.

Under our reality assumptions \eqref{eq:Reality_Condition}, a Riemannian supersymmetric solution has a 
Dirac spinor $\epsilon$ satisfying \eqref{eq:Euclidean_KSE_epsilon}. 
This spinor generates an identity structure in four dimensions.  Indeed, using
the normalized chiral projections 
\begin{equation}
	\eta_\pm \equiv \frac{\epsilon_\pm}{\sqrt{S_\pm}} \, , \qquad \text{where} \qquad \epsilon_\pm \equiv \Pnew_\pm \epsilon \, , \quad S_\pm \equiv \epsilon_\pm^\dagger \epsilon_\pm \, ,
\end{equation}
we can construct an orthonormal frame $\{ \E^a\}$ via
\begin{equation}
\label{eq:Canonical_Frame}
	\ii \E^4 - \E^3 \equiv \eta_-^\dagger \gamma_{(1)} \eta_+ \, , \qquad \ii \E^1 - \E^2 \equiv (\eta^c_-)^\dagger \gamma_{(1)} \eta_+ \, ,
\end{equation}
where recall that $\eta^c \equiv - \cC^{-1} \eta^*$. The relation with the Killing spinor bilinears is then given by \eqref{eq:IdentityStructure}, 
along with
\begin{align}
	\overline{\epsilon} \gamma_{(2)} \epsilon &= - S ( \E^1 + \ii \E^2) \wedge ( \E^4 + \ii \cos\theta \, \E^3) \, , \nn
\label{eq:Additional_Relations_Frame_Bilinears} 
	\overline{\epsilon} \gamma_{(2)}\gamma_5 \epsilon &= - S ( \E^1 + \ii \E^2) \wedge ( \cos\theta \, \E^4 + \ii \E^3)\,,
\end{align}
where we recall $\bar\epsilon\equiv \epsilon^T\cC$.
As already remarked in section \ref{subsec:Preliminaries}, the functions $S$, $P$ and $\theta$, and the 
bilinear forms in \eqref{eq:Bilinears}, are globally defined on the spacetime $M$, whereas the frame \eqref{eq:Canonical_Frame} degenerates on the locus of fixed points, $M_0$, where the Killing vector has vanishing norm $\|\xi\|^2=0$. Therefore, the comments and expressions that follow should be interpreted as holding on $M\setminus M_0$.

Looking at \eqref{diffSP}, since $K$ is closed and $\xi^\flat$ is dual to a Killing vector, we can define two coordinates $y$ and $\psi$ such that
\begin{equation}
\label{eq:E3_E4}
	\E^3 = \frac{1}{S \sin\theta} \dd \left( \frac{1}{y} \right) \, , \qquad \E^4 = S \sin\theta \, (\dd\psi + \newphi) \, , 
\end{equation}
where $\newphi$ 
is a local basic one-form, and we have defined $\psi$ to be a local coordinate such that $\xi = \partial_\psi$. If we choose a gauge such that $\partial_y \hook A^I = 0$, we can also introduce a local complex coordinate $u$ and a local function $V$ such that
\begin{equation}
\label{eq:E1_E2}
	\E^1 + \ii \E^2 = \frac{2 \, \e^{V/2}}{y^2 S \sin\theta} \dd u \, .
\end{equation}
Therefore, locally we can write the line element of a supersymmetric solution as
\begin{equation}
\label{eq:Local_SUSY_LineElement}
	\dd s^2 = S^2 \sin^2\theta ( \dd\psi + \newphi)^2 + \frac{1}{y^4 S^2 \sin^2\theta} \left( \dd y^2 + 4 \e^{V} \, \dd u \, \dd\ubar \right) \, .
\end{equation}
This highlights that the metric on $M\setminus M_0$ is locally a fibration over an (orbifold) base $B$. We can also write
the metric as
\begin{align}
	\dd s^2 &= \|\xi\|^2 (\dd\psi + \newphi)^2 + \frac{1}{\|\xi\|^2} \gamma_{ij} \dd x^i \dd x^j \, ,\nn
	&= \|\xi\|^2 \eta^2 + \frac{1}{\|\xi\|^2} \gamma_{ij} \dd x^i \dd x^j \, ,
\end{align}
where 
$\eta \equiv \dd\psi + \newphi$
and
\begin{align}
\gamma_{ij} \dd x^i \dd x^j = \frac{1}{y^4} \left( \dd y^2 + 4 \e^{V} \, \dd u \, \dd\ubar \right)\,,
\end{align} is a line element
on $B$. We choose the orientation $\vol_g = \|\xi\|^{-2} \eta \wedge \vol_\gamma$, which means that if $\alpha_k$ is a basic $k$-form
\begin{equation}
\label{eq:HodgeDual_BasicForm}
	*_g \alpha_k = \|\xi\|^{2(k-1)}\eta \wedge *_\gamma \alpha_k \, , \qquad *_g(\alpha_k \wedge \eta) = - \|\xi\|^{2(k-2)} *_\gamma \alpha_k \, .
\end{equation}
We also remark that we can take the line element \eqref{eq:Local_SUSY_LineElement} to be associated with 
asymptotically locally hyperbolic solutions with conformal boundary located at $y=0$ (in particular $S\sim 1/y$ at this boundary; see appendix \ref{app:holrenn}).
The form of the gauge fields, with $\cL_\xi A^I=0$, is determined by \eqref{gaugehookxi}:
\begin{equation}
\label{eq:LocalForm_AI}
	A^I = \left( - \Phi_0^{I} + c^I \right) (\dd\psi + \newphi) + a^I \, ,
\end{equation}
where {$c^I $ are constants and $a^I$ are basic one-forms}.

The differential constraints \eqref{diffSP} are satisfied by the real ``Dirac'' bilinears, but additional information can be gleaned from the equations satisfied by the complex bilinears, which again can be derived from \eqref{eq:Euclidean_KSE_epsilon} together with the fact that
\begin{equation}
	S \sin\theta ( \E^1 + \ii \E^2) = \ii \overline{\epsilon}\gamma_{(1)} \gamma_5 \epsilon \, .
\end{equation}
Taking the exterior derivative we find 
\begin{align}
\label{eq:HR_dE1iE2}
	\left( \dd - \frac{\ii}{2} \zeta_I A^I \wedge \right) \left( S \sin\theta ( \E^1 + \ii \E^2) \right) &= \frac{\ii}{\sqrt{2}} \zeta_I S ( \E^1 + \ii \E^2)\wedge	\left[ (\widetilde{L}^I - L^I) ( \E^4 + \ii \cos\theta \, \E^3) \right. \nn
	& \qquad \qquad \left. + (\widetilde{L}^I + L^I) ( \cos\theta \, \E^4 + \ii \E^3) \right] \, .
\end{align}
Considering its components, in a gauge where $\partial_y \hook A^I = 0$, we find
\begin{align}
\label{eq:BPS_Local_dyV}
	\frac{y}{4} \partial_y V &= 1 - \frac{1}{y S^2 \sin^2\theta} \frac{\zeta_I}{2\sqrt{2}} \left( C^I + \widetilde{C}^I \right) \, ,  \nn
	\zeta_I a^I_{\ubar} &= - \ii \partial_{\ubar}V \, .
\end{align}
In general, the form of each of the gauge fields 
$A^I$ in \eqref{eq:LocalForm_AI} is not fixed by the local geometry. By contrast, however,
the R-symmetry gauge field $A^{R}$ in \eqref{rsymgf}, under which the spinor is charged, is fixed. Indeed using
the second equation in \eqref{eq:BPS_Local_dyV} we have
\begin{equation}
\label{eq:AR_LocalForm}
	A^{R} \equiv \frac{1}{2} \zeta_I A^I = \frac{1}{2} \left[ \left( - \zeta_I \Phi_0^{I} + 4 Q \right) (\dd\psi + \newphi) +  ( \ii \partial_u V \, \dd u - \ii \partial_{\ubar}V \, \dd \ubar ) \right] \, ,
\end{equation}
where $Q$ is the charge \eqref{eq:SpinorRcharge} of the spinor under the Killing vector.

Two useful equations describing derivatives of the line element \eqref{eq:Local_SUSY_LineElement} can be found from \eqref{diffSP} and \eqref{dxiflatfirst}
that will be useful in appendix \ref{app:holrenn}.
For the first, we observe that by definition
\begin{equation}
	\dd\eta = (S \sin\theta)^{-4} \xi\hook(\dd\xi^\flat \wedge \xi^\flat) \, ,
\end{equation}
and we can take $\dd\xi^\flat$ from \eqref{dxiflatfirst}. We then observe that the gauge field contribution only consists of the basic part of the (anti-)self-dual projections, defined via $(F^I)_T \equiv F^I + ( \xi \hook F^I) \wedge \eta$, and moreover \eqref{eq:HodgeDual_BasicForm} implies that for an (anti-)self-dual projection
\begin{equation}
	\left( F^I_{[\pm]} \right)_T = \pm \left( S \sin\theta \right)^{-2} *_\gamma \left( \xi \hook F^I_{[\pm]} \right) \, .
\end{equation}
The latter can be expressed using \eqref{diffSP}, and combining everything we find
\begin{equation}
\label{eq:BPS_deta}
	\dd \eta = \frac{2}{S^3 \sin^3\theta} *_\gamma \left[ \frac{\zeta_I }{\sqrt{2} S \sin\theta } \left( \widetilde{C}^I - C^I \right) \dd \left( \frac{1}{y} \right)  - S(\dd\theta - \ii \cA \, \sin\theta ) \right] \, .
\end{equation}

For the second, we need the second derivative of $V$ in order to compute the curvature of $\gamma_{ij}$, for which we use $\dd U_{[\pm]}$ from \eqref{eq:dWU} and the local form of $F^R$ from \eqref{eq:AR_LocalForm}
\begin{align}
\label{eq:BPS_Local_duubarV}
	\partial^2_{u\ubar}V  &= - 2y^2 \partial_y \left[\left( \frac{y}{4}\partial_y V - 1 \right) \frac{2\e^V}{y^3} \right] - 2 \cV \, \frac{\e^V}{y^4 S^2 \sin^2\theta} \nn
	&= - \e^V \left[  \partial^2_{yy}V + \frac{1}{4} (\partial_y V)^2 + \frac{12}{y^2} \left( \frac{y}{4} \partial_y V - 1\right)^2 + 2 \cV \, \frac{1}{y^4 S^2 \sin^2\theta} \right] \, .	
\end{align}

Finally, there are useful equations involving the scalar fields that we also use in 
appendix \ref{app:holrenn}. We only need the $\E^3$ component of the second and third equations in \eqref{algSP}. 
Since $\E_3 = - y^2 S \sin\theta \, \partial_y$ and $\xi = S \sin\theta \, \E_4$
we have
\begin{align}
	0 &= - y^2 S\sin\theta (S - P) \partial_y z^i - \sqrt{2} S\sin\theta \cI_{IJ} \cG^{i\tilde{j}} \nabla_{\tilde{j}}\tL^I F^{J-}_{34} + \frac{\zeta_I}{\sqrt{2}} S \sin\theta \cG^{i\tilde{j}} \nabla_{\tilde{j}}\tL^I \, , \nn
	0 &= - y^2 S\sin\theta (S + P) \partial_y \tz^{\tilde{i}} + \sqrt{2} S\sin\theta \cI_{IJ} \cG^{\tilde{i} j } \nabla_j L^I F^{J+}_{34} + \frac{\zeta_I}{\sqrt{2}} S \sin\theta \cG^{\tilde{i} j} \nabla_j L^I \, .
\end{align}
Then, writing 
\begin{equation}
\label{eq:FI_LocalForm}
	F^I = \eta \wedge \dd\Phi_0^{I} +  f^I\, ,
\end{equation}
where $f^I=\dd a^I$ is basic,
and using \eqref{eq:HodgeDual_BasicForm}, we obtain
\begin{equation}
\begin{split}
	F^{I\pm}_{34} &= \frac{1}{2} y^2 \left[ \partial_y \Phi_0^{I} \mp S^2\sin^2\theta (*_\gamma f^I)_y \right]\, .
\end{split}
\end{equation}
Thus,
\begin{align}
\label{eq:BPS_ScalarEqn_1}
	0 &= y^2 (S-P) \cG_{\tilde{i}i} \partial_y  z^i + \frac{1}{\sqrt{2}} \cI_{IJ} \nabla_{\tilde{i}}\tL^I y^2 \left[ \partial_y \Phi_0^{J} + S^2\sin^2\theta (*_\gamma f^J)_y \right] - \frac{\zeta_I}{\sqrt{2}} \nabla_{\tilde{i}}\tL^I \, , \nn
	0 &= y^2 ( S + P) \cG_{i\tilde{i}} \partial_y \tz^{\tilde{i}} - \frac{1}{\sqrt{2}} \cI_{IJ} \nabla_i L^I y^2 \left[ \partial_y \Phi_0^{J} - S^2\sin^2\theta (*_\gamma f^J)_y \right] \nonumber\\ 
& \quad - \frac{\zeta_I}{\sqrt{2}} \nabla_i L^I  \, .
\end{align}

\section{Minimal gauged supergravity}
\label{mingaugedsugraapp}
We can obtain minimal $D=4$,  $\mathcal{N}=2$ Euclidean gauged supergravity by setting all the vector multiplets to zero.
This can be achieved, consistently with our conventions, by setting 
\begin{align}
	X^0= \tX^0=\frac{1}{2^{3/2}}\,,\quad\zeta_0= 4 \,, \quad \mathcal{F}(X^0)= -2\ii (X^0)^2=-\frac{\ii}{4}\,,\quad
\end{align}
so that 
$\mathcal{K}=0$, $\mathcal{V}=- 6$, $\mathcal{I}_{00}=-4$ and $\mathcal{R}_{00}=0$. 
We also note that $u=\frac{1}{4}$, so
\begin{equation}
	\ii \cF(u) = \frac{1}{8} \, , \qquad \ii \frac{\partial \cF(u)}{\partial u} = {1} \, . 
\end{equation}
Writing the gauge field as $A\equiv A^0$, we then obtain from \eqref{themainaction} the action
\begin{align}
I	&= - \frac{1}{16\pi G_4} \int  \left( R + 6 - F^2\right)\vol_4  \, ,
\end{align}
with $F^2=F_{\mu\nu}F^{\mu\nu}$
and from \eqref{eq:Euclidean_KSE_epsilon}, \eqref{eq:Euclidean_KSE_tepsilon}
the supersymmetry transformations
\begin{align}
\label{eq:Euclidean_KSE_epsilonminimal}
	0 &= \nabla_\mu \epsilon  -  \ii  A_\mu  \epsilon +  \frac{1}{2}  \gamma_\mu  \epsilon +\frac{\ii}{4}F_{\nu\rho} \gamma^{\nu\rho}\gamma_\mu \epsilon \, ,\nn
		0 &= \nabla_\mu \tepsilon  +  \ii  A_\mu  \tepsilon +  \frac{1}{2}  \gamma_\mu  \tepsilon -\frac{\ii}{4}F_{\nu\rho} \gamma^{\nu\rho}\gamma_\mu \tepsilon \, .
\end{align}
We also note that the R-symmetry gauge field is defined as $A^R = \tfrac{1}{2}\zeta_IA^I=2A$ and that the $AdS_4$ vacuum solution
has unit radius, $\ell=1$.

From \eqref{PhiF} we have the equivariantly closed form
\begin{equation}\label{PhiFapp}
\Phi_{(F)}\equiv F-P\,,
\end{equation}
and from \eqref{phiecform}, \eqref{Phidef}
\begin{equation}\label{phiecformapp}
    \Phi=\Phi_4+\Phi_2+\Phi_0\, ,
\end{equation}
with
\begin{align}\label{Phidefapp}
\Phi_4&= 3\smallspace \vol_{4}+F\wedge*F\, ,\nn
\Phi_2
&= U+SF-P *F\nn
&=-\frac{1}{2}*\dd\xi^\flat-2*F P\,,\nn
\Phi_0&=-SP\,.
\end{align}
For the case of real metric and gauge field and $\tepsilon=\epsilon^c$ (as in section \ref{realitycondssec})
the expression for $\Phi$ is in precise agreement with
\cite{BenettiGenolini:2023kxp}; here we have shown that the equivariantly closed form still exists when
these conditions are relaxed. The equivariant closure for $\Phi_{(F)}$ is equivalent to $\xi\hook F=-\dd P$, a condition
which was noted in (3.7) of \cite{BenettiGenolini:2019jdz} in the real setting.

The result for the free energy \eqref{Fgrav} now reads
\begin{align}\label{Fgravappc}
	I_{\mathrm{OS}}^{\mathrm{FP}} &= \frac{\pi}{2G_4} \bigg[ \mp\sum_{{\rm nuts}_\pm}  \frac{ (b_1 \mp b_2)^2 }{4b_1 b_2} 
	  + \sum_{{\rm bolts}_\pm} \left( -2\kappa  \mathfrak{p}_\pm\pm \frac{1}{4} \int_{\text{bolt}_\pm} c_1(L) \right)
  \bigg] \, .
\end{align}
The R-symmetry gauge field $A^R = 2A$ has flux through a bolt constrained by \eqref{eq:R_Symmetry_Constraint_Bolt}, which reads
\begin{align}
	\mathfrak{p}^R_\pm = 2 \mathfrak{p}_\pm
	&\equiv 2 \frac{1}{4\pi}\int_{\Sigma_\pm} F\nn
	& = \frac{\kappa}{2} \int_{\Sigma_\pm} \left[ \pm c_1(L) - c_1(T\Sigma_\pm) \right] \, ,
\end{align}
(and matches (3.74) of \cite{BenettiGenolini:2019jdz} with $\kappa = +1$). We also note that 
eliminating $\mathfrak{p}_\pm$ from \eqref{Fgravappc} is in agreement with (11) of \cite{BenettiGenolini:2023kxp}.

Finally, for minimal gauged supergravity from \eqref{eq:sigmaI} we 
obtain a relation between the boundary geometry and the bilinear $P$:
\begin{equation}
	\ii \sigma^0=\frac{\ii}{4\pi}P|_{S^1_\mathrm{UV}}  = \frac{1}{8\pi } *_2 \dd\eta_{(0)}|_{S^1_\mathrm{UV}} \, .
\end{equation}

\section{Evaluation of boundary terms to action}
\label{app:holrenn}

\subsection{Supersymmetric holographic renormalization with scalars}
\label{app:SUSYHolRen}

We assume the Euclidean gauged supergravity theory admits an $AdS_4$ vacuum solution with unit radius. This
corresponds to a critical point of the potential with constant scalars, for which the potential has the value $\cV_*$ (here and in the following we will denote quantities evaluated at this critical point with a $*$).
Taking the trace of Einstein's equation \eqref{eq:4d_N2_EinsteinEOM} then gives
\begin{equation}
	\cV_* = - 6 \, .
\end{equation}
This solution is supersymmetric provided $\nabla_i W = \nabla_{\widetilde{i}}\widetilde{W} = 0$, which from
\eqref{curlyvpot}, \eqref{wsuperpots}
implies that $\cV_* = - 3 \zeta_I \zeta_J L^I \widetilde{L}^J$ and hence  $\zeta_I \zeta_J L^I_* \widetilde{L}^J_* = 2 $.
In fact, without loss of generality we will assume that 
\begin{equation}
\label{eq:Vacuum_Assumptions}
	\zeta_I L^I_* = \zeta_I \widetilde{L}^I_* = \sqrt{2} \, , \qquad \left( \nabla_i L^I \right)_* = \left( \nabla_{\widetilde{i}} \widetilde{L}^I \right)_* \in \R \, ,  \qquad z^i_* = \widetilde{z}^{\widetilde{i}}_* = 0 \, . 
\end{equation}
The first two conditions can be imposed using the Abelian symmetry within the K\"ahler transformations described around \eqref{phaseabelianrots}, and the last one is obtained by a simple field redefinition, shifting the value of the physical scalars at the critical point of the potential. 
We now consider supersymmetric solutions that asymptotically approach this $AdS_4$ critical point and implement holographic renormalization.

We begin by introducing a cut-off at $y=\cutoff>0$, where $y$ is the coordinate introduced in section \ref{app:LocalForm_SUSY_Sols}. This gives us a manifold $\cutoffspace$ with boundary $\cutoffbdry \cong \partial M$ with induced metric on $\cutoffbdry$ given by
\begin{equation}
\label{eq:hInduced_Cutoff}
	h = S^2\sin^2\theta (\dd\psi + \newphi)^2 + \frac{4\e^V}{\delta^2 S^2 \sin^2\theta} \dd u \dd\ubar\,.
\end{equation}
The bulk action $I_{\rm OS}$ is evaluated using the fixed point formula, and, as discussed in section \ref{subsec:Using_BVAB}, 
gives $I_{\mathrm{OS}}=I^{\mathrm{FP}}_{\mathrm{OS}}+I^{\partial M}_{\mathrm{OS}}$, with $I^{\mathrm{FP}}_{\mathrm{OS}}$ the bulk fixed point contribution and $I^{\partial M}_{\mathrm{OS}}$ the boundary contribution
\begin{equation}\label{appbdbulkicont}
	I_{\mathrm{OS}}^{\partial M} = - \frac{\pi}{2G_4} \frac{1}{(2\pi)^2} \int_{\cutoffbdry} \eta \wedge \left( \Phi_2 + \Phi_0 \, \dd\eta \right) \, .
\end{equation}
To this we add the Gibbons--Hawking--York term needed to yield a well-defined variational problem
\begin{equation}
	I_{\rm GHY} =  - \frac{1}{8\pi G_4} \int_{\cutoffbdry} K \, \vol_h \, , 
\end{equation}
where $K$ is the trace of the extrinsic curvature computed using the outward-pointing normal to $\cutoffbdry$. The resulting boundary contribution is divergent in the limit $\cutoff\to 0$, so we add counterterms in order to cancel the divergences \cite{Emparan:1999pm}. Due to the presence of scalars, there is the possibility of adding Weyl-invariant finite counterterms, so we fix the form of the counterterms to preserve supersymmetry \cite{Freedman:2013oja}
\begin{equation}
	I_{\rm ct} = - \frac{1}{8\pi G_4} \int_{\cutoffbdry} \left( - \frac{1}{2} R + \cW  \right) \vol_h \, ,
\end{equation}
where $R$ is the Ricci scalar of the induced metric $h$ on $\cutoffbdry$ and $\cW \equiv - \sqrt{2} \e^{\cK/2} \sqrt{W \widetilde{W}} $ is the real superpotential defined in \eqref{eq:Euclidean_Superpotentials}.\footnote{The form of this term has been proposed in \cite{Freedman:2013oja} for a specific ansatz, and has been supported by numerous studies, including \cite{Freedman:2016yue, Halmagyi:2017hmw, Gauntlett:2018vhk, Bobev:2018wbt, Arav:2018njv, Bobev:2020pjk, BenettiGenolini:2020kxj}. We also note that in some cases it has been suggested to change the Ricci scalar counterterm $- R/2$ with $\cW R/4$, but we have checked that the difference between the two vanishes upon taking $\cutoff \to 0$.}

The boundary contribution
\begin{equation}
\label{eq:I_PartialM}
	I^{\partial M} \equiv \lim_{\cutoff\to 0} \left( I^{\partial M}_{\rm OS} + I_{\rm GHY} + I_{\rm ct} \right)
\end{equation}
is finite. However, the quantity $I \equiv I^{\rm FP}_{\rm OS} + I^{\partial M}$ is not the ``gravitational free energy'', in the sense that it does not correspond to the generating functional of the dual supersymmetric conformal field theory. Indeed, as pointed out in section \ref{subsec:Using_BVAB}, the gauged supergravity theories we consider include scalar fields that are dual to scalar conformal primary operators $\cO_{\Delta}$ of both scaling dimensions $\Delta_+ = 2$ and $\Delta_- = 1$. In particular, in order to be consistent with supersymmetry, we should take \cite{Freedman:2013oja, Bobev:2020pjk, Zan:2021ftf}
\begin{equation}
	\text{$z^i - \tz^i$ dual to $\cO_{\Delta_+}$} \, , \qquad \text{$z^i + \tz^i$ dual to $\cO_{\Delta_-}$} \, .
\end{equation}
Correspondingly, we can impose canonical Dirichlet boundary conditions for $z^i - \tz^i$, i.e., the source for $\cO_{\Delta_+}$ is proportional to the leading term in the expansion of $z^i - \tz^i$. We should also impose alternate boundary conditions for $z^i + \tz^i$ so that the source for $\cO_{\Delta_-}$ is proportional to the quantity canonically conjugate to the leading term in the expansion $z^i + \tz^i$ \cite{Klebanov:1999tb}.\footnote{One might naively expect that the variable canonically conjugate to the leading term in the expansion $z^i + \tz^i$ would be the first subleading term. However, as we shall see, the presence of the supersymmetric counterterms $I_{\rm ct}$ modifies this expectation.\label{footnote:Subleading}} 

Concretely, with our assumptions the scalars have the following asymptotic expansion near the conformal boundary
\begin{align}
	h_{ij} &= \frac{1}{y^2} g_{{\rm bdry},ij} + o\left( \frac{1}{y^2} \right) \,  ,\nn
	z^i &= y \smallspace z^i_{(1)} + \frac{y^2 }{2} \smallspace z^i_{(2)} + o\left( y^2 \right) \, , \nn
	\tz^{\tilde{i}} &= y \smallspace \tz^{\tilde{i}}_{(1)} + \frac{y^2 }{2} \smallspace \tz^{\tilde{i}}_{(2)} + o\left( y^2 \right) \, .
\end{align}
The leading-order contributions are the boundary metric $g_{\rm bdry}$ and the scalars $z^i_{(1)}$ and $\tz^{\tilde{i}}_{(1)}$, and we define the canonically conjugate variables
\begin{equation}
\label{eq:Defn_Pi_TildePi}
	\Pi_i \equiv \frac{1}{\sqrt{g_{\rm bdry}}} \frac{\delta I}{\delta z^i_{(1)}} \, , \qquad \widetilde{\Pi}_{\tilde{i}} \equiv \frac{1}{\sqrt{g_{\rm bdry}}} \frac{\delta I}{\delta \tz^{\tilde{i}}_{(1)}} \, .
\end{equation}
The on-shell action $I$ is a functional of $z^i_{(1)}$ and $\tz^{\tilde{i}}_{(1)}$, but we are looking for the dual to the field theory generating functional, which is a functional of the sources for $\cO_{\Delta_+}^i$ and $\cO_{\Delta_-}^i$, namely $z^i_{(1)} - \tz^{i}_{(1)}$ and $\Pi_i + \widetilde{\Pi}_{i}$. To construct this, and thus obtain the supersymmetric gravitational free energy, we should perform a Legendre transform, by suitably adding a boundary term and considering
\begin{align}\label{eq:J_app}
	J &\equiv I - \frac{1}{2} \sum_i \int_{\partial M} \left( z^i_{(1)} + \tz^{i}_{(1)} \right) \left( \Pi_i + \widetilde{\Pi}_{i} \right) \vol_{g_{\rm bdry}}\,.
	\end{align}
Then, on-shell we have 
\begin{align}
	F_{\rm grav} \equiv J_{\rm OS}&= I^{\rm FP}_{\rm OS} + I^{\partial M} - \frac{1}{2} \sum_i \int_{\partial M} \left( z^i_{(1)} + \tz^{i}_{(1)} \right) \left( \Pi_i + \widetilde{\Pi}_{i} \right) \vol_{g_{\rm bdry}} \, .
\end{align}

We shall now study the asymptotic expansion of the fields for a general supersymmetric solution and compute the various boundary contributions to 
$J_{\rm OS}$.
Recall that there is potential ambiguity in the definitions of the equivariant completion of each polyform, in the sense that
each lower-degree form can be changed by the addition of a globally defined closed basic form. Remarkably, we show that the canonical expressions \eqref{defphioiflux} and \eqref{Phidef} have the property
\begin{equation}
\label{eq:HolRen_Bdry_Statement}
	I^{\partial M} = \frac{1}{2} \sum_i \int_{\partial M} \left( z^i_{(1)} + \tz^{i}_{(1)} \right) \left( \Pi_i + \widetilde{\Pi}_{\tilde{i}} \right) \vol_{g_{\rm bdry}} \, .
\end{equation}
That is, the canonical expressions have the property
that the net contribution to the gravitational free energy 
only comes from the fixed point set of the action of the R-symmetry Killing vector
and we have
\begin{align}
F_{\rm grav} = I^{\rm FP}_{\rm OS}\, .
\end{align}

\subsection{Summary of the expansion}

We assume that the various quantities describing a solution admit an analytic expansion near the conformal boundary at $\{ y = 0\}$. Therefore, in a neighbourhood of the conformal boundary we can write
\begin{align}
	\theta(y,u , \ubar) &= \theta_{(0)} (u,\ubar) + y \, \theta_{(1)}(u,\ubar) + \frac{y^2}{2} \theta_{(2)}(u,\ubar) + o(y^2) \, , \nn
	S(y,u , \ubar) &= \frac{1}{y}S_{-1}(u,\ubar) + S_{(0)} (u,\ubar) + y \, S_{(1)}(u,\ubar) + \frac{y^2}{2} S_{(2)}(u,\ubar) + o(y^2) \, , \nn
	\newphi(y,u , \ubar) &= \newphi_{(0)} (u,\ubar) + y \, \newphi_{(1)}(u,\ubar) + \frac{y^2}{2} \newphi_{(2)}(u,\ubar) + o(y^2) \, , \nn
	V(y,u , \ubar) &= V_{(0)} (u,\ubar) + y \, V_{(1)}(u,\ubar) + \frac{y^2}{2} V_{(2)}(u,\ubar) + o(y^2) \, ,  \nn
	a^I(y,u,\ubar) &= a^I_{(0)}(u,\ubar) + y \, a^I_{(1)}(u,\ubar) + \frac{y^2}{2} \, a^I_{(2)}(u,\ubar) + o(y^2) \, , \nn
	z^i(y,u , \ubar) &= y \, z^i_{(1)}(u,\ubar) + \frac{y^2 }{2} \, z^i_{(2)}(u,\ubar) + o\left( y^2 \right) \, , \nn 
	\tz^{\tilde{i}}(y,u , \ubar) &= y \, \tz^{\tilde{i}}_{(1)}(u,\ubar) + \frac{y^2 }{2} \, \tz^{\tilde{i}}_{(2)}(u,\ubar) + o(y^2) \, .
\end{align}
Here recall that $S$ and $\theta$ are functions defined as spinor bilinears by \eqref{eq:Bilinears} and \eqref{eq:IdentityStructure}, $\newphi$ is a local basic one-form defined by \eqref{eq:E3_E4}, and $V$ is a local function defined by \eqref{eq:E1_E2}. These objects determine the supersymmetric line element via \eqref{eq:Local_SUSY_LineElement}. The local basic one-form $a^I$ determines the basic component of the gauge fields $A^I$ \eqref{eq:LocalForm_AI}, and finally $z^i$ and $\tz^{\tilde{i}}$ are the scalar fields. We shall insert this asymptotic expansion into equations obtained from the Killing spinor equations and find relations between the expansion coefficients by solving the equations order by order. More precisely, we shall consider the equations for the metric \eqref{eq:BPS_Local_dyV}, \eqref{eq:BPS_deta} and \eqref{eq:BPS_Local_duubarV}, and the equations \eqref{eq:BPS_ScalarEqn_1} for the scalar fields.

\subsubsection{Leading order}

At leading order, we find that the line element \eqref{eq:Local_SUSY_LineElement} takes the form\begin{equation}\label{eq:ConformalBoundary_Metric0}
	\dd s^2 \sim \frac{\dd y^2}{y^2} + \frac{1}{y^2} \left[ \left( \dd\psi + \newphi_{(0)} \right)^2 + 4 \e^{V_{(0)}} \dd u \dd\ubar  \right]\, .
\end{equation}
Therefore, a supersymmetric solution has a conformal boundary $\partial M$ on which we can define a metric conformal to\footnote{Note that the Killing vector $\xi$ has unit norm for this metric. To apply the formalism of this appendix in a specific setting, one may need to do a coordinate transformation of the form $y\to y f(u,\bar u)$ in order to change the boundary metric to this conformal representative.}
\begin{equation}
\label{eq:ConformalBoundary_Metric}
	\dd s^2_{\rm bdry} =\eta_{(0)}^2+ 4 \e^{V_{(0)}} \dd u \dd\ubar \, ,
\end{equation}
where we denote $\eta_{(0)} \equiv \dd\psi + \newphi_{(0)}$, which is further constrained by
\begin{equation}
\label{eq:ConformalBoundary_detaConstraint}
	\dd\eta_{(0)} = \left( \theta_{(1)} - \ii \cA_{y(0)} \right) 4\ii \e^{V_{(0)}} \, \dd u \wedge \dd\ubar \, .
\end{equation}
Assuming that the orbits of $\xi$ close (otherwise, see footnote \ref{footnote:Orbits}), we can write $\partial M$ as the total space of a circle bundle $\cL$ over $\Sigma_2$ (which is generically an orbifold) with
\begin{equation}
\label{eq:BaseMetric_app}
	\dd s^2_2 = 4\e^{V_{(0)}} \, \dd u \dd \overline{u} \, , \qquad R_{(2d)} = - \e^{-V_{(0)}} \partial^2_{u\overline{u}} V_{(0)} \, \qquad \vol_2 = 2\ii \e^{V_{(0)}} \, \dd u \wedge \dd \overline{u} \, .
\end{equation}
Here, and in the following, the subscript $(0)$ indicates the leading-order term in the expansion of a composite quantity, such as a function of the scalars. In this case, $\cA_{y(0)}$ is the leading-order term in the expansion of the K\"ahler connection obtained Euclideanizing \eqref{eq:Lorentzian_KahlerHodgeConnection} as described in section \ref{subsec:EuclideanTheory}. Namely
\begin{equation}
	\cA_{y(0)} = - \frac{\ii}{2} \left[ \left(\partial_i \cK\right)_{(0)} z^i_{(1)} - \left( \partial_{\tilde{i}} \cK \right)_{(0)} \tz^{\tilde{i}}_{(1)} \right] \, . 
\end{equation}
On the other hand, since $z_{(0)}^i$ and $\tz_{(0)}^{\tilde{i}}$ vanish, the other components of the K\"ahler connection vanish at the leading order: $\cA_{u(0)} = \cA_{\ubar(0)}=0$. For the STU model, the K\"ahler potential is \eqref{eq:cK_STU}, and $\left(\partial_i \cK\right)_{(0)} = \left( \partial_{\tilde{i}} \cK \right)_{(0)} = \cA_{y(0)}= 0$.

The leading order of the expansion of the gauge fields can be obtained from the expansion of the bilinears and scalar fields. We first note that the lowest component of the equivariantly closed form $\Phi^I_{(F)}$, $\Phi^I_0$ defined in \eqref{defphioiflux}, has a finite boundary limit as $y\to 0$
\begin{align}
\label{eq:PhiI0_app}
	\Phi^I_0 &= \sqrt{2} \left( 2 L^I_{(0)} \left( \theta_{(1)} - \ii \cA_{y(0)} \right) + \left( \nabla_i L^I \right)_{(0)} \left( z^i_{(1)} - \tz^{\tilde{i}}_{(1)} \right) \right) + o(1) \nn
	&= \sqrt{2} \left( L^I_{(0)} *_2 \dd \eta_{(0)} + \left( \nabla_i L^I \right)_{(0)} \left( z^i_{(1)} - \tz^{\tilde{i}}_{(1)} \right) \right) + o(1) \, ,
\end{align}
where in the second line we have used \eqref{eq:ConformalBoundary_detaConstraint} and referred to the metric \eqref{eq:BaseMetric_app} on the two-dimensional base.
In particular we see that this gauge-invariant expression is determined both by the boundary geometry as well as the scalar mass deformations $( z^i_{(1)} - \tz^{\tilde{i}}_{(1)}  )$.
From this, using the expression \eqref{eq:LocalForm_AI} for $A^I$, we find
\begin{align}
	A_{(0)}^I &= \left[ - \sqrt{2} \left( L^I_{(0)} *_2 \dd\eta_{(0)} + \left( \nabla_i L^I \right)_{(0)} \left( z^i_{(1)} - \tz^{\tilde{i}}_{(1)} \right) \right) + c^I \right] \eta_{(0)} + a^I_{(0)} \, ,
\end{align}
where $c^I$ are constants and $a^I_{(0)}$ are basic one-forms.
Recall that by assumption
\begin{equation}
	L^I_{(0)} = \tL^I_{(0)} = L^I_* \, , \qquad \left( \nabla_i L^I \right)_{(0)} = \left( \nabla_{\tilde{i}} \tL^I \right)_{(0)} = \left( \nabla_{\tilde{i}} \tL^I \right)_* \, , 
\end{equation}
with the vacuum values defined around \eqref{eq:Vacuum_Assumptions}, so they satisfy
\begin{equation}
	\zeta_I L^I_{(0)} = \sqrt{2} \, , \qquad \zeta_I \left( \nabla_i L^I \right)_{(0)} = 0 \, .
\end{equation}
In particular, this provides us with the expansion of the R-symmetry gauge field \eqref{eq:AR_LocalForm} 
\begin{equation}
\label{eq:AR0_LocalForm} 
	A^R_{(0)} =  \left( - *_2 \dd\eta_{(0)} + 2Q \right) \eta_{(0)} + \frac{\ii}{2} \left( \partial_u V_{(0)} \, \dd u - \partial_{\ubar}V_{(0)} \, \dd \ubar \right) \, ,
\end{equation}
where $Q$ is the charge \eqref{eq:SpinorRcharge} of the spinor under the Killing vector.
Therefore, the boundary data is comprised of $V_{(0)}$, $\theta_{(1)}$, $\newphi_{(0)}$ (constrained by \eqref{eq:ConformalBoundary_detaConstraint}), $a^I_{(0)}$, $z_{(1)}^i$ and $\tz_{(1)}^{\tilde{i}}$. 
{In fact, for the purposes of holographic comparisons discussed in section \ref{subsec:Holography}, we consider non-compact $\R^2$ submanifolds in the bulk and a particularly relevant quantity is the restriction of $\Phi^I_0$ (given in \eqref{eq:PhiI0_app})
at the UV boundary $S^1_{\rm UV}$, which is denoted by $\sigma^I$ in \eqref{eq:Defn_sigma_y}.}
Notice also that $A^R_{(0)}$ is completely fixed in terms of the geometry on $\partial M$ and the choice of $\eta_{(0)}$, 
up to the gauge dependent choice of $Q$ \cite{Klare:2012gn, Closset:2012ru}. 
In particular, its curvature has the form
\begin{equation}
\label{eq:FR_Leading_App}
	F^R_{(0)} = \eta_{(0)} \wedge \dd *_2 \dd \eta_{(0)} + \frac{1}{4} \left(  R_{(2d)} - 8 \lVert \dd \eta_{(0)} \rVert^2 \right) \vol_2 \, .
\end{equation}

In expanding the scalar sections we have taken advantage of the fact that they are covariantly holomorphic/anti-holomorphic (see \eqref{nablaLdef}), so
\begin{align}
	\partial_i L^I &= \nabla_i L^I - \frac{1}{2}\partial_i \cK \, L^I \, , &\qquad \partial_{\tilde{i}}L^I &= \nabla_{\tilde{i}}L^I + \frac{1}{2} \partial_{\tilde{i}} \cK \, L^I = \frac{1}{2} \partial_{\tilde{i}} \cK \, L^I \, , \nn
	\partial_{\tilde{i}} \tL^I &= \nabla_{\tilde{i}} \tL^I - \frac{1}{2}\partial_{\tilde{i}} \cK \, \tL^I \, , &\qquad \partial_{i} \tL^I &= \nabla_{i}\tL^I + \frac{1}{2} \partial_{i} \cK \, \tL^I = \frac{1}{2} \partial_{i} \cK \, \tL^I \, ,
\end{align}
and in turn
\begin{align}
	L^I &= L^I_{(0)} + y \left[ \left(\nabla_i L^I\right)_{(0)} z^i_{(1)} - L^I_{(0)} \ii \cA_{y(0)} \right] + o(y) \, ,\nn
\label{eq:LItilde}
	\tL^I &= L^I_{(0)} + y \left[ \left(\nabla_i L^I\right)_{(0)} \tz^{i}_{(1)} + L_{(0)}^I \ii \cA_{y(0)} \right] + o(y) \, .
\end{align}
Therefore,
\begin{align}
	\zeta_I L^I &= \sqrt{2} - \sqrt{2} \ii \cA_{y(0)}  + o(y) \, , \qquad \zeta_I \tL^I = \sqrt{2} + \sqrt{2} \ii \cA_{y(0)} + o(y) \, .
\end{align}
Using the homogeneity of the prepotential and the definitions of the various quantities, we also find the following useful expressions for the leading order 
terms in the expansion in the $y$ coordinates:
\begin{align}
	\ii \cF(L_{(0)}) &= \frac{1}{4} \, , \nn
	\left( \cG_{i\tilde{j}} \right)_{(0)} &= -2\ii \cF_{IJ}(L_{(0)}) \left( \nabla_i L^I \right)_{(0)} \left( \nabla_{\tilde{j}}\tL^J \right)_{(0)} \, , \nn
	\left( \cN_{IJ} \right)_{(0)} &= - \cF_{IJ}(X_{(0)}) + \frac{ \cF_I(X_{(0)}) \cF_J(X_{(0)}) }{ \cF(X_{(0)}) } = - \left(\tilde{\cN}_{IJ}\right)_{(0)} \, , \nn
	\label{eq:LeadingOrder_KineticMatrix}
	\left( \cR_{IJ}\right)_{(0)} &= 0 \, , \qquad \left( \cI_{IJ}\right)_{(0)} = - \ii \left( \cN_{IJ} \right)_{(0)} \, .
\end{align}
While deriving the leading order of $\cG_{i\tilde{j}}$, we also find expressions for $(\partial_i \cK)_{(0)}$ and $(\partial_{\tilde{i}} \cK)_{(0)}$ directly from their definition obtained by the Euclideanized version of \eqref{symconlorentzian}:
\begin{align}
\label{eq:pdi_mcK}
	\partial_i \cK &= - \ii \e^{\cK} \partial_i X^I \left( \cF_I(\tX) + \cF_{IJ}(X)\tX^J \right) \, , \nn
	 \partial_{\tilde{i}} \cK &= - \ii \e^{\cK} \partial_{\tilde{i}} \tX^I \left( \cF_I(X) + \cF_{IJ}(\tX) {X}^J \right)  \, ,
\end{align}
namely
\begin{align}
	\left(\partial_i \cK\right)_{(0)} &= - 2\ii \e^{\cK_{(0)}} \left( \partial_i X^I \right)_{(0)} \cF_I({X}_{(0)}) \, , \nn 
	 \left( \partial_{\tilde{i}} \cK \right)_{(0)} &= - 2\ii \e^{\cK_{(0)}} \left( \partial_{\tilde{i}} \tX^I \right)_{(0)} \cF_I(X_{(0)}) \, .
\end{align}
In turn, these also imply that
\begin{equation}
\label{eq:cFI_NablaXI}
	\cF_I(X_{(0)}) \left( \nabla_i X^I \right)_{(0)} = 0 = \cF_I(X_{(0)}) \left( \nabla_{\tilde{i}} \tX^I \right)_{(0)} \, .
\end{equation}

\subsubsection{Subleading orders}

As we proceed further in the asymptotic expansion of the quantities relevant for the computation of $J$ in \eqref{eq:J_app}, the expressions become more unwieldy. 

First, we focus on the scalars. To find expressions for the terms in their asymptotic expansion, it is useful to expand \eqref{eq:dLI} and its analogue for $\tL^I$, namely
\begin{align}
\dd L^I&=\nabla_i L^I \diff z^i-\ii\cA L^I\,,\nn
	\dd \tL^I &= \nabla_{\tilde{i}}\tL^I \, \dd \tz^{\tilde{i}} + \ii \cA \tL^I \, .
\end{align}
Along the $\dd y$ direction, they provide expressions for $L^I_{(2)}$ and $\tL^I_{(2)}$, and for our purposes we need their sums with the gauge couplings. For the latter, we can find useful relations on the derivative of the scalars by expanding \eqref{eq:BPS_ScalarEqn_1}
\begin{equation}
\label{eq:Subleading_BPS_ScalarEqn}
	\zeta_I \left( \nabla_i L^I \right)_{(1)} = \sqrt{2} \left(\cG_{i\tilde{j}}\right)_{(0)} \tz^{\tilde{j}}_{(1)} \, , \qquad \zeta_I \left( \nabla_{\tilde{j}} \tL^I \right)_{(1)} = \sqrt{2} \left(\cG_{\tilde{j}i}\right)_{(0)} z^{i}_{(1)} \, .
\end{equation}
These allow us to find
\begin{align}
	\zeta_I L^I_{(2)} &= \sqrt{2} \left( \cG_{i\tilde{j}} \right)_{(0)} z^i_{(1)} \tz^{\tilde{j}}_{(1)} - \sqrt{2} \ii \cA_{y(1)} - \sqrt{2} \left( \cA_{y(0)} \right)^2 \, , \nn
	\zeta_I \tL^I_{(2)} &= \sqrt{2} \left( \cG_{i\tilde{j}} \right)_{(0)} z^i_{(1)} \tz^{\tilde{j}}_{(1)} + \sqrt{2} \ii \cA_{y(1)} - \sqrt{2} \left( \cA_{y(0)} \right)^2 \, .
\end{align}
In turn, we can substitute these values in the expansion of the scalar potential $\cV$ defined in \eqref{eq:EuclideanScalarPotential}
\begin{equation}
	\cV = - 6 + \frac{y^2}{2} \, \left[ - 8 \left( \cG_{i\tilde{j}} \right)_{(0)} z^{i}_{(1)} \tz^{\tilde{j}}_{(1)} \right] + o(y^2) \, .
\end{equation}
Moreover, using \eqref{eq:LItilde}, \eqref{eq:LeadingOrder_KineticMatrix} and \eqref{eq:cFI_NablaXI}, we find that
\begin{equation}
	\ii \cF(L) = \frac{1}{4} + y \, \left( - \frac{1}{2}\ii \cA_{y(0)} \right) + o(y) \, , \qquad \ii \cF(\tilde{L}) = \frac{1}{4} + y \, \left( \frac{1}{2}\ii \cA_{y(0)} \right) + o(y) \, .
\end{equation}

With these results we can tackle the divergences in $I^{\partial M}$ as defined in \eqref{eq:I_PartialM}, which we repeat here
\begin{equation}\label{appbd7againn}
	I^{\partial M} = \lim_{\cutoff\to 0} \left( I^{\partial M}_{\rm OS} + I_{\rm GHY} + I_{\rm ct} \right) \, ,
\end{equation}
and we write the contribution $I^{\partial M}_{\rm OS}$ in \eqref{appbdbulkicont} as
\begin{align}\label{appbbreakdown}
	I^{\partial M}_{\rm OS} &\equiv I^{\partial M}_{\xi} + I^{\partial M}_{\rm gauge} + I^{\partial M}_{\rm scalars}\,, \nn
	I^{\partial M}_{\xi} &\equiv - \frac{\pi}{2G_4}\frac{1}{(2\pi)^2} \int_{\cutoffbdry} \eta \wedge * \left( - \frac{1}{2} \dd\xi^\flat \right) \, , \nn
	I^{\partial M}_{\rm gauge} &\equiv - \frac{\pi}{2G_4}\frac{1}{(2\pi)^2} \int_{\cutoffbdry} \eta \wedge \left[ - \frac{1}{2} ( \cI_{IJ} *F^I - \ii \cR_{IJ} F^I) \Phi^{J}_0 \right] \, , \nn
	I^{\partial M}_{\rm scalars} &\equiv - \frac{\pi}{2G_4}\frac{1}{(2\pi)^2} \int_{\cutoffbdry} \eta \wedge \dd\eta \, \e^{\cK} \Big[ \ii\cF(X) (S-P)^2 + \ii\tilde{\cF}(\tX)(S+P)^2 \nn
	& \qquad \qquad \qquad \qquad \qquad \qquad \qquad - \frac{1}{2} \ii(\cF_I \tX^I + \widetilde{\cF}_I X^I) (S^2 - P^2) \Big] \, .
\end{align}
Here we used the expression for $\Phi_2$ given in \eqref{phi2xiappbexp}.

This way of splitting the terms in the action simplifies the computation, since $I^{\partial M}_{\xi}$ combines naturally with the extrinsic curvature in the GHY term. That is, we have
\begin{align}
	\eta \wedge * \dd\xi^\flat \rvert_{\cutoffbdry} &= - \eta \wedge *_\gamma \dd\log \|\xi \|^2 \rvert_{\cutoffbdry} \nn
	&= - \left( \frac{\e^V}{y^2} \partial_y \log ( S^2 \sin^2\theta ) \right) \Bigg\rvert_{\cutoffbdry} \ii \, \eta \wedge \dd u \wedge \dd \ubar \, ,
\end{align}
and
\begin{equation}
\label{eq:Kvolh}
	K \, \vol_h \rvert_{\cutoffbdry} = - \delta^2 S \sin\theta \, \partial_y \left( \frac{2\e^V}{y^4 S \sin\theta} \right) \Bigg\rvert_{\cutoffbdry} \, \ii \, \eta \wedge \dd u \wedge \dd\ubar \, ,
\end{equation}
where we have used the outward pointing normal $n = - \sqrt{g_{yy}} \, \dd y$, and $g_{yy}$ refers to the line element \eqref{eq:Local_SUSY_LineElement}. We can then write the combination
\begin{align}
	I^{\partial M}_{\xi} + I_{\rm GHY} &= \frac{1}{8\pi G_4} 2\cutoff^2 \int_{\cutoffbdry} S^2 \sin^2\theta \, \partial_y \left( \frac{\e^V}{ y^4 S^2 \sin^2 \theta } \right) \Bigg\rvert_{\cutoffbdry} \ii \, \eta \wedge \dd u \wedge \dd \ubar \, .
\end{align}
For the gauge field term, $I^{\partial M}_{\rm gauge}$, recall that the gauge field curvature has the form \eqref{eq:FI_LocalForm} and its Hodge dual is
\begin{equation}
	* F^I = \|\xi \|^{-2} *_\gamma \dd\Phi_0^{F^I} + \eta \wedge *_\gamma f^I \, ,
\end{equation}
so that the combination relevant for $I^{\partial M}_{\rm gauge}$ is
\begin{equation}
\label{eq:ContributionIgauge_Form}
	\eta \wedge \left( \cI_{IJ} * F^I - \ii \cR_{IJ} F^I \right) = \cI_{IJ} \|\xi\|^{-2} \eta \wedge *_\gamma \dd\Phi_0^{F^I} - \ii \cR_{IJ} \eta \wedge f^I \, .
\end{equation}
To leading order, $\eta \wedge f^I = (\dd\psi + \newphi_{(0)}) \wedge f^I_{(0)}$, which together with \eqref{eq:LeadingOrder_KineticMatrix} means that the second term above only contributes at $o(1)$, so we can ignore it. Moreover, the first term contributes only a finite piece, so we postpone its analysis.

Finally, we consider the term $I^{\partial M}_{\rm scalars}$: interestingly, we find that the second line makes no
contribution since
\begin{equation}
	\ii(\cF_I \tX^I + \widetilde{\cF}_I X^I) = o(y^2) \, .
\end{equation}
In fact this term also makes no contribution for the fixed points. 

At this stage, if we substitute all of the results obtained until now into \eqref{appbd7againn}, excluding the
(finite) contribution from $I^{\partial M}_{\rm gauge}$ in \eqref{appbbreakdown},  we find that all
divergences cancel, and we are left with a finite expression
\begin{align}\label{appbmostofit}
	\lim_{\delta\to 0} &\left( I^{\partial M}_{\xi} + I^{\partial M}_{\rm scalars} + I_{\rm GHY} + I_{\rm ct} \right) \nn
	&= \frac{1}{8\pi G_4} \int_{\partial M} \frac{1}{4} \left[ -2 V_{(1)}  \left( \left( \cG_{i\tilde{j}} \right)_{(0)}  z^i_{(1)} \tz^{\tilde{j}}_{(1)} + ( \theta_{(1)} - \ii \cA_{(0)} )^2 \right)-4 \ii \cA_{y(1)}  \theta_{(1)} \right. \nn
	& \qquad \qquad \qquad \quad \left. + 4 \theta_{(1)}  \theta_{(2)} + 2 \left( \theta_{(1)} - \ii \cA_{y(0)} \right) \left( \cG_{i\tilde{j}} \right)_{(0)} \left( z^i_{(1)} z^j_{(1)} - \tz^{i}_{(1)} \tz^{j}_{(1)} \right)  \right. \nn
	& \qquad \qquad \qquad \quad \left. - 4 \ii \cA_{y(0)}  \left( \theta_{(2)} - 4 \ii \cA_{y(1)} \right) -8 \left( \cG_{i\tilde{j}} \right)_{(1)}  z^i_{(1)} \tz^{\tilde{j}}_{(1)}  \right. \nn
	& \qquad \qquad \qquad \quad \left. +2 \sqrt{2} \zeta_I \left( \left( \nabla_i L^I \right)_{(2)} z^i_{(1)} +  \left( \nabla_{\tilde{i}} \tL^I \right)_{(2)} \tz^{\tilde{i}}_{(1)} \right) \right. \nn
	& \qquad \qquad \qquad \quad \left. - \sqrt{2} \zeta_I ( L^I_{(3)} + \tL^I_{(3)}) \right] 2\ii \e^{V_{(0)}} \, \dd\psi \wedge \dd u \wedge \dd\ubar \, .
\end{align}
For the remaining $I^{\partial M}_{\rm gauge}$ contribution, we note that expanding \eqref{eq:ContributionIgauge_Form} involves the contractions $(\cI_{IJ})_{(0)}L^I_{(a)} L^J_{(b)}$ for $a, b=0,1,2$. These simplify using \eqref{eq:LeadingOrder_KineticMatrix} and the previously found expressions for $L^I_{(a)}$. In particular, we have
\begin{align}
	(\cI_{IJ})_{(0)} (\nabla_iL^I)_{(0)} (\nabla_{\tilde{j}}\tL^J)_{(0)} &= - \frac{1}{2} (\cG_{i\tilde{j}})_{(0)} \, , \qquad\qquad\qquad(\cI_{IJ})_{(0)} L^I_{(0)}L^J_{(0)} = - \frac{1}{2} \, , \nn
	(\cI_{IJ})_{(0)} L^I_{(0)} \left( \nabla_i L^J \right)_{(1)} &= - \frac{1}{2} (\cG_{i\tilde{j}})_{(0)} \tz^{\tilde{j}}_{(1)} \, , \qquad (\cI_{IJ})_{(0)} L^I_{(0)} \left( \nabla_i L^J \right)_{(0)} = 0 \, , \nn
		(\cI_{IJ})_{(0)} L^I_{(0)} \left( \nabla_{\tilde{j}} \tL^J \right)_{(1)} &= - \frac{1}{2} (\cG_{i\tilde{j}})_{(0)} z^{i}_{(1)} \, .
\end{align}
Moreover, we need to find $\zeta_I \left( \nabla_i L^I \right)_{(2)}$ and $\zeta_I L^I_{(3)}$ (and the corresponding tilded expressions). As before, we find these expressions going one order down in the expansion of \eqref{eq:BPS_ScalarEqn_1} and \eqref{eq:dLI} (and the corresponding tilded expressions). Substituting the result, 
we obtain $I^{\partial M}_{\rm gauge}$ and then combining with
\eqref{appbmostofit} we find the finite boundary contribution of the on-shell action is given by
\begin{align}
\label{eq:IpartialM}
	I^{\partial M} &= \frac{1}{8\pi G_4} \int_{\partial M} \Bigg[ \frac{1}{2\sqrt{2}} \ii \cF_{IJ}(L_{(0)}) \left( \nabla_i L^I \right)_{(0)} \left( *_\gamma f^J \right)_{y(2)}  ( z^i_{(1)} + \tz^i_{(1)} ) \nn
	& \qquad + \frac{1}{2}  \left( \cG_{i\tilde{j}} \right)_{(0)} (\theta_{(1)} - \ii \cA_{y(0)} ) \left( z^i_{(1)} z^j_{(1)} -\tz^i_{(1)} \tz^j_{(1)} \right) \Bigg] 2\ii \e^{V_{(0)}} \, \dd\psi \wedge \dd u \wedge \dd\ubar \, ,
\end{align}
where we have written
\begin{equation}
	\left( *_\gamma f^J \right)_{y(2)} = - \ii  f^I_{(0)u\ubar} \e^{- V_{(0)}} \, .
\end{equation}

\subsubsection{Legendre transform}
\label{subsubsec:Legendre_Transform}

As discussed in section \ref{app:SUSYHolRen}, in order to obtain a gravitational free energy that corresponds to a supersymmetric generating functional we should perform a Legendre transform, exchanging $z^i_{(1)} + \tz^i_{(1)}$ for the canonically conjugate variables $\Pi_i + \widetilde{\Pi}_i$.

The latter are defined by \eqref{eq:Defn_Pi_TildePi}, or, equivalently
\begin{equation}
	\Pi_i = \lim_{\cutoffdue\to 0} \frac{1}{\cutoffdue^2 \sqrt{h}} \frac{\delta I}{\delta z^i} \, , \qquad \widetilde{\Pi}_{\tilde{i}} = \lim_{\cutoffdue\to 0} \frac{1}{\cutoffdue^2 \sqrt{h}} \frac{\delta I}{\delta \tz^{\tilde{i}}} \, .
\end{equation}
Using the results of the expansions summarised until here, we can confirm that $\Pi_i$ and $\widetilde{\Pi}_{\tilde{i}}$ are both finite.
The relevant linear combinations for our purposes are
\begin{align}\label{pisappb}
	\Pi_i + \widetilde{\Pi}_i &= \frac{1}{8\pi G_4} \Bigg[ \frac{1}{\sqrt{2}} \ii \cF_{IJ}(X_{(0)}) \left( \nabla_i L^I \right)_{(0)} \left( *_\gamma f^J \right)_{y(2)}  \nn
	& \qquad \qquad \quad + \left( \cG_{ij} \right)_{(0)} \left( \theta_{(1)} - \ii \cA_{y(0)} \right) \left( z^j_{(1)} - \tz^j_{(1)} \right) \Bigg] \, , \nn
	\Pi_i - \widetilde{\Pi}_i &= \frac{1}{8\pi G_4} \Bigg[ - \sqrt{2} \ii \cF_{IJ}(X_{(0)}) \left( \nabla_i L^I \right)_{(0)} \left( \partial_y \Phi_0^{F^J} \right)_{(0)} \nn
	& \qquad \qquad \quad - \left( \cG_{ij} \right)_{(0)} \left( \theta_{(1)} - \ii \cA_{y(0)} \right) \left( z^j_{(1)} + \tilde{z}^j_{(1)} \right) \Bigg] \, .
\end{align}
These are proportional, respectively, to the sources for the field theory operators $\cO^i_{\Delta_-}$, and to the one-point function of the field theory operators $\cO^i_{\Delta_+}$. More explicitly
\begin{align}
	\Pi_i - \widetilde{\Pi}_i &= \frac{1}{32\pi G_4} \Bigg[ 4 \left( \nabla_i L^I \right)_{(0)} \left( \cI_{IJ} \right)_{(0)} \left( \left( \nabla_{\tilde{i}} \tilde{L}^I \right)_{(1)} \tilde{z}^{\tilde{i}}_{(1)} - \left( \nabla_i L^I \right)_{(1)} z^i_{(1)} \right) \nn
		& + \left( \cG_{ij} \right)_{(0)} \big( V_{(1)} ( z^j_{(1)} - \tilde{z}^j_{(1)} )+ 2 \ii \cA_{y(0)}  ( z^j_{(1)} + \tilde{z}^j_{(1)} )+ 2 ( z^j_{(2)} - \tilde{z}^j_{(2)} ) \big) \Bigg]\,.
\end{align}
These generalize the expressions (3.57) in \cite{Bobev:2020pjk}, which by assumption have $\theta_{(1)} - \ii \cA_{y(0)} = 0$.
The Legendre transform term anticipated in \eqref{eq:J_app} is
\begin{align}
	\frac{1}{2} \int_{\partial M} &\left( \Pi_i + \tilde{\Pi}_i \right) \left( z^i_{(1)} + \tz^i_{(1)} \right) \vol_{g_{\rm bdry}} \nn
	 &= \frac{1}{8\pi G_4} \int_{\partial M} \Bigg[ \frac{1}{2\sqrt{2}} \ii \cF_{IJ}(X_{(0)}) \left( \nabla_i L^I \right)_{(0)} \left( *_\gamma f^J \right)_{y(2)} \left( z^i_{(1)} + \tz^i_{(1)} \right) \nn
	 & \ + \frac{1}{2} \left( \cG_{ij} \right)_{(0)} \left( \theta_{(1)} - \ii \cA_{y(0)} \right) \left( z^i_{(1)} z^j_{(1)} -\tz^i_{(1)} \tz^j_{(1)} \right) \Bigg] 2 \ii \e^{V_{(0)}} \, \dd\psi \wedge \dd u \wedge \dd \ubar  \, ,
\end{align}
which precisely agrees with $I^{\partial M}$ in \eqref{eq:I_PartialM}. We have thus proven \eqref{eq:HolRen_Bdry_Statement} and showed that 
all of the boundary contributions to the gravitational free energy cancel using the expressions \eqref{defphioiflux} and \eqref{Phidef} for the polyforms.

The computation for the boundary contribution to the on-shell action in this appendix has used the canonical expressions
for the equivariantly closed polyform expressed in terms of spinor bilinears in \eqref{Phidef}. It is interesting to see how this computation would change if we had used different expressions for $\Phi_2$ and $\Phi_0$ while maintaining equivariant closure
of $\Phi$. For example, we could change
\begin{equation}\label{changef2f0}
	\Phi_2 \to \Phi_2 + \zeta_2 \, , \qquad \Phi_0 \to \Phi_0 + c_0 \, ,
\end{equation}
where $\zeta_2$ is a closed basic two-form (note that this is more stringent than the conditions required below \eqref{eq:NonMinimalPhi2}; one can consider more general changes, but one should demand that they are globally defined), and $c_0$ is a constant. We find this would lead to the following change in the boundary contribution
\begin{equation}\label{newwbdynewec}
	I^{\partial M} \to I^{\partial M} - \frac{1}{8\pi G_4} \int_{\partial M} \eta_{(0)} \wedge \Bigg[ \zeta_2 + 2 c_0 (\theta_{(1)} - \ii \cA_{y(0)}) \vol_2 \Bigg] \, ,
\end{equation}
where
\begin{equation}
	\vol_2 = 2\ii \e^{V_{(0)}} \, \dd u \wedge \dd \ubar\,,
\end{equation}
is the volume of the two-dimensional metric \eqref{eq:2dMetric}. The total on-shell action, which is the integral of the top form
in $\Phi$, is obviously unchanged, which implies
that the additional term in \eqref{newwbdynewec} would necessarily be cancelled by a contribution from the fixed point set in the bolt
and we expand on this below.
It is appealing that the canonical expressions for the bilinears in \eqref{Phidef} have the feature that $I^{\partial M}=0$.

We use Stokes' theorem to explicitly see that there is a cancellation between the conformal boundary contribution from 
$\partial M$ in \eqref{newwbdynewec} and a boundary contribution from around the fixed point sets, 
that we call $\partial M_0$. 
We begin by recalling that on $M\setminus M_0$ the one-form
$\eta\equiv \diff\psi+\omega$ is a global one-form. 
Then on $M\setminus M_0$ we have
\begin{align}
0 = \diff\eta \wedge (\zeta_2 + c_0 \diff\eta) = \diff [\eta\wedge (\zeta_2 + 
c_0\diff\eta)]\, .
\end{align}
Here the first equation follows since $\xi$ contracted into $\zeta_2$ and 
$\diff\eta$ are both zero, so this four-form is identically zero 
on dimensional grounds. The second equation shows that 
it is globally exact on $M\setminus M_0$ and so Stokes' theorem then 
immediately gives
\begin{align}\label{boundariescancel}
0 = \int_{\partial M} \eta\wedge (\zeta_2 + 
c_0\diff\eta) + \int_{\partial M_0} \eta\wedge (\zeta_2 + 
c_0\diff\eta)\, .
\end{align}
For an alternative point of view, this result can be rephrased 
in terms of the fixed point theorem on a manifold with boundary that 
we described earlier. Recall in section \ref{subsec:Using_BVAB}
we introduced $\Phi_\xi\equiv -\eta\wedge (1-\diff\eta)^{-1}$ on $M\setminus M_0$ and
for an equivariantly closed form $\Phi$ on $M$ we wrote 
$\Phi=\diff_\xi(\Phi \Phi_\xi)$. We now apply this to the case 
$\Phi\equiv\zeta_2+c_0$, where the four-form component is just zero, 
and thus trivially integrates to zero on $M\setminus M_0$. 
Integrating by parts again gives \eqref{boundariescancel}.

\section{The analytic solutions of Freedman--Pufu}
\label{subsubsec:FP}

It is instructive to look at the explicit solution of \cite{Freedman:2013oja} which was constructed in the STU model and
with $SO(4)$ isometry. The ansatz for the Euclidean solution is
\begin{align}
\diff s^2&=\e^{2B(r)} \left[ \dd r^2 + r^2 \dd s^2(S^3) \right] \, ,\nn
z^i&=z^i(r)\,,\qquad
\tz^i=\tz^i(r)\,,
\end{align}
and vanishing gauge fields.
We need to solve the equations of motion \eqref{eq:4d_N2_EinsteinEOM} together with the Killing spinor equations \eqref{eq:Euclidean_KSE_epsilon} and \eqref{eq:Euclidean_KSE_tepsilon}. With vanishing gauge fields and the reality constraint \eqref{eq:Reality_Condition},\footnote{By contrast, the solutions in \cite{Freedman:2013oja} allowed for $z^i$, $\tz^i$ to be complex, a point we return to below.} notice that
\eqref{eq:Euclidean_KSE_epsilon} and \eqref{eq:Euclidean_KSE_tepsilon} are equivalent, which is consistent with the fact that effectively we are working with a truncation to an $\cN=1$ theory. 

In order to solve the spinor equations we pick a representation of the Clifford algebra
\begin{equation}
\label{eq:FP_CliffordAlgebra}
	\gamma^\mu = \begin{pmatrix}
	0 & {\sigma}^\mu \\ \overline{\sigma}^\mu & 0
	\end{pmatrix} \,, \qquad \gamma_5 = \begin{pmatrix}
	- \mathbb{I}_2 & 0 \\ 0 & \mathbb{I}_2
	\end{pmatrix} \, ,
\end{equation}
where $\sigma^\mu = (\vec{\sigma} , - \ii \mathbb{I}_2 )$, $\overline{\sigma}^\mu = ( \vec{\sigma}, \ii  \mathbb{I}_2)$
and
$\vec{\sigma}$ are the Pauli matrices. We then 
introduce an orthonormal frame 
\begin{equation}\label{fpframe}
	\e^i = r\smallspace \e^{B(r)} \hat{\e}^i \,, \qquad \e^4 = \e^{B(r)} \dd r \, ,
\end{equation}
where $\hat{\e}^i$ is a dreibein on $S^3$ and
write the four-component spinor as
\begin{equation}
	\epsilon = \begin{pmatrix}
		\epsilon_u \\ \epsilon_l
	\end{pmatrix} \, .
\end{equation}

After some work, we recover the two ``branches'' of supersymmetric solutions of \cite{Freedman:2013oja}.
For both branches the metric function is given by \begin{equation}
	\e^{2B} \equiv 4 \frac{( 1 + c_1 c_2 c_3)( 1 + c_1 c_2 c_3 r^4)}{(1-r^2)^2(1+ c_1 c_2 c_3 r^2)^2} \, ,
\end{equation}
where $c_i$ are real integration constants. The coordinate range for the radial coordinate is taken to be $r\in [0,1)$, so that the solutions are topologically $\R^4$ with the $AdS_4$ boundary at 
$r\to 1$. 
In addition, for the two branches we have
\begin{align}
\label{eq:FP_LowerBranch_Solution}
	\text{positive branch:} \quad 
	z^i &= - \frac{c_1c_2c_3 ( 1 - r^2)}{c_i(1 + c_1 c_2 c_3 r^2)} \, , \qquad
	\tz^{\tilde{i}} = \frac{c_i ( 1 - r^2)}{1 + c_1 c_2 c_3 r^2} \, , \nonumber \\
	\epsilon &= \frac{\left( 1 + c_1 c_2 c_3r^4 \right)^{1/4} }{\sqrt{ (1 - r^2)( 1 + c_1 c_2 c_3 r^2) } } \begin{pmatrix}
		r\chi_+ \\ - \ii \chi_+
	\end{pmatrix} \, , 
	\end{align}
	and
	\begin{align}
\label{eq:FP_UpperBranch_Solution}
	\text{negative branch:} \quad 
	z^i &= \frac{c_i ( 1 - r^2)}{1 + c_1 c_2 c_3 r^2} \, , \qquad
	\tz^{\tilde{i}} = - \frac{c_1c_2c_3 ( 1 - r^2)}{c_i(1 + c_1 c_2 c_3 r^2)} \, , \nonumber \\
	\epsilon &= \frac{\left( 1 + c_1 c_2 c_3r^4 \right)^{1/4} }{\sqrt{ (1 - r^2)( 1 + c_1 c_2 c_3 r^2) } } \begin{pmatrix}
		\chi_- \\ - \ii r \chi_-
	\end{pmatrix} \, , 
\end{align}
where $\chi_\pm$ are Killing spinors on $S^3$ satisfying
\begin{equation}
	\nabla_i^{(S^3)} \chi_\pm = \pm \frac{\ii}{2} \sigma_i \chi_\pm \, .
\end{equation}
Notice that the scalar fields $z^i \leftrightarrow \tz^{\tilde{i}}$ are exchanged in the two branches, but the supersymmetric structure is different.
Also, by examining the scalars at $r=0$ we see that $|c_i|<1$.  For convenience, we normalize the Killing spinors as
\begin{equation}
\label{eq:FP_NormalizationKS}
	\chi^\dagger_\pm \chi_\pm = \sqrt{1+c_1c_2c_3} \, .
\end{equation}

It is now straightforward to compute the bilinears \eqref{eq:Bilinears}. 
For the scalar bilinears we find
\begin{align}\label{spexps}
	S^{(\pm)} &= \frac{(1+r^2) (1 + c_1c_2c_3 r^4)^{1/2}}{(1-r^2)(1 + c_1c_2c_3r^2)} \sqrt{1+c_1c_2c_3}  \, , \nn
	P^{(\pm)}  &= \pm\frac{(1 + c_1c_2c_3 r^4)^{1/2}}{1 + c_1c_2c_3r^2} \sqrt{1+c_1c_2c_3}  \, .
\end{align}
For the R-symmetry vector $\xi$ we compute for the two branches
\begin{align}
\xi^{(\pm)}=-\frac{1}{1+c_1 c_2 c_3}\chi_\pm^\dagger\sigma^i\chi_\pm \hat {\rm E}_i\,,
\end{align}
where $\hat {\rm E}_i$ are the vector fields on the $S^3$ dual to the frame $\hat{\mathrm{e}}^i$ for $S^3$ in \eqref{fpframe}.
A special feature of this class of solutions is that both branches are in fact $\frac{1}{2}$-BPS. For a choice of 
Killing spinor given by a specific Killing spinor $\chi_{\pm}$ on $S^3$ we obtain a Killing vector for the $\pm$ branches
that lies in the $\mathfrak{su}(2)_\pm$ subalgebra of the $\mathfrak{su}(2)_+\times \mathfrak{su}(2)_-\subset \mathfrak{so}(4)$ isometry algebra of the round $S^3$.
Notice that the solutions have more Killing vectors than those that can be constructed out of the Killing spinors;  
the action is equivariantly closed with respect to each of them (as shown in appendix \ref{localnosusy}). However, it is important in order to use the formula \eqref{Fgrav} to use the supersymmetric Killing vector $\xi$. 
Proceeding, we observe that the square norm is
\begin{equation}
	\left\| \xi^{(\pm)} \right\|^2 = \frac{4 \mathit{r}^2 \left( 1 + c_1 c_2 c_3 r^4 \right)}{\left( 1 - r^2 \right)^2 \left( 1 + c_1 c_2 c_3 r^2 \right)^2} (1 + c_1c_2c_3) \, .
\end{equation}
Clearly the fixed point set of any of the above supersymmetric Killing vectors is a nut located at the origin $r=0$, where 
$\left\| \xi^{(\pm)} \right\|^2=0$. From 
\eqref{eq:FP_LowerBranch_Solution} and \eqref{eq:FP_UpperBranch_Solution} we immediately see that the two branches of solutions
are in fact characterized by the $\pm$ chirality of the Killing spinors at the nut. From \eqref{spexps} we also see that 
$P^{(\pm)} \rvert_{r=0} = \pm S^{(\pm)} \rvert_{r=0}$, as expected.

From the knowledge of the bilinears and the scalars, we can compute the boundary data introduced in section \ref{sec:uvirrels}. In particular
\begin{equation}
\begin{aligned}
	4 \pi \ii \sigma^1_{(\pm)} &= \mp \frac{(1+c_1)(1+c_2)(1+c_3)}{1 + c_1c_2c_3} \, , &\qquad 4 \pi \ii \sigma^2_{(\pm)} &= \mp \frac{(1+c_1)(1-c_2)(1-c_3)}{1 + c_1c_2c_3} \, , \\
	4 \pi \ii \sigma^3_{(\pm)} &= \mp \frac{(1-c_1)(1+c_2)(1-c_3)}{1 + c_1c_2c_3} \, , &\qquad 4 \pi \ii \sigma^4_{(\pm)} &= \mp \frac{(1-c_1)(1-c_2)(1+c_3)}{1 + c_1c_2c_3} \, . 
\end{aligned}
\end{equation}
These are indeed constant and satisfy \eqref{eq:Rsymmetry_sigmaI_S3_Simpler2} with the provision that the two different branches of solutions induce the Hopf and anti-Hopf fibrations on $S^3$, that is
\begin{equation}
	\sqrt{2}\pi \ii \sigma^I = \mp L^I_* + \frac{1}{2} \left( \nabla_i L^I \right)_* \left( z^i_{(1)} - \tz^i_{(1)} \right) \, .
\end{equation}
As mentioned in section \ref{subsec:Using_BVAB} and reviewed in detail in appendix \ref{app:SUSYHolRen}, the bulk scalar fields are dual to scalar field theory operators with conformal dimensions $\Delta = 1, 2$. In particular, the sources for the boundary operators of dimension $2$ are proportional to\footnote{Here it is convenient to use the normalization \eqref{eq:FP_NormalizationKS}, which guarantees that $r=1-y+o(y^2)$.}
\begin{align}
	z_{(1)}^1 - \tz_{(1)}^1 &= \mp 2 \frac{c_1 + c_2 c_3}{1+c_1 c_2 c_3} \,, \nn 
	z_{(1)}^2 - \tz_{(1)}^2 &= \mp 2 \frac{c_2 + c_3 c_1}{1+c_1 c_2 c_3} \, , \nn 
	z_{(1)}^3 - \tz_{(1)}^3 &= \mp 2 \frac{c_3 + c_1 c_2}{1+c_1 c_2 c_3} \, .
\end{align}
On the other hand, the sources for the boundary operators of dimension $1$ are proportional to the quantity canonically conjugate to $z_{(1)}^i + \tz_{(1)}^i$. These are computed using holographic renormalization in appendix \ref{subsubsec:Legendre_Transform}. In the absence of gauge fields, the result is simply
\begin{align}
	\Pi_i + \widetilde{\Pi}_i &= \frac{1}{8\pi G_4} ( \cG_{ij})_{(0)} ( \theta_{(1)} - \ii \cA_{y(0)}) \left( z_{(1)}^j - \tz_{(1)}^j\right) \\
	&= \mp \frac{1}{8\pi G_4} \delta_{ij} \left( z_{(1)}^j - \tz_{(1)}^j\right) \, , 
\end{align}
where in the second line we have used $( \cG_{ij})_{(0)} = \delta_{ij}$
and $\cA_{y(0)}=0$ for the STU model (see \eqref{eq:STU_KahlerStuff}), and that one finds $\theta_{(1)}=\mp 1$ for the two branches of solutions. Notice that the sources for the $\Delta=1$ and $\Delta=2$ operators are proportional to each other.

We can also check that our formula \eqref{FP} for the on-shell action agrees with the expression given in \cite{Freedman:2013oja}. 
For the positive branch we should use $b_1 = - b_2$ and substitute $\tz^{\tilde{i}}$ in \eqref{eq:FP_LowerBranch_Solution} into
$\ii \cF(u_+)\rvert_{r=0}$. For the negative branch 
we should use $b_1 = + b_2$ and substitute ${z}^{{i}}$ in \eqref{eq:FP_UpperBranch_Solution} into
$\ii \cF(u_-)\rvert_{r=0}$. In both case we obtain the final result
\begin{equation}
\label{eq:FP_Final_Fgrav}
	F_{\rm grav} = \frac{\pi}{2 G_4} \frac{(1-c_1^2)(1-c_2^2)(1-c_3^2)}{(1+c_1c_2c_3)^2} \, ,
\end{equation}
which matches (6.21) of \cite{Freedman:2013oja}.

We highlight that in the construction of the explicit solutions in \cite{Freedman:2013oja}, the constants $c_i$ arise as integration constants and can be taken to be complex. In our approach using localization, because of the assumption of the reality constraints in \eqref{eq:Reality_Condition}, we have to restrict $c_i\in\R$. On the other hand, the result \eqref{eq:FP_Final_Fgrav} obtained from \eqref{FP} and the solutions \eqref{eq:FP_LowerBranch_Solution}, \eqref{eq:FP_UpperBranch_Solution} reproduce the results in \cite{Freedman:2013oja} even if $c_i\in\mathbb{C}$, which is suggestive that the localization approach can be extended when the reality constraints are relaxed. We discuss further examples in section \ref{sec:six}.

\bibliographystyle{utphys}
\bibliography{biblio}{}

\end{document}